\documentclass[letterpaper,12pt,titlepage,oneside,final]{book}
 
\newcommand{\href}[1]{#1} % does nothing, but defines the command so the print-optimized version will ignore \href tags (redefined by hyperref pkg).
\usepackage{ifthen}
\newboolean{PrintVersion}
\setboolean{PrintVersion}{false}
\usepackage{amsmath,amssymb,amstext} % Lots of math symbols and environments
\usepackage[pdftex]{graphicx} % For including graphics N.B. pdftex graphics driver
\graphicspath{ {./Figures/} }

\usepackage{rotating}

\usepackage[pdftex,pagebackref=false]{hyperref} % with basic options
\hypersetup{
    plainpages=false,       % needed if Roman numbers in frontpages
    unicode=false,          % non-Latin characters in Acrobat’s bookmarks
    pdftoolbar=true,        % show Acrobat’s toolbar?
    pdfmenubar=true,        % show Acrobat’s menu?
    pdffitwindow=false,     % window fit to page when opened
    pdfstartview={FitH},    % fits the width of the page to the window
    pdfnewwindow=true,      % links in new window
    colorlinks=true,        % false: boxed links; true: colored links
    linkcolor=blue,         % color of internal links
    citecolor=green,        % color of links to bibliography
    filecolor=magenta,      % color of file links
    urlcolor=cyan           % color of external links
}
\ifthenelse{\boolean{PrintVersion}}{   % for improved print quality, change some hyperref options
\hypersetup{	% override some previously defined hyperref options
    citecolor=black,%
    filecolor=black,%
    linkcolor=black,%
    urlcolor=black}
}{} % end of ifthenelse (no else)

\usepackage[automake,toc,abbreviations]{glossaries-extra} % Exception to the rule of hyperref being the last add-on package
\usepackage{enumitem}
\let\origdoublepage\cleardoublepage
\newcommand{\clearemptydoublepage}{%
  \clearpage{\pagestyle{empty}\origdoublepage}}
\let\cleardoublepage\clearemptydoublepage

\usepackage{hyperref}
\usepackage{mleftright}

\newglossaryentry{computer}
{
name=computer,
description={A programmable machine that receives input data,
               stores and manipulates the data, and provides
               formatted output}
}

\newglossary*{nomenclature}{Nomenclature}
\newglossaryentry{dingledorf}
{
type=nomenclature,
name=dingledorf,
description={A person of supposed average intelligence who makes incredibly brainless misjudgments}
}

\newabbreviation{aaaaz}{AAAAZ}{American Association of Amateur Astronomers and Zoologists}

\newglossary*{symbols}{List of Symbols}
\newglossaryentry{rvec}
{
name={$\mathbf{v}$},
sort={label},
type=symbols,
description={Random vector: a location in n-dimensional Cartesian space, where each dimensional component is determined by a random process}
}
\makeglossaries
\newcounter{numquote}
\newenvironment{lquote}{%
  \refstepcounter{numquote}%
  \quote}{\unskip~\thenumquote\endquote}

\usepackage{subfig}

\usepackage{arabtex}
\usepackage{utf8}
\setcode{utf8}

\begin{document}

%----------------------------------------------------------------------
% FRONT MATERIAL
% title page,declaration, borrowers' page, abstract, acknowledgements,
% dedication, table of contents, list of tables, list of figures, nomenclature, etc.
%----------------------------------------------------------------------
% T I T L E   P A G E
% -------------------
% Last updated October 23, 2020, by Stephen Carr, IST-Client Services
% The title page is counted as page `i' but we need to suppress the
% page number. Also, we don't want any headers or footers.
\pagestyle{empty}
\pagenumbering{roman}

% The contents of the title page are specified in the "titlepage"
% environment.
\begin{titlepage}
        \begin{center}
        \vspace*{1.0cm}

        \Huge
        {\bf Information Retrieval with Entity Linking}

        \vspace*{1.0cm}

        \normalsize
        by \\

        \vspace*{1.0cm}

        \Large
        Dahlia Shehata \\

        \vspace*{3.0cm}

        \normalsize
        A thesis \\
        presented to the University of Waterloo \\ 
        in fulfillment of the \\
        thesis requirement for the degree of \\
        Master of Mathematics \\
        in \\
        Computer Science \\

        \vspace*{2.0cm}

        Waterloo, Ontario, Canada, 2022 \\

        \vspace*{1.0cm}

        \copyright\ Dahlia Shehata 2022 \\
        \end{center}
\end{titlepage}

% The rest of the front pages should contain no headers and be numbered using Roman numerals starting with `ii'
\pagestyle{plain}
\setcounter{page}{2}

\cleardoublepage % Ends the current page and causes all figures and tables that have so far appeared in the input to be printed.
 \begin{center}\textbf{Author's Declaration}\end{center}
  
 \noindent
% I hereby declare that I am the sole author of this thesis. This is a true copy of the thesis, including any required final revisions, as accepted by my examiners.
This thesis consists of material all of which I authored or co-authored: see Statement of Contributions included in the thesis. This is a true copy of the thesis, including any required final revisions, as accepted by my examiners.

  \bigskip
  
  \noindent
I understand that my thesis may be made electronically available to the public.

\cleardoublepage

% Statement of contributions
% -------------------------------

\begin{center}\textbf{Statement of contributions }\end{center}

Part of the research conducted in Chapter \ref{Chapter3} is based on the published work \cite{me} co-authored with Negar Arabzadeh and Charles Clarke.

For the work in \cite{me}, Negar Arabzadeh trained the BERT classifier of \ref{chp3:bertClassifier} presented in Chapter \ref{Chapter3}, and did the editing.

Charles Clarke ran parameter tuning for the Pseudo Relevance Feedback (PRF) model presented as part of \ref{sparse_technique} in Chapter \ref{Chapter3}, reviewed \cite{me}, and provided supervision.

\cleardoublepage

% A B S T R A C T
% ---------------

\begin{center}\textbf{Abstract}\end{center}

Despite the advantages of their low-resource settings, traditional sparse retrievers depend on exact matching approaches between high-dimensional bag-of-words (BoW) representations of both the queries and the collection. As a result, retrieval performance is restricted by semantic discrepancies and vocabulary gaps. On the other hand, transformer-based dense retrievers introduce significant improvements in information retrieval tasks by exploiting low-dimensional contextualized representations of the corpus. While dense retrievers are known for their relative effectiveness, they suffer from lower efficiency and lack of generalization issues, when compared to sparse retrievers. For a lightweight retrieval task, high computational resources and time consumption are major barriers encouraging the renunciation of dense models despite potential gains. In this work, I propose boosting the performance of sparse retrievers by expanding both the queries and the documents with linked entities in two formats for the entity names: 1) explicit and 2) hashed. A zero-shot end-to-end dense entity linking system is employed for entity recognition and disambiguation to augment the corpus. By leveraging the advanced entity linking methods, I believe that the effectiveness gap between sparse and dense retrievers can be narrowed. Experiments are conducted on the MS MARCO passage dataset using the original qrel set, the re-ranked qrels favoured by MonoT5 and the latter set further re-ranked by DuoT5. Since I am concerned with the early stage retrieval in cascaded ranking architectures of large information retrieval systems, the results are evaluated using recall@1000. The suggested approach is also capable of retrieving documents for query subsets judged to be particularly difficult in prior work. In addition, it is demonstrated that the non-expanded and the expanded runs with both explicit and hashed entities retrieve complementary results. Consequently, run combination methods such as run fusion and classifier selection are experimented to maximize the benefits of entity linking. Due to the success of entity methods for sparse retrieval, the proposed approach is also tested on dense retrievers. The corresponding results are reported in MRR@10.

\cleardoublepage

% A C K N O W L E D G E M E N T S
% -------------------------------

% \begin{center}\textbf{Acknowledgements}\end{center}

% % I would like to thank all the little people who made this thesis possible.
% First and foremost, I would like to praise the Almighty Allah for all His blessings during my studies.

% I would like to express my sincere gratitude to my supervisor Charles Clarke for his support and guidance throughout my Master's program. 

% I would like to thank my readers Gordon Cormack and Mark Smucker for their feedback, my colleagues Negar, Nicole and Linhnhi for their help and encouragement.

% Finally, I am mostly grateful to my parents and my sister yasmine for their constant love and backing during the whole journey of my studies. I am also grateful to my friend Menna
% for her encouragement and support.

% This research was enabled in part by support provided by Compute Ontario (\href{https://www.computeontario.ca/}{computeontario.ca}) and the Digital Research Alliance of Canada 
% (\href{https://alliancecan.ca/en}{alliancecan.ca})

% \begin{RLtext}

% \centerline{(
%  وَقَالَ رَبِّ أَوْزِعْنِي أَنْ أَشْكُرَ نِعْمَتَكَ الَّتِي أَنْعَمْتَ عَلَيَّ وَعَلَى وَالِدَيَّ وَأَنْ أَعْمَلَ صَالِحًا تَرْضَاهُ 
% }
% \centerline{وَأَدْخِلْنِي بِرَحْمَتِكَ فِي عِبَادِكَ الصَّالِحِينَ) [سورة النمل: ٩١]}

% \end{RLtext}
% \cleardoublepage

% D E D I C A T I O N
% -------------------

% \begin{center}\textbf{Dedication}\end{center}

% \centerline{\textit{To my parents and yasmine.}}

% \cleardoublepage

% T A B L E   O F   C O N T E N T S
% ---------------------------------
\renewcommand\contentsname{Table of Contents}
\tableofcontents
\cleardoublepage
\phantomsection    % allows hyperref to link to the correct page

% L I S T   O F   F I G U R E S
% -----------------------------
\addcontentsline{toc}{chapter}{List of Figures}
\listoffigures
\cleardoublepage
\phantomsection		% allows hyperref to link to the correct page

% L I S T   O F   T A B L E S
% ---------------------------
\addcontentsline{toc}{chapter}{List of Tables}
\listoftables
\cleardoublepage
\phantomsection		% allows hyperref to link to the correct page

% Change page numbering back to Arabic numerals
\pagenumbering{arabic}

%----------------------------------------------------------------------
% MAIN BODY
% We suggest using a separate file for each chapter of your thesis.
% Start each chapter file with the \chapter command.
% Only use \documentclass or \begin{document} and \end{document} commands in this master document.
% Tip: Putting each sentence on a new line is a way to simplify later editing.
%----------------------------------------------------------------------
%======================================================================
\chapter{Introduction}
%======================================================================

Multi-stage ranking pipelines represent a pivotal transition in Information Retrieval (IR). Early stage retrieval, also known as the recall stage or first stage, aims to find all potentially relevant documents to a query from large collections using inexpensive and efficient ranking models. The retrieved candidate document pool is then forwarded to later re-ranking stages that employ more complex rankers~---~often neural architectures based on contextualized pre-trained transformers~---~for refinement and pruning. This cascaded ranking pipeline has proved to be highly practical in both academia \cite{optimcascade, Wang2011ACR} and industry \cite{ecommerce, wang2020cold, chen2019behavior}. 

The objective of the first stage is to efficiently recall a large pool of documents related to the information need. Sparse retrievers such as BM25, with WAND query processing \cite{wand}, have long prevailed over other retrievers in this stage thanks to their simple logic, inverted index mechanism for large-scale corpora, low requirement of training data, generalization capabilities across different datasets, cost-efficiency, scalability and lower latency. Nonetheless, classical sparse retrievers suffer from the longstanding vocabulary mismatch problem \cite{mismatchavoidance, mismatchir} since they calculate the relevance score by relying on heuristics defined over the exact lexical matching between the queries and the collection. Traditional term-based retrievers use sparse, high-dimensional, bag-of-words (BoW) representations to perform the matching neglecting vocabulary ambiguities and semantic nuances such as synonymy and polysemy. They also fail to capture document semantics because they ignore order dependencies between the terms \cite{semanticmatch}.
Due to these shortcomings, there has been increased research interest in adopting dense retrievers for first-stage ranking.

Pre-trained transformer-based dense retrievers offer a significant performance improvement by mapping queries and documents to dense low-dimensional embedding-based contextualized representations in order to softly match query-document pairs beyond the explicit text surface form \cite{gao2021complementing, khattab2020colbert}. Despite their ability to outperform sparse retrievers, dense retrievers require greater computational resources and a large training corpus, with perhaps hundreds of thousands of labels \cite{msmarco}. In addition to their inability to detect token-level matches \cite{arabzadeh2021predicting}, they also struggle with higher latency issues, lack of generalization \cite{generalization} and lower efficiency compared to classical sparse models. This efficiency-effectiveness trade-off between sparse and dense retrievers often limits the adoption of the costly dense retrievers to later re-ranking stages, with a relatively smaller number of retrieved documents, while sparse retrievers such as BM25 are prioritized in the first stage of cascaded ranking systems. Although some approaches suggest leveraging dense retrievers in the early ranking stage, these methods are usually conditioned by hybrid paradigms to maintain efficiency, either by supplementing the sparse retriever by semantic information generated by a dense retriever \cite{gao2021complementing}, interpolating relevance scores of both retrievers \cite{Kuzi2020LeveragingSA}, \cite{lin2020distilling}, \cite{luan2021sparse},
or intelligently selecting the best retriever using a trained classifier \cite{arabzadeh2021predicting}.

Efforts were made to overcome the limitations of sparse retrievers such as query expansion \cite{qexpansion}, document expansion \cite{nogueira2019document}, \cite{Nogueira2019FromDT}, topic models \cite{topicmodel}, translation models \cite{translation} and term dependency models \cite{termdependence}. Nonetheless, advances in this area were relatively slow in contrast with later re-ranking stages that experienced numerous transformations in the last decade \cite{Cai2021SemanticMF}. 
In another context, entity linking, an important task of NLP, has been revolutionized in terms of scalability, efficiency and accuracy by recent advances in pre-trained transformer-based architectures. In this work, I aim to leverage the development of entity linking systems to expand the collection with relevant entity names in an attempt to reduce imminent semantic gaps preventing document retrieval for later re-ranking stages.

The objective is to demonstrate that in the ``age of muppets'' \cite{zhang-etal-2021-learning-rank}, sparse retrievers can still hold a solid performance boosted by the novel semantic linking systems. Hence, it is possible to shrink the effectiveness gap between sparse and dense retrievers. The effect of corpus expansion with linked entities is experimented using both sparse and dense retrievers. As a sparse retriever, BM25 is chosen, being a BOW state-of-the-art TF-IDF-like retrieval model and a strong baseline that is hard to beat. As for dense retriever, the STAR-ADORE \cite{dphard} pipeline is employed thanks to its efficient performance compared to its dense counterparts.

The experiments are conducted on the MS MARCO passage dataset \cite{msmarco} focusing on the early stage retrieval and using three types of relevance judgments: the original qrels, a set generated from matches favoured by MonoT5, and the later set further re-ranked by DuoT5 \cite{Pradeep2021TheED}. The proposed methods also retrieve relevant query-document matches that were not identified in the non-expanded version of the three so-called Chameleons sets of obstinate queries from MS MARCO~\cite{hardqueries}. My best reported results with the sparse retriever beat standard BM25 results by adopting two methods: classifier selection and Reciprocal Rank Fusion (RRF) \cite{rrf} between the original and the entity-aware runs. While I was successful in improving the recall of BM25 with corpus augmentation, there was no significant change in the overall performance of the dense retriever.

To emphasize the worth of the problem under study. Let's examine an example from the MS MARCO dataset. Below is one of those queries labelled as \textit{``hard"} and the corresponding answer passage that was retrieved by BM25 as a potential match.
\begin{lquote}
\textbf{Query:} \textit{who are in the eagles} 

\textbf{Passage: }\textit{The scratch of an eagle in a dream means a sickness. A killed eagle in a dream means the death of a ruler. If a pregnant woman sees an eagle in her dream, it means seeing a midwife or a nurse. In a dream, an eagle also may be interpreted to represent a great ruler, a prophet or a righteous person. Eagle Dream Explanation  The eagle symbolizes a strong man, a warrior who can be trusted neither by a friend nor by a foe. Its baby is an intrepid son who mixes with rulers.}
\label{quote:one}
\end{lquote}

Since BM25 is a BOW sparse retriever, it selects the above passage as an answer due to the shallow exact match of the word \textit{``eagle"} in the text of both the query and the passage. Nonetheless, one can clearly see that the passage is unrelated to the given query. Augmenting the query text with a semantic description of the word \textit{``eagle"} can help reduce the ambiguity and improve the retrieval's accuracy. Hence, I approach the problem under consideration by expanding the corpus (i.e. queries and passages) of meaningful short descriptions (i.e. linked entity names) in an attempt to reduce possible vocabulary gaps and improve recall.

\section{Contributions}

My contributions can be summarized as follows: 1) Selection of a fast end-to-end encoder-based zero-shot entity linking model used for short questions and extending the model's functionality to fit longer passages. 2) Wikification of the MS MARCO passage dataset (queries and passages) using the modified entity linking system. 3) Query and document expansion using retrieved entity names in two forms: a) explicit and b) hashed. 4) Run selection between the non-expanded and the two entity-equipped runs (word and hashed forms) to determine the maximum recall gain achieved by entity linking with BM25. 5) Run fusion achieving higher recall@1000 results in comparison with the original BM25 on MS MARCO development set and the three sets of hard queries \cite{hardqueries}. 6) Experimentation with three different types of qrels as a ground truth: original qrels, MonoT5 qrels and DuoT5 qrels to cover for possible shortcomings in the original set. 7) Computation of the maximum possible recall gain that BM25 can achieve with the help of linked entities using a perfect hypothetical selection method between runs. 8) Study and evaluation of the entity linking effect using dense retrievers (i.e. STAR-ADORE pipeline).

\section{Thesis Organization}

The thesis is organized as follows. Chapter \ref{Chapter2} gives an overview of the history of sparse and dense retrievers, and presents a general literature review of the entity linking problem including its sub-tasks: entity recognition and entity disambiguation. This chapter also summarizes prior research that was conducted at the intersection of entity linking and information retrieval. The proposed methods are presented for both metric and dataset choices, sparse and dense retrievers selection, entity linking system modifications, corpus expansion, run selection and fusion strategies in Chapter \ref{Chapter3}. Experiments and results are demonstrated in Chapter \ref{Chapter4}. Research gaps and open problems are also discussed. Chapter \ref{Chapter5} concludes the thesis with a reflection on possible future works.

%======================================================================
\chapter{Background}
\label{Chapter2}
%======================================================================
\section{Preliminary Terminology}

\begin{enumerate}
    \item \textbf{Entity}
    
    According to \cite{ROSALESMENDE}, there is no universal consensus on the description of an \textit{entity} in the entity linking field. More generally, an entity is any concrete or abstract object with an identifiable presence and independent existence. In the context of entity linking, identifying an object as an \textit{entity} solely relies on the underlying knowledge base (KB) \cite{moller2021survey}.

     \item \textbf{Named Entity}
     
     Named Entity is a specific entity that can be referred to using rigid designators \cite{MONTECOOK} such as proper names. Named entities can consist of single words or multi-word expressions. 

      \item \textbf{Entity Mention}
      
      It is a raw text segment that designates an entity and is used to associate with the actual entity name in the knowledge base (KB) during the entity linking process.
    
       \item \textbf{Knowledge Base (KB)}

       As per the definition in \cite{kbms}, a knowledge base is a “representation of heuristic and factual information, often in the form of facts, assertions and deduction rules”. Similar to the \textit{entity}’s definition, there is no universal agreement on the definition of a knowledge base. Nonetheless, a KB can be considered as a rich repository storing complex structured and unstructured information in the form of entities, related attributes and mutual relationships \cite{kgir}. Among the notable examples of KBs, one can cite Wikidata \cite{wikidata}, DBpedia \cite{dbpedia}, YAGO (Yet Another Great Ontology) \cite{yago}, ReadTheWeb \cite{readtheweb}, Freebase \cite{freebase}, Probase \cite{Wu2012ProbaseAP}, and KnowItAll \cite{knowitall}. It is possible that an entity mention does not map to a particular entity name in a given KB, and that is why knowledge base population and enrichment is an active research area \cite{Shen2015EntityLW}. It is also worth mentioning that a KB is not always graph-structured, hence not any KB is a knowledge graph (KG) while the converse is true \cite{moller2021survey}.
       
        \item \textbf{Knowledge Graph (KG)}

        Similar to \textit{entity} and \textit{Knowledge base}, there is no uniform definition for a KG in literature. According to Paulheim et al. \cite{Paulheim2017KnowledgeGR}, ``a knowledge graph (i) mainly describes real-world entities and their interrelations, organized in a graph, (ii) defines possible classes and relations of entities in a schema, (iii) allows for potentially interrelating arbitrary entities with each other and (iv) covers various topical domains.” Färber et al. \cite{faerber} define a KG as a Resource Description Framework (RDF) graph consisting of triples of subject, predicate and object. Such a definition is refuted by Ehrlinger and Wöß \cite{Ehrlinger2016TowardsAD} since it restricts KG’s definition to one data model. Although Pujara et al. \cite{pujara} do not provide a formal definition, they explicitly state some of KG’s characteristics. According to their perspective, KGs encompass three fact types: about entities, their corresponding labels and possible relations. The work of Ji et al. \cite{kgsurvey} perceives a KG as a structured representation of facts in the form of entities, relationships and semantic descriptions. Ehrlinger and Wöß \cite{Ehrlinger2016TowardsAD} argue that representing a KG as graph-based knowledge representation is insufficient for an adequate usage of KGs since this definition lacks the minimum requirements a KG needs to satisfy. Although the term \textit{``knowledge Graph”} is not new in literature, its popularity has grown after Google introduced Knowledge Graph in 2012 \cite{Ehrlinger2016TowardsAD, moller2021survey}. In general, a KG represents entities, attributes, and relations through nodes and edges in a graphical structure. Entities are typically represented as vertices, while relationships are represented as edges \cite{kgir}. All KGs are one representation format of KBs. Among the KG known implementations, one can mention Wikidata, DBPedia, Freebase, YAGO, Microsoft’s Satori, Yahoo’s Spark and Google’s Knowledge Vault as described in the work of Ehrlinger and Wöß \cite{Ehrlinger2016TowardsAD}.
    
        \item \textbf{Question Answering (QA)}

        Question Answering (QA) is the process of finding concrete answers to natural language queries. On another hand, Information retrieval(IR) is concerned with fetching documents encompassing the answers to a query regardless of their exact location in the documents \cite{Moldovan2002OnTR}. While IR is originally considered a super-set of the QA task, they have a similar objective in my case since my experiments are conducted on the MS MARCO passage dataset. The passages act as both the retrieved documents (IR) and the exact answers to the user's query (QA).
    
        \item \textbf{Ad-hoc Retrieval}

        It is the standard text-based information retrieval task where a user expresses his information need from a large document collection through a query formulated in natural language. As a result, the search is initiated by information systems with the objective of retrieving the most relevant documents satisfying the user's query.

        \item \textbf{Cascaded Ranking Model}
        
        Cascaded ranking architectures, also known as retrieval and re-ranking pipelines, arise from the need to simultaneously optimize the effectiveness and the efficiency of end-to-end information systems with large document collections \cite{Wang2011ACR}. While efficiency is achieved with the advances in evaluation and caching strategies, effectiveness results from complex ranking functions. Due to the decoupled nature of efficiency and effectiveness, recent advancements in information systems design attempt to balance the trade-off by constructing a cascade of increasingly sophisticated ranking functions that gradually refine the candidate set of retrieved documents, hence maximizing the overall result quality and minimizing the retrieval latency.
    
        \item \textbf{First Stage Ranking}

        This is the first stage of the retrieval and re-ranking pipeline that currently achieves state-of-the-art performance on many information retrieval benchmarks.
        Given a query, the first stage ranking (i.e. the retrieval stage) aims to retrieve ``all potentially relevant documents from a large corpus" according to \cite{Cai2021SemanticMF}. The size of the retrieval set can range between millions and billions. As a result, efficiency is crucial in this stage. The result set is then passed to further subsequent ranking stages in a pipeline. The effectiveness of this stage solely depends on the model choice. A wrong choice may block relevant documents from later stages at the very beginning.

        \item \textbf{Re-ranking Stage}

        The re-ranking stage(s) (a.k.a the later stages) of the cascaded ranking architectures prune and refine the ranked result set that was retrieved from the previous stage. Since only a smaller subset of the results is fed into this stage, the adopted models are usually built using sophisticated ranking architectures to ensure state-of-the-art performance. As per \cite{Cai2021SemanticMF}, the goal is to make ``as many relevant documents as possible top" the final re-ranked list.

        \item \textbf{Indexing}

        Information Retrieval systems require an index in order to maintain the storage and the retrieval speed in large repositories. The indexing scheme choice is pivotal for efficiency. The goal is not to review the different types of indexing strategies found in literature. However, there are two main indexing techniques that are typically employed in information retrieval systems according to the work of Guo et al. \cite{Cai2021SemanticMF}: 
        (1) The Inverted Index: one of the simplest yet efficient indexing schemes. To construct an inverted index, the collection is first parsed and tokenized in order to build the index. The inverted index speeds up the retrieval by fetching the top $K$ similar documents corresponding to a user's query \cite{Cai2021SemanticMF}.
        (2) The Dense Vector Index: which relies on Approximate Nearest Neighbor (ANN) search algorithms. Most of the recent semantic retrieval models represent documents as dense vectors. Consequently, the inverted term index that worked efficiently with the sparse document-term matrix is no longer adequate. ANN mechanism sacrifices a slight drop in precision in favour of major speed improvement \cite{Cai2021SemanticMF}.
        Further information about the different indexing methods can be found in \cite{Malki2016ComprehensiveSA, Roshdi2015ReviewIR, Cai2022SemanticMF}.

        \item \textbf{Pseudo-Relevance Feedback (PRF)}

        Pseudo-Relevance Feedback (PRF), also known as blind relevance feedback, is a method used to improve search engine results \cite{prfirgen}. It was introduced by Robertson and Spärck Jones \cite{Robertson1994SimplePA}. PRF's mechanism relies on expanding the original query set with extra information from a previous search result.  The extracted information is usually the $m$ terms having high Offer Weights (OW) from the top $R$ ranked documents where the Offer Weight is computed as the dot product between term relevance weight and the document frequency  \cite{Zeng2019BM25PR}. There are different implementations of PRF \cite{Zeng2019BM25PR, charles}. In this work, I use the implementation of Büttcher et al. \cite{charles}.
    
\end{enumerate}

%====================================================================================================================================================================

% \section{Research Context}

% In this work, the aim is to benefit from the development the entity linking systems have witnessed in the last decade; and from the omnipresence of fast, efficient and effective end-to-end entity linking tools in order to improve early-stage retrieval. I am concerned with boosting the recall of the initial set of candidate passages/documents in cascaded ranking architectures of large information retrieval systems. To give a background context, the following surveys \cite{Cai2021SemanticMF, ir_survey1, pretrained22, Greengrass2000InformationRA, Nyamisa2017ASO, Sharma2013ASO, Zhang2016NeuralIR} review the history of information retrieval methods: classical traditional models and neural retrievers including sparse, dense and hybrid. 
%  These reviews \cite{li2020survey, ling-etal-2015-design, moller2021survey, Oliveira2021TowardsHE, sevgili2021neural, sharmaamrita, shen2021entity, Shen2015EntityLW, yadav2019survey, roy2021recent} study literature exclusively dedicated to the entity linking task including. The second half is more aligned with my research purpose since it studies prior works exploiting linked entities for information retrieval purposes.

\section{Research Context}

In this work, the aim is to benefit from the development the entity linking systems have witnessed in the last decade; and from the omnipresence of fast, efficient and effective end-to-end entity linking tools in order to improve early-stage retrieval. I am concerned with boosting the recall of the initial set of candidate passages/documents in cascaded ranking architectures of large information retrieval systems. In this chapter, I briefly review the history of information retrieval methods: classical and neural retrievers with the help of six main comprehensive surveys \cite{Cai2021SemanticMF, ir_survey1, pretrained22, Greengrass2000InformationRA, Nyamisa2017ASO, Sharma2013ASO, Zhang2016NeuralIR}. Without deviating from the research goal emphasized earlier, I also think that it is important to provide the readers with enough background of the current research position with respect to the entity linking task. 
I claim that this overview is essential to justify my choice of the tool employed to identify and extract entity mentions from the target corpora. From another perspective, it consolidates my research cause of leveraging entities in retrieval systems by presenting the current state-of-the-art methods in this field, and the potential advancement entity linking can offer when exploited in natural language processing/information retrieval (NLP/IR) related domains. As a result, I cover the entity linking related works in two folds. The first part briefly reviews literature exclusively dedicated to the entity linking task through the lens of previous related surveys and reviews \cite{li2020survey, ling-etal-2015-design, moller2021survey, Oliveira2021TowardsHE, sevgili2021neural, sharmaamrita, shen2021entity, Shen2015EntityLW, yadav2019survey, roy2021recent}. The second half is more aligned with my research purpose since it studies prior works exploiting linked entities for information retrieval purposes.

%====================================================================================================================================================================
\section{Retrieval Methods}

Numerous prior works \cite{Cai2021SemanticMF, ir_survey1, pretrained22, Greengrass2000InformationRA, Nyamisa2017ASO, Sharma2013ASO, irbooklin, Zhang2016NeuralIR} have comprehensively studied the classification of information retrievers. For example, one of the earliest works in this context is the survey of Faloutsos and Oard \cite{ir_survey1}. The latter categorizes the main retriever types into two main classes: traditional types (i.e. full scanning, signature files, inverted index, vector modeling and clustering), and semantic methods such as NLP-based parsers, Latent Semantic Indexing (LSI)  and neural models. Greengrass \cite{Greengrass2000InformationRA} divides early retrieval methods into statistical and semantic techniques. According to \cite{Greengrass2000InformationRA}, statistical methods cover boolean approaches; vector space modeling (VSM) such as LSI, similarity measures and n-gram vectorization; and probabilistic methods such as Bayesian networks and Binary Independence Model (BIM).  In contrast, the semantic methods rely on NLP-related methods, clustering, query expansion, result fusion and user interaction. Also tailored to the classical models, the work of Sharma and Patel \cite{Sharma2013ASO} highlights that the ``fundamental" retrievers incorporate probabilistic, boolean and vector space models. The authors also review two indexing schemes: inversion indices and signature files. Additionally, Nyamisa et al. \cite{Nyamisa2017ASO} base their classification of the traditional retrievers on the dimensionality degree of the models. The authors explore two types of retrievers according to this categorization: mathematical-based first dimension models and model-properties-based second dimension models. Examples from the first type include boolean models and their extension, fuzzy retrieval models, VSM with variations, LSI, BIM, probabilistic relevance models and uncertain inference models \cite{Nyamisa2017ASO}. The second category covers models with no, immanent or transcendent term interdependencies \cite{Nyamisa2017ASO}.  More recently, Zhang et al. \cite{Zhang2016NeuralIR} review neural network architectures for ad-hoc retrieval and study the notable query expansion approaches. The review of Fan et al. \cite{pretrained22} is oriented towards pre-trained models employed in IR. The authors categorize existing retrieval methods in this context into sparse, dense and hybrid retrievers. They also explore the current research concerning negative sampling strategies, joint learning approaches, and generalization capabilities. In addition, they classify the pre-trained models adopted in the re-ranking stage into discriminative, generative and hybrid ranking models. Furthermore, they investigate pre-trained architectures for query processing (i.e. expansion, reformulation, rewriting etc.), user understanding and document summarization \cite{pretrained22}. Aligned with the prior work, Lin et al. \cite{irbooklin} present an overview of transformer-based text architectures dedicated to re-ranking in multi-stage cascaded IR systems, and dense retrievers that are employed for direct ranking. This work also reviews existing literature focusing on refining and learning query and document representations. The comprehensive review of Guo et al. \cite{Cai2021SemanticMF} offers a more thorough study combining classical and advanced IR methodologies. The survey divides existing works in the IR domain into classical, semantic and neural retrieval covering sparse, dense and hybrid paradigms. The authors also address indexing techniques, first stage ranking pre-training methods, learning strategies and evaluation benchmarks.
In general, retrieval methods have been revolutionized in the last few decades starting from the classical term-based retrieval methods based on the bag-of-words (BOW) representations of documents and queries. Then come semantic retrieval with query reformulation and document expansion to compensate for the shallow matching shortcomings of the term-based models. Finally, neural semantic retrieval models with transformer-based architectures bring a new era to end-to-end ranking systems thanks to their ability of low dimensional vector representations and deep semantics learning. 

The next few lines present a synthesis of the classification of the retrieval methods combining the previously described surveys \cite{Cai2021SemanticMF, ir_survey1, pretrained22, Greengrass2000InformationRA, Nyamisa2017ASO, Sharma2013ASO, Zhang2016NeuralIR}. In general, retrieval methods can be classified into two fundamental classes: 1) Classical and 2) Neural approaches.

\subsection{Classical Term-based Retrieval}

The classical term-based retrieval methods rely on the sparse BOW representations of both the queries and the documents. The text of a query or a document is represented as a container of its terms ignoring grammatical rules and the relative order between sentence words. Usually, the inverted index is the most efficient index for this type of retriever. Nonetheless, these models suffer from the vocabulary mismatch problem due to the shallow lexical matching \cite{me}. Classical term-based techniques can be divided into two main categories: statistical and semantic approaches.

\subsubsection{Statistical Methods}

Statistical methods are also known as mathematical-based models and first dimension models. This type of method relies on applying statistical methods to the pre-processed text words \cite{Greengrass2000InformationRA}. Five main statistical approaches are considered: boolean, fuzzy retrieval, vector space, probabilistic and learning to rank models.

\begin{enumerate}[label*=\arabic*.]
    \item \textbf{Boolean Model}

    The boolean model has been studied in previous reviews \cite{Greengrass2000InformationRA, Sharma2013ASO, Nyamisa2017ASO}. It is a classical bag-of-words model that is based on set theory and boolean algebra, and also suffers from the hard matching limitations \cite{Nyamisa2017ASO}. One can distinguish two types of boolean models: standard and extended where term weights are taken into account \cite{Greengrass2000InformationRA}.
    
    \item \textbf{Fuzzy Retrieval Model}

    The fuzzy retrieval model is rarely mentioned in prior works: only one survey \cite{Nyamisa2017ASO} addresses this model among those under study. It is based on fuzzy logic and related membership functions. Per the description in \cite{Nyamisa2017ASO}, the fuzzy set theory permits the ``manipulation" of term weights in an IR-based system.
    
    \item \textbf{Vector Space Model}

    The Vector Space Model (VSM) \cite{Salton1975AVS} is one of the earliest term-based models that have represented the text of each query or a document as a sparse vector in a common multi-dimensional space. VSM-related literature has been thoroughly studied in prior surveys \cite{ir_survey1, Greengrass2000InformationRA, Sharma2013ASO, Nyamisa2017ASO, Cai2021SemanticMF, pretrained22}. Each dimension in the high-dimensional vector maps to a particular vocabulary term. The term-document matrix is then generated by stacking document vectors in a matrix format where rows represent term vectors across all the documents, and columns represent document vectors across all the terms. Each dimension weight is calculated using functions such as term frequency (TF), inverse document frequency (IDF) or a combined TF-IDF \cite{Ramos2003UsingTT, Nyamisa2017ASO, Cai2021SemanticMF}. The relevance score between a query vector and a document vector is determined using a similarity measure \cite{Sharma2013ASO, Nyamisa2017ASO} such as cosine similarity, Jaccard distance, Kullback-Leibler divergence (KLD) or Euclidean distance. The resultant score is then used to identify the top relevant documents to the given query. The work of Greengrass \cite{Greengrass2000InformationRA} presents a detailed description of the methods concerning building the vector space model, normalizing the term vectors, choosing the term vector weighting scheme, and calculating the similarity between query-document pairs. Two common types of VSM are  the Generalized Vector Space Model (GVSM) and the Enhanced Topic-Based Vector Space Model \cite{Nyamisa2017ASO}. Apart from the n-gram-based VSM that was employed in literature on multiple occasions \cite{Greengrass2000InformationRA}, Latent Semantic Indexing (LSI) \cite{Deerwester1990IndexingBL} is considered to be the most popular VSM \cite{Nyamisa2017ASO, Greengrass2000InformationRA, ir_survey1}.

    \item \textbf{Probabilistic Model}

    The probabilistic retrieval model can be considered as an ``instantiation of VSM with different weighting schemes" according to Guo et al. \cite{Cai2021SemanticMF}. Probabilistic models have been investigated in four reviews \cite{Sharma2013ASO, Greengrass2000InformationRA, Nyamisa2017ASO, Cai2021SemanticMF} among the ones under consideration. They are one of the earliest model types that rank documents in decreasing order of their calculated relevance probability to a user's query \cite{Sharma2013ASO}. 
    A number of previous works \cite{probirs, probirs2, FUHR198955} have reviewed specifically probabilistic models in IR. I leave the references to the readers for further exploration. But in general, probabilistic models can be further divided into Bayesian models, language models and uncertain inference models.

     \begin{enumerate}[label*=\arabic*.]
            \item \textbf{Bayesian Probability Model}

            Bayesian probability model, also known as Probabilistic Relevance Model \cite{bm25beyond}, relies on  the conditional probability. This category of models follows Bayes Theorem of calculating the posterior probability as the fraction of the product of the likelihood and prior probability over the evidence\footnote{\url{https://machinelearningmastery.com/bayes-theorem-for-machine-learning/}}.
            In IR, the prior probability refers to that of the document relevance, and the posterior probability designates that of relevance as explained in \cite{Greengrass2000InformationRA}. Examples of this type include: 1)
            Binary Independence Model (BIM) \cite{Robertson1976RelevanceWO} is one of the first original models that were introduced in the probabilistic IR field. It maps the queries and documents to binary vectors where 0 or 1 refer to the term's absence or presence respectively.  The main drawbacks of this model according to \cite{Cai2021SemanticMF, Nyamisa2017ASO, Greengrass2000InformationRA} are (a) the assumption that the documents are binary vectors: only the absence or the presence of a term is stored regardless of other important document-related information; (b) the assumption that the terms are independently distributed in the sets of both relevant or irrelevant documents.  Due to these shortcomings, a number of extensions were suggested to compensate for BIM constraints. 2) BM25 is a modified extension to BIM that is also based on the probabilistic retrieval framework \cite{bm25beyond}. In addition to the term's presence or absence, BM25 considers document length, and both document and term frequencies. BM25 is currently a strong BOW representative of the classical term-based models.
            
            \item \textbf{Language Model}

            Also based on probability distributions, language modeling (LM) for IR has been also studied in prior works \cite{lmiro, lmir2, lmir3}. The core idea of this modeling approach is to construct an LM for each document based on the BOW assumption. The documents are then ranked according to the likelihood of generating the required query \cite{Cai2021SemanticMF}. The results in \cite{lmiro} have demonstrated that the language-model-generated term weights outperform the typical TF-IDF weights. Existing literature exploring LM for IR is summarized in \cite{Nyamisa2017ASO, Cai2021SemanticMF}.

            \item \textbf{Uncertain Inference Model}

            The uncertain Inference Model is also a type of probabilistic model. According to \cite{Nyamisa2017ASO}, the uncertain inference model is concerned with ``formally defining" the relationship between queries and documents in the IR field. The core idea of this model is that the query is converted to a group of assertions based on which an IR system should be able to deduce if these assertions are valid given a particular document \cite{Nyamisa2017ASO}.
            
        \end{enumerate}

    \item \textbf{Learning To Rank}

    Learning to rank (LTR) methods are feature-based retrieval models. Features can vary between query-based (such as the length of a query), document-based (such as the length of a document) or a combination of the two mentioned features \cite{pretrained22}. As per the classification of Fan et al. \cite{pretrained22}, LTR methods are classified into pointwise, pairwise and listwise approaches according to the number of documents taken into consideration.
    
\end{enumerate}

\subsubsection{Semantic Methods}

In order to improve classical term-based retrieval and compensate for the vocabulary mismatch issue, early research studies have been oriented towards enriching the sparse BOW query and/or document representations with additional information, either from external resources or from within the collection. Most of these enriching methods are still adopted in conjunction with the classical term-based models representing text in sparse high-dimensional vectors. Hence, they are easily embedded into the inverted index for an effective retrieval \cite{Cai2021SemanticMF}. The concepts for some of the augmentation techniques are briefly reviewed in the next lines.

\begin{enumerate}[label*=\arabic*.]

\item \textbf{NLP-based}

Natural Language Processing (NLP) retrieval methods refer to matching queries and documents based on the syntactic and semantic aspects of query and document texts. Related works are explored in prior reviews  \cite{ir_survey1, Greengrass2000InformationRA}. Greengrass \cite{Greengrass2000InformationRA} identifies distinct levels of NLP-based IR methods: morphological, lexical, syntactic, semantic, discourse and pragmatic levels. Methods in this context vary between sentence identification, parsing, word sense disambiguation (WSD), concept matching, part-of-speech identification, stemming, stop-words removal, semantic description and information extraction \cite{ir_survey1, Greengrass2000InformationRA}.

\item \textbf{Clustering}

Clustering of documents is considered an early semantic retrieval method. It is comprehensively reviewed in \cite{ir_survey1, Greengrass2000InformationRA}. Clustering is a form of classification that distributes documents into clusters or containers based on their intrinsic features such as topics, statistical characteristics or even the writing language. According to \cite{Greengrass2000InformationRA}, clustering methods can be divided into hierarchical, heuristic and incremental approaches. It is also worth mentioning that the work of Faloutsos and Oard does not consider clustering as a semantic method contrary to Greengrass's survey. I follow the latter's convention in classifying the retrieval-related clustering as semantic approaches.

\item \textbf{Result Fusion}

Merging multiple retrieval results together is also considered one of the earliest semantic retrieval techniques. Works reviewing previous literature in this context are not numerous. Only the survey of Greengrass \cite{Greengrass2000InformationRA} addresses fusion as a semantic retrieval method among the considered reviews. According to his perception, result fusion approaches can be grouped into three classes: result fusion from multiple collections, result fusion from different IR methods and fusion of results generated by multiple versions of the same IR method \cite{Greengrass2000InformationRA}.

\item \textbf{User Interaction}

Similar to result fusion, there are not many reviews exploring prior works leveraging user interaction as a semantic retrieval method. Greengrass \cite{Greengrass2000InformationRA} addresses interaction from the user's perspective and reviews methods focusing on searching retrieved documents, browsing collections and interactively searching collections with a determined information need (i.e. directed searching). The widely known example of interactive and directed searching is Relevance Feedback \cite{cRocchio}, which is also considered as a query expansion method (as will be later explained). Ruthven \cite{interactiveir} investigates existing user-interaction-based IR methods for query formulation and reformulation, complex query languages, clustering, categorization, automated search and assistance, and implicit feedback and evidence.

\item \textbf{Query Processing and Reformulation}

Query processing methods for early semantic retrieval include query rewriting, query expansion and query refinement.

\begin{enumerate}[label*=\arabic*.]

\item \textbf{Query Rewriting}

 Query rewriting aims to reformulate the query under consideration into a easy familiar format by substituting vague terms with clear self-contained words in order to improve the exact matching quality between the queries and the documents \cite{pretrained22}. 
 Greengrass \cite{Greengrass2000InformationRA} perceives query reformulation as part of two previously explained semantic retrieval methods: 1) NLP-based (such as stemming and removing stop-words) and 2) user-interaction-based methods (Such as Relevance Feedback).
 In general, research conducted for query reformulation incorporates both classical and neural approaches.
 Although outside the scope of early semantic methods,
 Fan et al. \cite{pretrained22} review prior literature adopting pre-trained models for query rewriting. Similarly, Zhang et al. \cite{Zhang2016NeuralIR} neural architectures for query auto-completion which is also considered a type of query reformulation.

\item \textbf{Query Expansion}

Query expansion is the process of augmenting the original query text with meaningful terms as a means of improving the effectiveness of document retrieval. The relevant terms are usually selected to reduce the semantic gaps between a query and the potentially relevant documents. Query expansion has been rigorously studied in prior works \cite{Greengrass2000InformationRA, Zhang2016NeuralIR, Cai2021SemanticMF, pretrained22, irbooklin}. Few Surveys \cite{pretrained22, Zhang2016NeuralIR, irbooklin} among the mentioned ones explore specifically neural transformer-based models for query expansion, which are outside of the traditional semantic approaches. I leave them to the readers for exploration.
As highlighted in \cite{Cai2021SemanticMF}, the existing query expansion techniques are categorized into two classes: global \cite{qiuyonggangf} and local methods \cite{CheiangModelbased}. Figure \ref{fig:qe} summarizes the classification of query expansion approaches. On one hand, the global methods reformulate the query by implicitly selecting expansion terms from large corpora or from hand-crafted resources such as WordNet \cite{Voorhees1994QueryEU}. The global analysis can further be divided into four classes depending on data sources and query terms \cite{Azad2019QueryET}: (1) linguistic-based \cite{QEG}, (2) corpus-based \cite{nastsecv}, (3) search log-based \cite{Hua2013ClickageTB}, and (4) web-based \cite{karanmaladen}. On the other hand, Local analysis chooses the expansion terms from the retrieved result set as a response to the original non-expanded query. The logic behind local analysis depends on the fact that since the retrieved documents were deemed relevant to the user's information need, then the terms in these documents should also be relevant. Two main methods are used for local query expansion: (1) Relevance Feedback (RF) \cite{cRocchio} and (2) Pseudo Relevance Feedback (PRF) \cite{Cao2008SelectingGE}. The Relevance Model \cite{rm3} was among the first proposed approaches for PRF, and still holds a state-of-the-art PRF performance compared to other more recent methods \cite{Cai2021SemanticMF}. Despite the performance improvement introduced by query expansion methods, their performance is inconsistent across all the queries \cite{querycontextir}. In addition, PRF is prone to the query drift problem \cite{QEoptimization, Cai2021SemanticMF}.
Prior works \cite{Azad2019QueryET, Carpineto} review intensively query expansion methods, classifications and applications in IR.

\begin{figure}
    \centering
    \includegraphics[scale=0.3]{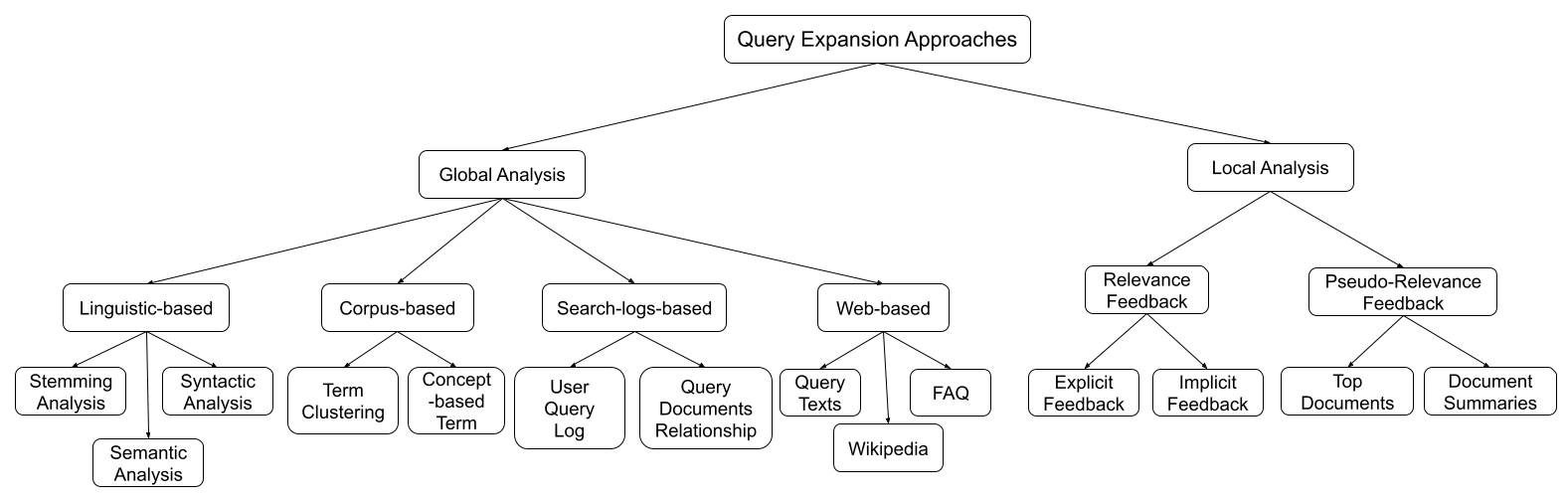}
    \caption{Query Expansion Methods classified per data sources (adapted from \cite{Azad2019QueryET})}
    \label{fig:qe}
\end{figure}

 \item \textbf{Query Refinement}

Query refinement designates term re-weighting per the definition of Greengrass \cite{Greengrass2000InformationRA}. The latter thoroughly reviews relevant literature to this context. Term re-weighting usually follows query rewriting or expansion procedures \cite{Greengrass2000InformationRA}.

\end{enumerate}

\item \textbf{Document Expansion}

Instead of query augmentation, document expansion is performed by enriching the documents themselves with additional term-based information before the indexing procedure. In general, document expansion is less explored in literature when compared to query expansion due to the cost of index regeneration every time the expansion technique changes \cite{Cai2021SemanticMF}. Among prior works that have studied document expansion for IR, Efron et al. \cite{deefron} attempt to compensate for the short texts focusing on a single topic by submitting all the documents in the corpus as pseudo-queries and performing lexical or temporal expansion according to the result set analysis. In addition to the typical document retrieval, some prior works explore different methods for identifying the required documents needed for expansion. For example, Kurland and Lee \cite{Kurland2004CorpusSL} propose a framework for incorporating both corpus-structure information, using overlapping clusters, and individual-document information with the objective of determining similar documents. Liu and Croft \cite{Liu2004ClusterbasedRU} also experiment with cluster-based retrieval using language modeling for IR. Previous literature \cite{sherman, dewordnet} explore document expansion with external collections \cite{Cai2021SemanticMF} such as WordNet \cite{dewordnet} and Wikipedia \cite{sherman}. It is also important to highlight that in this context, I am reviewing document expansion as one of the earliest semantic retrieval methods. As a result, I defer the talk about neural-based document expansion methods such as doc2query \cite{nogueira2019document} and docTTTTTquery \cite{Nogueira2019FromDT} to \ref{sparse_retrival}.

\item \textbf{Term Dependency Modeling}

A major disadvantage of the classical term-based retrieval methods is their inability to capture term order, hence the ``relevance of consecutive terms" between matching queries and documents is ignored \cite{Cai2021SemanticMF}. Term dependency modeling aims to overcome this limitation by integrating term dependencies as latent variables in the representation functions of both the queries and the documents \cite{Cai2021SemanticMF}. Gao et al. \cite{termdependence} build on the unigram-based language model by representing the term dependencies of each query as an undirected graph. Xu et al. \cite{relevnceranking} adopt relevance ranking using term dependency information  (i.e. n-grams and n-dependent terms) by proposing a number of kernel functions as ranking models such as BM25 Kernel, LMIR Kernel and KLDKernel. Srikanth and Srihari \cite{Srikanth2003IncorporatingQT} integrate term dependencies by proposing two new language models for query retrieval: Maximum-Bigram-based and Concept-based. Metzler and Croft \cite{metzlermarkov} develop a framework for term dependency modeling through Markov Random Fields (MRF). Text features such as single terms, and ordered and unordered phrases are incorporated into the dependence model in order to improve its effectiveness. The work of Shi and Nie \cite{shilixin} accounts for each term's strength, utility and impact on the retrieval effectiveness by introducing weighted term dependency modeling.

\item \textbf{Topic Modeling}

Topic modeling for IR is based on extracting semantic relationships between the terms to discover latent topics in texts. Hence, the matching between a query and a document is conducted on their topic representations.
Yi and Allan \cite{xingir} present a comprehensive study of leveraging topic modeling for IR.
In general, topic models can be classified into two sets: probabilistic and non-probabilistic models according to \cite{Cai2021SemanticMF}. Probabilistic topic models are generative statistical models \cite{xingir}. Each topic is represented as a probabilistic distribution over the words of the underlying vocabulary, while each document in the collection is represented as a probabilistic distribution over the topics \cite{Cai2021SemanticMF}.  Latent Dirichlet Allocation (LDA) \cite{lda} and probabilistic Latent Semantic Indexing (pLSI) \cite{pla} are the most known examples of probabilistic models. In contrast,  non-probabilistic models are usually generated using matrix factorization approaches. Examples in this category include Latent Semantic Indexing (LSI) \cite{Deerwester1990IndexingBL}, its variant: Regularized Latent Semantic Indexing (RLSI) \cite{rlsi} and Non-Negative Matrix Factorization (NNMF) \cite{nnmf}. Although LSI has been previously classified as a statistical-based vector space model, it has been also employed as a well-known topic model.
The application of topic modeling for IR is conducted in two forms. One approach is to map both the queries and the documents to a topic space, then the relevance evaluation is computed using the topic representations. In this context, Wei and Croft \cite{topicmodel} propose an LDA-based retrieval model where queries and documents are defined using their topic distributions. The second approach relies on joining the topic models with term-based ones. Hofmann \cite{pla} linearly combines the cosine similarities generated by both topic models (i.e. variants of Probabilistic Latent Semantic Analysis PLSA \cite{plsa}) and term-based models (i.e. VSM). In addition, Ormeño et al. \cite{info12090360} extend LDA-based IR strategies to overcome the parametric sensitivity of LDA using three different ensemble strategies: disjoint partitions, bagging-based corpus sampling and boosting-based corpus sampling.
In addition to exploiting topic modeling for document retrieval, it is also used for query expansion \cite{xingir}.
Despite the rich literature on topic modeling for IR, it is important to mention that the improvement gain introduced by topic modeling is usually minor, and that topic modeling has to be used in conjunction with the classical term-based models to achieve this improvement \cite{Cai2021SemanticMF}. Term-based baselines still outperform the standalone topic models \cite{Atreya2011LatentSI}.

\item \textbf{Translation Modeling}

Statistical translation models are also leveraged for IR in an attempt to overcome the vocabulary mismatch problem of the term-based approaches. A translation model for IR views the queries and documents as two texts in different languages \cite{Cai2021SemanticMF}. Using translation models for retrieval requires learning translation probabilities of the queries to the corresponding relevant documents using supervised machine learning approaches \cite{Aissa2018ARL}. The most notable work in this context is the one of Berger and Lafferty \cite{translation}. The latter introduce the first document-to-query translation model. In order to evaluate a document's relevance to a query, they estimate the probability that the query could be a potential translation of the document, and include this result in the form of a prior distribution over the documents. Liu et al. \cite{liuyuqi} revisit the work of Berger and Lafferty \cite{translation} and replicate the same experiments with MS MARCO passage dataset. They also replace the sum of translation probabilities in the prior work with MaxSim operator of ColBERT \cite{khattab2020colbert}. One of the challenges of the application of translation models in ad-hoc IR is the model estimation without training data. To overcome this issue, Liu et al. \cite{liuyuqi}  employs a larger dataset which is the MS MARCO passage as previously mentioned, and confirm that Berger and Lafferty's model is very effective if equipped with sufficient data. In addition, Karimzadehgan and Zhai \cite{Karimzadehgan2010EstimationOS} exploit normalized mutual information between words to estimate word-to-word translation probabilities. They also suggest regularizing self-translation probabilities to improve the model's retrieval performance. The latter was suggested as a possible solution for the known problem of the self-translation probability estimation in translation modeling for IR. The main difference between conventional statistical translation and translation for IR is that the same language is employed to express both the queries and the documents. Hence, both the target and the source languages are similar \cite{Cai2021SemanticMF}. Generally, The probability of self-translation is supposed to be large in order to correspond to a matching case in the retrieval domain. However, if the self-translation probability is too large, it may reduce the translation effectiveness since other probabilities will be too small in comparison, and they may lose their values. In contrast, if the self-translation probability is too small, this will reduce the effectiveness of the exact matching, hence the retrieval performance \cite{bruce2, Cai2021SemanticMF}. In addition to \cite{Karimzadehgan2010EstimationOS}, other research works \cite{AxiomaticAnalysis, Gao2010ClickthroughbasedTM} have also studied this problem and proposed variants of solutions. From another perspective, statistical translation modeling has been leveraged for query expansion purposes in a number of prior works \cite{bruce2, Gao2012TowardsCT, Riezler2010QueryRU}.  

\end{enumerate}

\subsection{Neural Semantic Retrieval}

The last decade has witnessed a radical breakthrough in deep learning methods which has affected nearly all NLP-related tasks including IR. Deep sophisticated contextualized architectures were leveraged to represent  documents and/or queries in low-dimensional dense vectors, and used to build end-to-end systems that are able to capture semantic and syntactic features. Deep-learning-based retrieval systems achieve the current state-of-the-art performance in IR. To allow a speedy retrieval, textual representations are learned via neural networks, then pre-computed and saved \cite{Cai2021SemanticMF}. This computation depends on the type of the semantic retrieval method. In general, semantic retrieval methods can be classified into three classes: sparse, dense and hybrid retrieval methods. I present an overview of the idea and the notable related works in each of these classes. Among the retrieval surveys under study, only these works \cite{Cai2021SemanticMF, pretrained22, Zhang2016NeuralIR} address deep neural retrieval methods. This review is conducted through the lens of the mentioned works and is not exhaustive. Readers can refer to the original works for further information about model architectures and experimental settings.

\subsubsection{Sparse Retrieval}
\label{sparse_retrival}

The sparse methods represent queries and documents with sparse BOW vectors where a certain number of dimensions is actually used \cite{Cai2021SemanticMF}. Sparse retrieval aims to compensate for the semantic gaps of the classical term-based retrieval methods using four approaches according to \cite{pretrained22}: 1) re-weighting terms with contextual semantics using neural models, which is known as \textit{Neural Weighing Schemes}, 2) mapping query and document texts into a latent space using neural networks, namely \textit{Sparse Representation Learning}. 3) augmenting documents with additional semantic terms, known as \textit{Document Expansion}. 4) learning term weights of all the expanded documents, instead of one document, is achieved through \textit{Combined Expansion and Re-weighting}. I give an overview of the four methods in the next lines. Empirical results show that sparse techniques are able to improve retrieval effectiveness. In addition, they are compatible with the inverted index of the classical methods. Prior works \cite{Cai2021SemanticMF, pretrained22} reviews existing literature for each of these categories in detail. In the next few lines, I list the most notable works. 

Neural weighing explored in literature relies on neural networks designed to learn term weights based on the semantic context instead of employing fixed heuristic functions \cite{Cai2021SemanticMF}. Instead of the traditional retrievers' term dependency approach, pre-trained contextualized term embeddings are leveraged for improved effectiveness \cite{pretrained22}. There is a rich body of literature that explores variants of approaches for term reweighing such as DeepTR \cite{Zheng2015LearningTR}, TDVs \cite{tdvs}, DeepCT \cite{Dai2019ContextAwareST, Dai2020ContextAwareTW}, and  HDCT \cite{Dai2020ContextAwareDT}. 

Sparse representation learning, the second approach studied in literature for sparse retrieval, represents queries and documents as contextualized sparse vectors capturing useful semantics. As a result, documents and queries are mapped to a latent space where each dimension is learnt by neural models with no defined concept. Inverted indices are employed to look up for the learnt sparse embeddings where each index unit designates a latent concept. In this category, one can cite the work of Salakhutdinov and Hinton \cite{SALAKHUTDINOV2009969} where semantic hashing is introduced. The model employs a deep auto-encoder architecture to capture document-term information. Jang et al. \cite{Jang2021UHDBERTBU} propose ultra-high dimensional (UHD) BERT-based \cite{bert} embeddings with high sparsity allowing for binarized representations for efficient storage and search. Zamani et al. \cite{Zamani2018FromNR} build a Standalone Neural Ranking Model (SNRM) by introducing a sparsity feature to learn hidden sparse representations queries and documents. These latent representations are able to uncover semantic relationships between a query and its relevant documents.

Document expansion is the third approach used for sparse retrieval. It relies on enriching document text with extra terms to alleviate semantic disambiguation. As previously underlined for document expansion as an early semantic classical method, document augmentation is not frequent in literature as query expansion \cite{pretrained22}. For example, doc2query \cite{nogueira2019document} trains a sequence-to-sequence (seq2seq) network on relevant pairs of queries and documents. For each document in the collection, the trained model generates corresponding queries that are used for document expansion. Finally, the augmented collection is indexed, and BM25 is employed to retrieve relevant documents to a given query. DocTTTTTquery \cite{Nogueira2019FromDT}, an extension to doc2query, substitutes the seq2seq model with the pre-trained T5 \cite{t5} model achieving a larger performance improvement over its predecessor. In addition, Yan et al.  \cite{Yan2021AUP} propose a Unified Encoder-Decoder network (UED) to simultaneously refine document expansion and document ranking tasks.

The final sparse method depends on weighted document expansion with additional terms using neural sequence-to-sequence (seq2seq) models before indexing. It combines the last two approaches together to learn vocabulary weights instead of a particular document \cite{pretrained22}.
Prior works that have developed a unified framework to concurrently learn term weights and expand documents such as DeepImpact \cite{Mallia2021LearningPI} and SparTerm \cite{sparterm}.

\subsubsection{Dense Retrieval}

Despite the disadvantage of time consumption and the costly computational resource requirement, dense retrieval is known to outperform sparse retrieval methods by a significant margin. The dense retrieval mechanism relies on the dual-encoder architecture, also known as Siamese network \cite{siamese}, consisting of twin networks to independently learn the embeddings of the queries and the documents \cite{Cai2021SemanticMF, pretrained22}. Queries and documents are represented as low-dimensional dense vectors in order to capture global semantics and external context from the input texts. Approximate Nearest Neighbor (ANN) algorithms are often employed to index the learned dense representations \cite{Cai2021SemanticMF, pretrained22}. Dense retrieval models can be divided into two categories depending on the form of the learnt document representations: term-level (Multi-vector) representation learning and document-level (Single-vector) representation learning. The next lines list briefly the known models in each category. For further information, the readers can refer to the original works and surveys \cite{Cai2021SemanticMF, pretrained22}.

Term-level representation learning methods, also known as Multi-vector representation \cite{pretrained22}, represent the queries and the documents in the form of sequences of term embeddings. The overall relevance between a query and a document is computed as an aggregation of the term-level matching relevance scores \cite{Cai2021SemanticMF}. Word embeddings are considered the most common approach for term-level representation methods \cite{pretrained22}. In this context, prior works have experimented with both fixed and contextualized embeddings. Among the works leveraging fixed embeddings, one can mention DESM \cite{Mitra2016ADE}. As for the contextualized embeddings that have recently dominated the NLP world \cite{Cai2021SemanticMF}, prior literature introduces models like DC-BERT \cite{Zhang2020DCBERTDQ}, ColBERT \cite{khattab2020colbert}, COIL \cite{Gao2021COILRE} and PreTTR \cite{MacAvaney2020EfficientDR}. Furthermore, a number of works \cite{Lee2020ContextualizedSR, Seo2018PhraseIndexedQA, phraseseo} extend the term-based approach to explore phrase-level representations for documents such as n-grams. In such paradigms, the query is typically one sentence and is represented as a single vector because of its short length \cite{Cai2021SemanticMF}. 

Document-level representation learning methods, also known as Single-vector representation \cite{pretrained22}, learn coarse-grained global representations of queries and documents by integrating their semantics into dense vectors \cite{Cai2021SemanticMF}. One of the early approaches to produce query and document embeddings is to accumulate their corresponding term encodings using heuristic functions \cite{pretrained22}. Examples in this category include models such as FV \cite{Clinchant2013AggregatingCW}, PV \cite{le2014distributed}, NVSM \cite{Gysel2018NeuralVS}, SAFIR \cite{Agosti2020LearningUK}, DPR \cite{dpr}, ConvDR \cite{Yu2021FewShotCD} and RepBERT \cite{Zhan2020RepBERTCT}. In addition, there is a line of previous works \cite{Liu2016ConstrainingWE, taminelynda} leveraging external knowledge to enhance embedding learning. Another research approach \cite{Tahami2020DistillingKF, distillation} relies on model distillation from its complex architecture (such as term-level multi-vector representation learning) to a document-level single-vector-based learning \cite{Cai2021SemanticMF, pretrained22}. Besides one global query/document representation, a number of models, such as Poly-encoders \cite{Humeau2020PolyencodersAA} and Multi-Vector BERT (ME-BERT) \cite{Luan2021SparseDA}, exploit multiple encoders for queries and documents in order to capture the diverse aspects (especially of a document) into multiple embeddings.

\subsubsection{Hybrid Retrieval}

While sparse retrieval preserves low-level features and performs hard matching, dense retrieval encodes semantic encoding information and adopts a soft matching approach \cite{Cai2021SemanticMF, pretrained22}. Hybrid architectures attempt to get the best of both worlds to ensure the efficiency and the effectiveness of the end-to-end pipeline. Hybrid retrieval methods determine specific functions for both queries and document representations. Sparse and dense encodings are simultaneously learnt for later matching \cite{Cai2021SemanticMF, pretrained22}. Despite the performance gains of the hybrid paradigm, it introduces an overall retrieval complexity and requires larger space \cite{Cai2021SemanticMF}. Among the hybrid models that leverage both the sparse and dense approaches, one can mention GLM \cite{Ganguly2015WordEB}, EPIC \cite{MacAvaney2020ExpansionVP}, DenSPI \cite{phraseseo}, SPARC \cite{sparc} and CLEAR \cite{gao2021complementing}.

%===================================================================================================================

\section{Entity Linking}

Entity linking is the process of identifying relevant entity mentions in raw textual data and disambiguating these mentions by associating/matching them to their corresponding entries in a knowledge base (KB) or linking them to graph nodes in a knowledge graph (KG) \cite{sevgili2021neural}. In this work, I consider that the process of Named Entity linking (NEL) covers the 2 sub-tasks: Named Entity Recognition (NER) and Named Entity Disambiguation (NED). Hence, the focus is on the history of the holistic entity linking process in my literature review. To avoid terminological confusion, I use the term \textit{“entity linking”} to denote joint entity recognition and disambiguation. Nonetheless, a limited number of prior works consider a joint NER and NED since the majority of studies concentrate on the disambiguation procedure assuming that the mentions are provided by external entity recognizers.  I also use the terms KB and KG interchangeably acknowledging the fact that most of the modern KBs arrange information in graphical structures as underlined in \cite{sevgili2021neural}. The selection criteria of an entity mention depends solely on the underlying KB \cite{moller2021survey}. For example, in the case of Wikidata KB, any Wikidata item is considered an entity, and the entity linking process is referred to as \textit{Wikification} in this case. In this work, the methods are restricted to the Wikification process, but I do not limit my review of the related works to Wikidata. It is also important to mention that although I recognize the distinction between Named Entity linking (NEL) and Entity linking (EL) as highlighted in \cite{ling-etal-2015-design}, I use both terminologies interchangeably in this context. While I briefly cover cross-lingual architectures in the literature review, I also limit my methods to the English language and for general domain purposes.

With the huge growth of web knowledge and information in the form of natural language, the development of information extraction (IE) methodologies, and the evolution of reliable knowledge bases (KB), new methods have emerged to leverage KB’s entities in understanding natural language textual data through what is known as the entity linking process. The latter is considered a challenging task because of name variations, multi-worded mentions, syntax errors, contextual discrepancies, multiple surface forms (aliases, abbreviations, nicknames, spelling variations etc.) and the lexical ambiguity of entity mentions. Nonetheless, EL’s success in the development of many ``fields of knowledge engineering and data mining" \cite{shen2021entity} such as KB population, natural language understanding (NLU), semantic parsing, content analysis, question answering (QA), information integration and relation extraction \cite{shen2021entity, sevgili2021neural, Shen2015EntityLW} encourages to further explore the entity impact on information retrieval as a potential application.

Currently, the state-of-the-art deep-network-based entity linking systems with their sophisticated architectures have shown promising results over the outdated handcrafted techniques and the classical machine learning approaches that rely predominantly on shallow architectures and hand-engineered features \cite{sevgili2021neural}. Deep networks are famous for domain adaptation, representation transfer and their ability to learn sophisticated architectures, which alleviates the burden of domain-specific manual labour for the EL task. Nonetheless, the performance of entity linking systems is still bound by the used dataset and the domain type \cite{Shen2015EntityLW}. Variations of the general neural entity linking prototype across literature can be grouped by common modification classes which include holistic entity linking, joint entity linking and disambiguation, global linking, domain independence and cross-lingual entity linking \cite{sevgili2021neural}.

 A significant number of recent works \cite{ling-etal-2015-design, Shen2015EntityLW} review comprehensively the methods, challenges and solutions in this area of knowledge, whether these works focus on the standalone entity recognition module \cite{li2020survey, sharmaamrita, yadav2019survey, roy2021recent}, restrict the term “entity linking” to the entity disambiguation sub-task \cite{moller2021survey,Oliveira2021TowardsHE,shen2021entity,Shen2015EntityLW}, or address the joint task of entity recognition and disambiguation \cite{sevgili2021neural}. Few of these works prioritize the state-of-the-art solutions in their studies \cite{li2020survey,sevgili2021neural,shen2021entity}, while the others offer a more generalized perspective. As previously described, I consider entity linking as the combination of both entity recognition and entity disambiguation tasks. 
 Due to the shortage of prior works that address jointly the two tasks, I review survey papers that handle each sub-task separately as well as those that give a combined overview.

\subsection{Named Entity Recognition (NER)}

Entity recognition is an information extraction (IE) task \cite{Perera2020NamedER} that aims to identify and classify mentions from raw text into pre-categorized semantic types such as person name, names from general domains, location, time etc. It is considered a crucial ``pre-processing step for many downstream applications" such as IR and QA \cite{li2020survey}. Many tools are made available online by academia or industry projects for NER purposes \cite{sharmaamrita} such as Natural Language Toolkit (NLTK) \cite{nltk}, Polyglot \cite{alrfou2014polyglot}, Stanford CoreNLP \cite{Manning2014TheSC}, LingPipe \cite{Carpenter2004PhrasalQW}, AllenNLP \cite{gardner2018allennlp}, ScispaCy \cite{scispacy}, OSU Twitter NLP \cite{osutwitter}, Illinois NLP \cite{2018_lrec_cogcompnlp}, OpenNLP \cite{opennlp} and NeuroNER \cite{dernoncourt2017neuroner}. The work of Vychegzhanin and Kotelnikov \cite{Vychegzhanin2019ComparisonON} presents a comparative evaluation of different NER tools when applied to news articles judging aspects such as target domains, processing techniques, supported languages and recognized entity types; while Atdag and Labatut \cite{biotext} limit the comparative experiments to bibliographical texts. Table 2 in \cite{li2020survey} summarizes the popular off-the-shelf NER tools used for the English language. The NER task includes multiple steps that can be simply handled with the help of these tools. The main NER steps include tokenization, lemmatization, Part-of-Speech (POS) tagging, and chunking \cite{sharmaamrita, roy2021recent}.
NER systems are typically evaluated against human-based judgments \cite{li2020survey}. Related evaluation methods are based on either ``exact-match" or ``relaxed-match", and each class has its own metric set as highlighted in \cite{li2020survey}. NER datasets and related evaluation metrics are beyond the scope of this background review, but a number of prior works \cite{li2020survey,sharmaamrita,yadav2019survey,roy2021recent} serve as good references for this field and its recent trends. As comprehensively reviewed in previous surveys such as the work of Li et al. \cite{li2020survey}; Sharma et al. \cite{sharmaamrita}; and Yadav and Bethard \cite{yadav2019survey}, approaches to NER can be distinguished into six main classes: rule-based methods, unsupervised learning, supervised learning, semi-supervised learning, deep learning and hybrid approaches.

\subsubsection{Rule-based Methods}

Rule-based methods rely on hand-crafted lexical, syntactic and grammatical rules designed by computational linguists for very specific domains and/or patterns to identify entities \cite{li2020survey, sharmaamrita, roy2021recent}. They are also known as knowledge-based/ knowledge-aware systems since they are overfitted to a particular text corpus \cite{yadav2019survey}. One of the most established reviews, I can cite the work of Nadeau and Sekine \cite{Nadeau2007ASO} that discusses in detail the literature history of the methods encompassing hand-engineered rules till the machine learning era. The major shortcomings of these rules are being domain-specific, language-specific and time-consuming to handcraft; inability to transfer to other domains; low recall; and also bad precision in case of the non-exhaustive lexicon. Among the known rule-based NER systems, I can list LaSIE-II \cite{humphreys-etal-1998-university}, NetOwl Extractor \cite{krupka-hausman-1998-isoquest}, RENAR \cite{renar} and SRA \cite{aone-etal-1998-sra}. There is also a rich research work addressing rule-based NER in other languages such as Arabic \cite{arabicner,Salah2017ArabicRN,renar}; Chinese \cite{chinesener}; Greek \cite{Farmakiotou2000RULEBASEDNE} and Urdu \cite{urduner}.

\subsubsection{Unsupervised Learning}

Unsupervised machine learning approaches deal with unlabeled data making the learning behaviour increasingly unpredictable. Clustering-based ER systems are the most known types compared to other unsupervised learning methods \cite{li2020survey}. These systems rely on context similarity to extract entities from text clusters that are grouped based on lexical patterns. In general, the unsupervised methods have emerged to produce additional contextual features to be used along with other NER methods \cite{roy2021recent}. Some of the earliest works in this context suggest the usage of minimal training data. Collins and Singer \cite{Collins1999UnsupervisedMF} show that the use of annotated data decreases the supervision requirement to seven simple “seed” rules including orthography and entity context. They also devise two algorithms: the first exploits redundancy in contextual features for word-sense disambiguation, and the second builds on Adaboost algorithm \cite{adaboost}. Evans \cite{Evans2003AFF} introduces a system for ``Named Entity Recognition in the Open domain" (NERO) relying on hypernym clustering. Etzioni et al. \cite{etzioni} present the KNOWITALL system to automate NER from the web in an unsupervised and domain-independent manner leveraging pattern learning, subclass extraction and instance listing. By addressing the limitations in \cite{Collins1999UnsupervisedMF} and \cite{etzioni}, Nadeau et al. \cite{Nadeau2006UnsupervisedNR} propose a system that can recognize more than three entity types such as person, location, and organization, without human intervention or gazetteer generation. Lin et al. \cite{linwu} build a distributed version of a simpler K-Means clustering, and use the resulting clusters as features in discriminative classifiers to improve the performance of NER systems. Several other works \cite{Zhang2013UnsupervisedBN,shinyama-sekine-2004-named,collier-kim-2004-introduction,uzuner} exploit and even improve unsupervised learning techniques for NER purposes in different scientific domains.

\subsubsection{Supervised Learning}

Supervised learning models learn from labelled data samples to recognize patterns on unseen data. Feature engineering, an important requirement for supervised NER, is explored in a number of prior works \cite{Tkatchenko2012NamedER,Nadeau2007ASO,sharnagat2014named}. Features can be represented using boolean, numeric or nominal attributes \cite{li2020survey, sharmaamrita}; and they are usually categorized according to their type. They typically capture local and external knowledge. The former is extracted from a token (word) and the surrounding context, while the latter includes word clustering, phrasal clustering and encyclopedic knowledge.  Examples of features are case (i.e. upper, lower, camel, etc.), punctuation (i.e. internal apostrophe, hyphen or ampersand), digit patterns (i.e. cardinal, ordinal, etc.), morphology (i.e. prefix, suffix, stem, etc.), Part-of-Speech (POS) tagging (i.e. verb, noun, adverb, etc.), gazetteers (such as general dictionary, organization names, etc.) \cite{li2020survey, roy2021recent}. In the context of NER, feature-based supervised machine learning methods focus on sequence labelling and multi-class classification \cite{li2020survey}. According to \cite{li2020survey}, examples from literature employ Maximum Entropy Models (ME) \cite{Kapur1989MaximumentropyMI}, Naive Bayes classification, Support Vector Machines (SVM) \cite{svm}, Hidden Markov Models (HMM) \cite{Eddy1996HiddenMM}, Decision Trees \cite{decisiontrees}, and Conditional Random Fields (CRF) \cite{crf}. CRF is considered one of the most prominent supervised learning algorithms for NER \cite{roy2021recent}.
Supervised learning models usually fall into two categories: classification and regression \cite{sharmaamrita}

Among the related previous works in this context, Bikel et al. \cite{bikel-etal-1997-nymble} build the first NER system \textit{Nymble} based on HMM. This work is extended in \cite{Bikel2004AnAT} introducing IdentiFinder, a Hidden-Markovian-based NER model that is able to learn and recognize names, dates, times, and numerical quantities. Zhou and Su \cite{zhousu} propose an HMM-based chunk tagger to build a NER model that is able to integrate deterministic local word features such as capitalization, semantic features of important triggers, gazetteer features, and external macro context features. Malouf \cite{malouf} addresses the language-independent NER task by experimenting with both HMM and maximum entropy (ME) models. Szarvas et al. \cite{Szarvas2006AMN} introduce a multilingual NER system that classifies Hungarian and English by adopting AdaBoostM1 and the C4.5 decision tree algorithms. A significant number of works \cite{takeuchi, Li2004SVMBL, mcnamee} train variants of SVM classifiers using different feature classes: linguistic, orthographical and punctuational; and multiple window sizes. Contrary to SVM, CRF takes the context into consideration, hence it is preferred for NER systems in previous works. For example, McCallum et al. \cite{McCallum} propose a feature induction approach for CRFs, and present results on the CoNLL-2003 NER task, consisting of news articles with tagged entities. The work of Liu et al. \cite{hamner} attempts to bridge the gap between NER-based distantly supervised methods and supervised methods by extending the dictionary of automatically annotated data with ``headword based non-exact matching". In their experiments, they choose CRF as their primary model. A number of prior works also leverage CRF in other domains such as Biomedicine \cite{biomedicalner, SunilKumar}, and the drug industry \cite{Hettne2009ADT, rocktaschel-etal-2013-wbi}.

As shown, supervised models for NER are thoroughly explored in literature. Nonetheless, supervised learning is constrained by the limited availability of structured text used for feature learning. This is when semi-supervised learning for NER has come to light.

\subsubsection{Semi-Supervised Learning}

From its definition, semi-supervised learning aims to combine a small labelled dataset with a larger unlabeled dataset \cite{sharmaamrita}. The resultant dataset is then fed to the classifier to learn and make predictions on new examples.
In this context, Liao and Veeramachaneni \cite{liaowenhui} present a simple semi-supervised learning model relying on CRFs. The algorithm aims to provide high-precision labels to unannotated data by exploiting additional evidence unrelated to the classifier's features. Chen et al. \cite{rosener} propose a robust two-step semi-supervised NER approach ROSE-NER to handle noisy data. Thenmalar et al. \cite{thenmalar} use a bootstrapping-based NER approach for entity identification and classification. They use a small set of tagged data to extract context features and define a context pattern for each named entity category. They explore feature representations for both English and Tamil languages. Agerri and Rigau \cite{Agerri_2016} present a multilingual  NER system based on a robust feature set across languages and datasets by combining shallow local information with clustering semi-supervised features. They experiment with their methods on Spanish, Dutch, English, German and Basque. In addition to the aforementioned languages, Althobaiti et al. \cite{Althobaiti} introduce ASemiNER, the first semi-supervised pattern-based bootstrapping learning approach to Arabic NER, without the need for annotated training data or gazetteers.

\subsubsection{Deep Learning}

Deep-learning-based techniques achieve the current state-of-the-art in nearly all domains including entity linking. The key feature of Deep learning (DL) is its ability of representation learning and semantic contextualization via vector representation, embeddings and neural network architectures to learn complex features through non-linear activation functions with little to no feature engineering or domain expertise \cite{li2020survey}. Yadav and Bethard \cite{yadav2019survey} suggest that the neural-networks-based architectures should be classified according to the type of the word representations. Roy \cite{roy2021recent} classifies the NER methods adopted in this context based on the network architecture. Li et al. \cite{li2020survey} categorize the related techniques based on the deep learning stage in which these methods are deployed. The general pipeline of DL-based NER in literature, according to Li et al. \cite{li2020survey}, incorporates three main components: 1) Distributed input representations. 2) Context encoder. 3) Tag decoder. 

For the first component, DL allows a variant of distributed representations for inputs such as 1) Word-level representation \cite{li2020survey, roy2021recent, yadav2019survey} i.e. pre-trained word embeddings that can be fine-tuned on domain-specific datasets \cite{li2020survey}.  Word2Vec \cite{word2vec}, GloVe \cite{pennington-etal-2014-glove} and fastText \cite{fasttext} are the most known word embeddings according to \cite{li2020survey, roy2021recent}. 2) Character-level representation \cite{li2020survey, roy2021recent, yadav2019survey} has been helpful to capture ``sub-word-level" details such as suffixes and prefixes \cite{li2020survey} and internal semantic details \cite{roy2021recent} in addition to its ability of handling out-of-vocabulary (OOV) problem \cite{li2020survey, roy2021recent}. Convolutional-neural-network-based (CNN-based) \cite{li-etal-2017-leveraging, yangjie} and (recurrent-neural-network-based) RNN-based \cite{kuru-etal-2016-charner, tran-etal-2017-named} are the two most common architectures for extracting character-level representations across literature surveys \cite{li2020survey, roy2021recent, yadav2019survey}. 3) Combined or Hybrid scheme where additional lexical, linguistic and visual features are incorporated into the input vector in addition to the word-level and/or character-level information \cite{li2020survey, roy2021recent}. Combining word-level long with character-level representations was deemed effective as they compensate for each other's limitations as highlighted in \cite{li2020survey, roy2021recent, yadav2019survey}. The survey of Yadav and Bethard \cite{yadav2019survey} even explored literature hybrid models encompassing word, character and affix features simultaneously. Previous works leverage features like spelling and context features \cite{Huang2015BidirectionalLM}; POS tags and word shape features \cite{Wei2016DiseaseNE}; and orthographic features \cite{aguilar-etal-2017-multi}.

Concerning the second component of DL-based NER, prior works have also experimented different context encoder architectures or a combination of such models including convolutional neural networks (CNN) \cite{wangchunqi, tianyangHaoyan, ZHANG202213}, recurrent neural networks (RNN) with both variants: gated recurrent networks (GRU) \cite{8660795, gridachmourad} and long-short-term-memory (LSTM) \cite{chiujason, limsopatham-collier-2016-bidirectional}, recursive neural networks \cite{li-etal-2017-leveraging}, neural language models \cite{peters-etal-2017-semi, lei-etal-2020-neural, SHARMA2022101356} and deep transformer architectures such as Generative Pre-trained Transformer (GPT) \cite{Radford2018ImprovingLU}, Bidirectional Encoder Representations from Transformers (BERT) \cite{bert} and Embeddings from Language Model (ELMo) \cite{deepcontextualized}. Further information about these works is exhaustively reviewed in \cite{li2020survey, roy2021recent, yadav2019survey}.

The third and final component of DL-based NER models is the tag decoder. This module processes context-dependent representations and outputs a tag sequence corresponding to the input \cite{li2020survey}. There are four main architecture types found in literature: multi-layer perceptron followed by a softmax layer \cite{Strubell2017FastAA, li-etal-2017-leveraging}, conditional random fields (CRF) \cite{Strubell2017FastAA, deepcontextualized}, recurrent neural networks (RNN) \cite{Zhou2017JointEO, suncong}, and pointer networks \cite{Feifei}. The first three types are explored in detail in the surveys \cite{li2020survey, roy2021recent, yadav2019survey}, in contrast with pointer networks which were only mentioned in \cite{li2020survey}.

Only the work of Li et al. \cite{li2020survey} review the latest advancement of deep learning techniques for the NER task. Advances incorporate: (1) deep multi-task learning such as jointly performing POS, Chunk and NER; (2) deep transfer learning such as parameter-sharing approaches for ``cross-domain, cross-lingual, and cross-application" schemes. Zero-shot, one-shot and few-shot learning are also explored; (3) active learning by adopting uncertainty sampling strategies; (4) reinforcement learning and (5) neural attention for NER tasks. I refrain from diving into the details of each technique since my objective is to review the principal literature contributions in the NER task. However, the survey of Li et al. \cite{li2020survey} explores in detail prior works for each of the mentioned methods. 

\subsubsection{Hybrid Approach}

This method relies on merging two or more of the previously discussed NER-based techniques. The combined model poses extra complexities in terms of architecture, latency and required space. There have not been many references exploring the hybrid scheme: only the work of Sharma et al. \cite{sharmaamrita} reviews the usage of rule-based methods in conjunction with a learning-based approach.

\subsection{Named Entity Disambiguation (NED)}

Given a text collection, a knowledge base (KB) and a set of pre-identified entity mentions, entity disambiguation is the process of linking each retrieved textual mention to its corresponding entity in the knowledge base (KB) \cite{moller2021survey, shen2021entity, Oliveira2021TowardsHE}. The general architecture of the NED task encompasses four main modules as explained in \cite{sevgili2021neural, Shen2015EntityLW}: (1) candidate entity generation, (2) encoding with its two types: entity-mention encoding and entity-name encoding , (3) candidate entity ranking and (4) unlinkable mention prediction. 
I limit my explanation of these modules to a brief description of the main methods. 
The reader can refer to previous NED literature reviews  \cite{sevgili2021neural, Shen2015EntityLW, moller2021survey, Oliveira2021TowardsHE, shen2021entity} for further information on the related works in each category.

\subsubsection{Candidate Entity Generation}

First, the candidate entity generation phase is concerned with the distillation of the irrelevant entities from the KB and the retrieval of the set of recognized text mentions from the previous NER task. The entity generation methodologies depend on string comparison between the surface form of the entity mention and the entity name in a KB \cite{Shen2015EntityLW}. Three techniques are considered the most notable ones in this area according to \cite{Shen2015EntityLW, sevgili2021neural}: 1) name-dictionary-based where Wikipedia structure can serve as an example, 2) surface form expansion from the local document or expansion using aliases, and 3) search-engine-based where the whole web information is exploited for candidate entity generation. It is worth mentioning that recent zero-shot models with their domain-independent architectures do not rely on external knowledge for candidate entity generation \cite{sevgili2021neural}.

\subsubsection{Encoding}

Related literature focusing on entity linking encoding for the NED task is comprehensively reviewed in the surveys \cite{shen2021entity, moller2021survey, Oliveira2021TowardsHE, sevgili2021neural}. In general, encoding is divided into two classes: entity-mention-related and entity-name-related.

\begin{enumerate}
    
    \item \textbf{Entity Mention Encoding}

    Capturing context information is a core step in the entity disambiguation process. The encoded representation is used to assess the resemblance between entity candidates and the actual mentions \cite{shen2021entity}. As exhaustively explained and reviewed in the survey of Sevgili et al. \cite{sevgili2021neural}, the current trend is using an encoder to generate dense contextualized mention vector representations. Early encoding techniques focused on convolutional encoder networks and attention between entity embedding and contextual embedding. More recently, research is more oriented toward RNNs with its two variants GRU and LSTM, and with different settings: 1) unidirectional (left or right) and bidirectional; and 2) self-attention-based encoding methods using the outputs of a pre-trained BERT or a pre-trained ELMo.

    \item \textbf{Entity Name Encoding}

    The entity encoding step aims to encode semantic relations between entity candidates via vector representations. According to \cite{sevgili2021neural}, there are three main encoding-source-based approaches: (1) entity representations learnt from natural language processing algorithms developed for word embeddings such as word2vec \cite{newman-griffis-etal-2018-jointly, moreno}, (2) entity representations constructed from the entity relations in KGs \cite{Huang2015LeveragingDN, bperozzi}, (3) entity embeddings generated from training an encoder neural network \cite{nguyen-etal-2016-joint, Gupta2017EntityLV}. 
    The work of Shen et al. \cite{shen2021entity} categorizes the entity encodings into four main classes depending on the entity-related info to be encoded: surface form, entity description, entity context and entity type information. This work also reviews in detail literature conducted in each class.
    
\end{enumerate}

\subsubsection{Candidate Entity Ranking}
Next comes the candidate entity ranking phase. The objective of this stage is to assign a score to each entity from a list of entity candidates of a KG reflecting its relevance to the actual entity, given an entity mention and the related context \cite{sevgili2021neural}. Features that are considered useful to entity ranking include both context-independent features and context-dependent features: local and global \cite{Shen2015EntityLW}. Entities are disambiguated based on the neighbour words in the local approaches in contrast with the global disambiguation where interdependent semantic coherence plays the major role in entity disambiguation \cite{Shen2015EntityLW}. Previous works address global disambiguation using four main methods that are comprehensively reviewed in \cite{sevgili2021neural}: random-walk-based, maximization of Conditional Random Field (CRF) potentials, as a sequential decision task, or with the help of a neural component. The review of Sevgili et al. \cite{sevgili2021neural} is more inclined towards the mathematical modeling of the entity ranking phase. The work of Wang et al. \cite{Shen2015EntityLW} categorizes the mostly adopted techniques in the entity ranking phase using two different types of categorization. From one perspective, the first categorization divides the ranking methods into supervised and unsupervised methods. Among the supervised methods, Shen at al. \cite{Shen2015EntityLW} mention binary classification using Support Vector Machines (SVM), confidence-based methods, Vector Space Model (VSM) based methods, binary logistic classifiers, Naive Bayes classifiers and K-Nearest Neighbors classifiers; Learning to Rank methods; probabilistic methods; graph-based approaches and ensemble methods. On the other hand, the unsupervised ranking methods incorporate VSM-based and information retrieval-based methods such as KLD retrieval model. From another perspective, the second classification strategy classifies the entity ranking methods into three categories of ranking models: independent, collective and collaborative methods.

\subsubsection{Unlinkable Mention Prediction}
The final phase is the unlinkable mention prediction sub-task that has been reviewed in \cite{shen2021entity, Oliveira2021TowardsHE, sevgili2021neural, Shen2015EntityLW}. EL systems are supposed to handle the absence of a particular mention reference from the target KB \cite{sevgili2021neural}. Although some studies ignore the unlinkable problem for the sake of simplicity \cite{Shen2015EntityLW}, the majority of the related works employ a NIL threshold automatically learnt from the training data to predict the unlinkable entity mention \cite{shen2021entity, Oliveira2021TowardsHE, sevgili2021neural, Shen2015EntityLW}. In addition, a significant number of EL systems even employ supervised machine learning techniques or heuristic-based approaches to address this aforementioned problem. In general, there are three approaches found in literature to handle the NIL problems besides ignoring the issue according to \cite{sevgili2021neural, Shen2015EntityLW}: (1) In the case of unlinkable mentions, a best linking probability threshold is set such that its value is below the score with which the first unlinkable mention appeared, (2) introducing an extra ‘NIL’ entity in the ranking phase such that entity linking systems models can select the special entity as ``the best match" for the absent mention \cite{sevgili2021neural}, and (3) training a binary classifier that takes pairs of entity mention and entity names as inputs, in addition to extra features (i.e best linking score, NER systems detection flag, etc.) and outputs a decision whether a mention should be linkable or not.

\subsection{Advanced entity linking systems}

Having explained the core steps of each sub-task of entity linking systems and reviewed the most notable techniques experimented in the related works. It is important to note that recent literature addresses the entity linking task differently by introducing modifications to the general architecture. I only list two variations of these architectures since they are tightly related to the entity linking system selected to perform my experiments.

\subsubsection{Joint NER and NED}
Most of the previous work, as described in the last few lines, handles NER and NED separately. Newer approaches suggest a combined solution for the two sub-tasks done by the same model at the same time. 
Figure 5 in the work of Oliveira et al. \cite{Oliveira2021TowardsHE} presents a conceptual framework of the holistic entity linking process encompassing NER and NED.
Certainly, a joint pipeline poses extra complexities. Nonetheless, it can improve the quality of the entity linking process thanks to the dependence between the two main sub-tasks \cite{sevgili2021neural}, and may reduce error possibilities in the end-to-end pipeline \cite{shen2021entity}. Adopting the classification in \cite{sevgili2021neural}, models in this area are mostly probabilistic-graphical-dependent or neural-based. In general, recent works adopting a joint approach can be categorized into three classes: candidate-based, multitask learning and sequence labeling. In this first category, models \cite{Peters2019KnowledgeEC} usually filter all text spans in a passage or a document for entity mention recognition with pre-determined window size and using special heuristics. Mentions are then matched with a pre-built entity index before being ranked by a certain neural network to determine the entity mention ranking and status: linkable or not. The second approach aims to benefit from multi-task learning to offer a tighter connection between the two EL sub-tasks: NER and NED by using an LSTM network equipped with the attention mechanism that propagates its inner states generated during the entity recognition step to the next stage: candidate entity ranking \cite{martins-etal-2019-joint}. The final technique known as sequence labeling gets rid of the candidate generation step entirely by proposing a completely end-to-end mechanism where each text token is assigned an entity link \cite{Broscheit2019InvestigatingEK}. The surveys \cite{sevgili2021neural, shen2021entity, Oliveira2021TowardsHE} review intensively related works that are conducted in the domain of joint entity recognition and disambiguation.

\subsubsection{Domain Independence}
Entity linking is notorious for being ``data and domain-dependent" according to \cite{shen2021entity}.
Due to the insufficiency of labeled data, current research in entity linking is oriented towards domain-independent EL architectures.  In addition to its availability for a few domains, annotating data is a laborious and time-consuming task. Earlier studies leverage unsupervised and semi-supervised learning to address this issue \cite{sevgili2021neural}. More recent works suggest distant learning and zero-shot methods as feasible solutions to ensure model robustness \cite{shen2021entity, sevgili2021neural}. The distant learning methods \cite{lephong} rely primarily on surface matching approaches using unlabelled documents where the model learns to differentiate between positive and negative entity sets. The main drawback of such a technique is that it necessitates the existence of a KG describing the relations between entities or mention-entity priors calculated from Wikipedia entity hyperlinks in order to build the positive set by matching entity candidates to those in the KG \cite{sevgili2021neural}. Recently, proposed solutions use a zero-shot setting by training the EL system on a rich annotated dataset in order to adapt the system to newer domains with minimal or no information (i.e. domain-specific entity descriptions) \cite{moller2021survey, sevgili2021neural}. Zero-shot literature for EL has been reviewed in prior surveys \cite{moller2021survey, shen2021entity, sevgili2021neural}. Related methods \cite{Logeswaran2019ZeroShotEL} benefit from pre-computed representations of entity descriptions during the candidate inference stage. Different similarity measures are adopted across literature to calculate the matching between a mention representation and all the description representations such as BM25 information retrieval similarity function, Cosine similarity, dot-product etc \cite{Shen2015EntityLW}. Despite the success of zero-shot models, it is worth mentioning that complex neural models trained on a general-purpose dataset outperform the former methods.

\subsubsection{Cross-lingual Architectures}

Although Wikipedia pages are available in a variety of languages, English is still considered the most resource-rich language. As a result, English labeled data for EL systems is easily available and accessible compared to data in other languages. The purpose of cross-lingual architectures is to exploit supervision of the data annotated in the rich-resource languages for compensating labeling deficiency in other languages. Wikipedia inter-language links are widely used in cross-lingual supervision since they map equivalent pages in different languages to each other. Hurdles in cross-lingual EL systems are found in both entity recognition and entity disambiguation procedures. From one perspective, Wikipedia pages in low-resource languages may be short of mappings between entity mentions and entity links. In terms of NER, Cotterell and Duh \cite{cotterell-duh-2017-low} propose the transfer of BiLSTM-CRF with a character encoding network from a high-resource language. For NED, methods used for the candidate entity generation stage in literature varied between training translation models and utilizing translation dictionaries or neural character-level string matching models \cite{Shen2015EntityLW}. For candidate entity ranking, prior works use a variety of supplementary data like the Abstract Meaning Representation (AMR) \cite{Pan2017CrosslingualNT}, monolingual embeddings for joint words and mentions \cite{Tsai2016CrosslingualWU}, contextual embeddings, typing information, mention-entity prior probabilities, global context information \cite{upadhyay-etal-2018-joint}. Some works \cite{Sil2018NeuralCE, Tsai2016CrosslingualWU} propose zero-shot transfer from the rich-resource language, while others rely on pre-trained multilingual embeddings for entity ranking.

%===================================================================================================================

\section{Entity Linking for Information Retrieval} 
\label{entity_linking_for_IR}

Extracting entity mentions and linking them to the corresponding entity names in a KB with the objective of query and/or document expansion to overcome the limitations of exact-match-based methods (i.e. hard matching) such as vocabulary gaps, syntax errors, ambiguous synonyms and expression discrepancies between queries and documents is not a novel idea in literature. Nonetheless, little research was conducted in this area; and most of it was published over a decade ago or more. My objective is to exploit entity linking to improve the performance of document retrieval systems. This must not be confused with benefiting from information retrieval systems for the purpose of entity identification and extraction. The difference lies in the objective. While I aim to find the most relevant documents to a query with the help of additional information such as related entities, the second research direction targets entity retrieval, and hence the evaluation relies on entity relevance and not on document relevance.  

In addition to the classic bag-of-words (BoW) representation of queries and documents, there is a rich body of work in literature that explores other representational methods in the information retrieval field such as: (1) statistical translation language models \cite{maryamzhai,songfei} where the likelihood of the document-to-query translation is used for ranking purposes; (2) latent semantic and topic models \cite{pla,lda} where the matching takes place when a query and a document share the same set of latent topics. Closely related to the aforementioned representation type, Liu and Fang \cite{Liu2015LatentES} introduce a Latent Entity Space (LES) model where queries and documents are projected into a set of latent entities that is later used to estimate the document relevance; (3) bag-of-concepts using multilingual knowledge resources like Wikipedia for cross-language and multilingual information retrieval, hence the text is augmented with semantic-analysis-based features \cite{SORG201226,egozi}; (4) bag-of-entities (BoE) \cite{raviv,docretehsan,Ensan2018AdHR,xiongliu,Gonalves2018ImprovingAH} extracted with the help of automatic linking systems to represent both queries and documents. The latter are usually ranked according to the number of occurrences of query entities in text. Previous studies \cite{Gonalves2018ImprovingAH,duetrepr} have proved that a duet term-based and entity-based representations achieves better retrieval results when compared to either standalone BoW or BoE approaches. My approach is also a combination of both BoW and BoE representations.

Query expansion using terms generated from knowledge graphs has been intensively studied in previous works. Xiong and Callan \cite{queryfreebase} use a supervised model to combine information derived from Freebase KB descriptions and categories to select relevant terms for query expansion, while Dahir et al. \cite{Dahir2021QueryEB} experiment with linked data from DBpedia using different numbers of expansion terms. Other studies \cite{querywiki} and \cite{querywiki2} approach query expansion using pseudo-relevance feedback based on Wikipedia. The work of Krishnan et al. \cite{Krishnan_2017} suggests a diversified query expansion technique using semantic tools
by harvesting terms from the original queries and later prioritizing candidate entities using Wikipedia and pre-learnt distributional word embeddings. 

From another perspective, a number of prior works address document expansion approaches either using external collections to augment document representation \cite{sherman}, supplementary text sentences that are stochastically generated by a pre-trained language model \cite{jeong2021unsupervised}, or a set of queries predicted by a sequence-to-sequence transformer \cite{nogueira2019document} or T5 \cite{Nogueira2019FromDT} trained on query-documents pairs. The latter method is then extended to handle unlabeled datasets using domain transfer and weak supervision approaches \cite{Tang2020NeuralDE}. To the best of my knowledge, there is no prior work addressing document expansion with linked entities. On a related note, \cite{Gollub2018PseudoDF} evolves from the typical document expansion paradigm by introducing pseudo-descriptions which are explicit text fields for meta-data records justifying the reason a document is relevant to a query.
 
Among the relatively recent works that may be relevant to my research, I can distinguish the work of Ensan et al. \cite{docretehsan} where the authors introduce entity-based soft matching by proposing the Semantics-Enabled Language Model (SELM) for document retrieval based on the degree of relatedness of the meaning of the query and the documents. The motivation of this work is oriented to situations when exact matching between queries and documents is not possible, hence the need for a shared semantic space to perform the ranking. They use TAGME \cite{Ferragina2010TAGMEOA}, an entity linking tool that is able to augment raw text with hyperlinks to corresponding Wikipedia pages, in order to model queries and documents to sets of semantic concepts connected to each other based on relatedness in an undirected graph. By adopting a probabilistic model based on CRF, both queries and documents are represented as a set of concept nodes instead of the traditional bag-of-words representation; and the relatedness relations are represented as probability dependencies. Since their language model is able to retrieve distinct and non-overlapping documents when compared with other retrieval models, the authors encourage an interpolation scheme between different retrievers for better overall performance. In their subsequent work, Ehsan et al. \cite{Ensan2018AdHR} further explore the aforementioned idea by building a semantic retrieval framework to increase the relevant results in ad-hoc keyword-based information retrieval (IR) systems. The core of this framework is based on the previously built language model SELM equipped with extra two new semantic analysis/ entity linking configurations, besides TAGME, which are: (1) Explicit semantic analysis (ESA) \cite{esa} used to find semantic similarities between text passages by representing the text as a weighted concept vector from Wikipedia entries; (2) Paragraph2Vec \cite{le2014distributed} which also maps variable-length texts from Wikipedia entries to vector representations, and finds their relatedness degree. The separate integration module selects the best retriever (between SELM and keyword-based systems) based on a linearly weighted mixture model combining different retrieval systems. Experiments are conducted in comparison with two query expansion models as baselines: a variant of relevance model (RM3) \cite{rm3}, and entity query feature expansion (EQFE) \cite{eqfe} which is a retrieval model that expands queries by entity-related information such as names, anchors and categories; and scores the relatedness of each document to the given query based on these features. The results show that ESA outperforms the other semantic analysis systems across all measures over TREC Robust04 and ClueWeb09-B datasets. The contributions in \cite{Ensan2019RelevancebasedES} extend the previous two works by addressing research gaps. Although semantic-knowledge-based models depending on entities extracted from KGs were deemed effective for retrieval performance, these models suffer from topic drift issues due to the non-transitive nature of relatedness between entities. As a result, the authors of this paper introduce the Retrieval through Entity Selection (RES) method by proposing a relevance-based model for entity selection based on pseudo-relevance feedback (PRF) for query entity expansion and ad-hoc retrieval. RES relies on choosing candidate entities that are jointly related to mentions in the query and the top-ranked documents. Hence, a candidate entity that satisfies the following two conditions: being related to almost all text mentions in the query, and also related to a number of pseudo-relevant entities is prioritized by RES for query expansion compared to an entity that strictly satisfies one condition only. This approach cares about the quality of the retrieved entities to be used for query expansion, in contrast with my work where irrelevant entities are not considered harmful since I am interested in finding new documents that have not surfaced by the retriever using the non-expanded version of both the collection and the queries. The main drawback of RES is that it requires all query entities to be present in the graph cliques, so that the final entities chosen for query expansion are only those sharing the same semantics as the rest of the query entities. My work tries to overcome topic drift and matching inconsistency shortcomings in previous research by exploiting advances in recent entity linking models. Similar to \cite{Ensan2019RelevancebasedES}, I have also leveraged PRF as a comparative baseline to my methods.

Closely related to the discussed works and also aligned with mine, Gonçalves et al. \cite{Gonalves2018ImprovingAH} explore the value of entity information for improving ad-hoc retrieval of feature-based learning-to-rank search engines using the Washington Post news dataset. The authors create entity representations of both queries and collection using TAGME tool and employ Learning To Rank (LTR) methods \cite{ltr} to re-rank the initial runs for a thousand documents generated by BM25 using different combinations of LTR features. The objective is to evaluate the retrieval impact of the following document and query representations: BoE, BoW, and joint BoE/BoW where the combined representation is proved to be the best. The work of Xiong et al. \cite{duetrepr} also explores the word-entity duet representation using knowledge bases in ad-hoc retrieval. The authors of this work generate ranking features incorporating details from the word space, entity space and the cross-space connections through the KG describing four distinct types of query-document interaction: query words to document words, query entities to document words, query words to document entities, and query entities to document entities. An attention-based ranking model is then developed to exploit the duet representations in document ranking while discarding imminent noisy entities.
As shown in \cite{Gonalves2018ImprovingAH}, adopting LTR methods for semantic retrieval is another active research direction where semantic information is integrated into building and training ranking models. In this context, several research works leverage named entities \cite{nattiya} or semi-structured meta-data (i.e. controlled vocabularies, KB details, BM25 scores between queries and corresponding entities etc.) \cite{esdrank} as additional features for the purpose of learning ranking architectures. The empirical study in \cite{Ensan2017AnES} introduces and examines the effectiveness of neural embedding features based on word and document embeddings representing both queries and documents along with entity embeddings. This study also employs several LTR methods for document ranking using embedding-based features, keyword-based features and the interpolated version of features that show a significant improvement. Despite that this line of prior works adopts the same combined BoW and BoE approach, they use entities as extra features. In contrast, I expand BOW representations with term-based entities. 
% To the best of our knowledge, our methods have not been previously experimented in literature.

Although my contribution aligns with the previous works, it differs from them on many occasions. I use a state-of-the-art zero-shot-based one-pass end-to-end entity linking tool to expand both queries and documents. Expansion is based on a duet method leveraging both BoW and BoE representations, and takes place before the indexing. I do not use entities as additional features, but expand the text (BOW) with additional words (BOE). My experiments are conducted using both the word and the hashed versions of the linked entities, which was not explored before in literature to the best of my knowledge. Similar to the literature, my proposed methods also do care about the number of occurrences of entity mentions in both queries and documents. I employ BM25 as a sparse retriever for the early-stage retrieval experiments (which is also used in doc2query \cite{nogueira2019document} experiments).

%======================================================================
%======================================================================
\chapter{Methods}
\label{Chapter3}
%======================================================================

\section{Metrics}

There is a number of IR evaluation metrics used to assess the degree to which search results satisfy the user's information need represented by a query. Below is a brief overview of the most common metrics. I also identify the metrics employed in the experiments and justify this choice.

\subsection{Preliminaries}
\begin{itemize}
    \item \textbf{True Positive@K ($TP@K$):} The number of relevant documents that are retrieved among the top $K$ documents found by an IR system in response to a query.
    \item \textbf{True Negative@K ($TN@K$):} The number of non-relevant documents that are not retrieved among the top $K$ documents found by an IR system in response to a query.
    \item \textbf{False Positive@K ($FP@K$):} The number of non-relevant documents that are retrieved among the top $K$ documents found by an IR system in response to a query.
    \item \textbf{False Negative@K ($FN@K$):} The number of relevant documents that are not retrieved among the top $K$ documents found by an IR system in response to a query.
\end{itemize}

\subsection{IR Evaluation Metrics}
\begin{itemize}
    \item \textbf{Precision@K ($P@K$):} 
    Precision@K is a single-value metric. It evaluates the number of relevant documents to a query from the top $K$ retrieved results where $K \in [1, \text{size of the result set}]$. The main limitation of this metric is that the rank of the relevant results is not taken into consideration. $P@K$ can be calculated using the formula:
    \begin{equation}
     P@K = \frac{TP@K}{TP@K + FP@K}
    \end{equation}
    
    \item \textbf{Recall@K ($R@K$):}
    Recall@K is a single-value metric. It shows how many relevant documents are retrieved out of the actual top $K$ relevant documents to a query, whether retrieved by the underlying IR system or not. $R@K$ can be calculated using the formula:
    \begin{equation}
     R@K = \frac{TP@K}{TP@K + FN@K}
    \end{equation}
    
    \item \textbf{F-measure:}
    known as F1-score, is also a single-value metric. It represents the weighted harmonic mean of precision and recall. F-measure is considered a better measure for an overall performance evaluation since it takes into consideration both precision and recall, which are complementary metrics. F-measure is calculated as:
    \begin{equation}
        \text{F-measure} = \frac{2*(P@K)*(R@K)}{(P@K)+(R@K)}
    \end{equation}
    
    \item \textbf{Average Precision (AP):}
    is an order-aware metric. Since Precision and Recall are single-value metrics, they do not consider the order of the retrieved documents in the result set. One way to address this limitation is by computing the precision and recall at every position in the ranked document list. These values are then used to plot the precision as a function of the recall (i.e. $p(r)$), also known as the precision-recall curve. AP computes the average value of $p(r)$ over the recall interval $ r \in [0, 1]$. Simply put, AP is the area under the precision-recall curve.
    \begin{equation}
        AP = \int_{r=0}^{1} p(r) \,dr
    \end{equation}
    
    \item \textbf{Mean Average Precision (MAP):} 
    for a query set is the mean of the average precision for each query. It is also an order-aware metric as AP. If $Q$ denotes the total number of queries, MAP can be computed using the following formula:
    \begin{equation}
        MAP = \frac{\sum_{q=1}^{Q} AP(q)}{Q}
    \end{equation}
    
    \item \textbf{Mean Reciprocal Rank (MRR):} 
    is a statistical measure used for evaluating the list of the retrieved documents relevant to a query set and ordered by correctness. MRR is an order-aware metric.
    The reciprocal rank of a query result is calculated as the multiplicative inverse of the rank of the first relevant answer. Hence, the mean reciprocal rank is computed as the mean of the previously calculated value across all queries. If $Q$ denotes the total number of queries, $rank_{q}$ is the rank position of the first relevant document to a query $q$. MRR is calculated as follows:
    \begin{equation}
        MRR = \frac{1}{|Q|} \sum_{q=1}^{|Q|} \frac{1}{rank_{q}}
    \end{equation}
    
    where $|Q|$ is the length of the query set. MRR takes into consideration the first relevant result to a query and ignores all other relevant documents.
    
    \item \textbf{Cumulative Gain@K ($CG@K$)}
    is a graded relevance metric, which is different from the previously explained metrics that adopt the binary relevance scheme. $CG@K$ is simply the sum of the relevance scores of the top $K$ retrieved results.
    \begin{equation}
        CG@K = \sum_{k=1}^{K} relevance_{k}
    \end{equation}
    
    \item \textbf{Discounted Cumulative Gain@K ($DCG@K$):}
    Since $CG@K$ does not account for the rank of a relevant item, $DCG@K$ is introduced to evaluate the gain of a relevant document based on its rank in the result set. $DCG@K$ mechanism is based on logarithmically penalizing any highly relevant document appearing in a low position in the result set. The traditional $DCG@K$ is calculated using the following formula:
     \begin{equation}
        DCG@K = \sum_{k=1}^{K} \frac{relevance_{k}}{log_{2}(k+1)}
    \end{equation}
    An alternative $DCG@K$ formula imposes a larger penalty if relevant documents are ranked in lower positions. This formulation is preferred in industry to the traditional one.
    \begin{equation}
         DCG@K = \sum_{k=1}^{K} \frac{2^ {relevance_{k}} - 1}{log_{2}(k+1)}
    \end{equation}
    
    \item \textbf{Normalized Discounted Cumulative Gain@K ($NDCG@K$):} is employed to better compare the performance of IR systems since the result set for each query may vary in length across different queries and/or systems. In order to compute $NDCG@K$, all the relevant documents are sorted by their relative relevance score generating the maximum possible $DCG$ value at position $K$, what is known as the ideal $DCG$ at position $K$ (i.e. $IDCG@K$). For a given query, $NDCG@K$ is computed as:
    \begin{equation}
        NDCG@K = \frac{DCG@K}{IDCG@K}
    \end{equation}
\end{itemize}

\subsection{Metric Choice}

Two metric types are used in my experiments depending on the experiment type, its requirements and needs. The two evaluation metrics are recall@1000 and MRR@10 of the trec\_eval\footnote{\url{https://github.com/usnistgov/trec\_eval}} implementation. 
Since the first type of experiment focuses on early-stage retrieval, I employ the recall@1000 to determine the fraction of the relevant documents that are retrieved out of the top 1000 actual relevant documents. In order to create the best pool for re-ranking, I aim to maximize the recall for the first stage retrieval. At this stage, I am concerned with expanding the pool of relevant documents that are consequently pruned and refined in the later re-ranking stages.

The second type of experiment evaluates the end-to-end IR system using dense retrieval. As a result, I use the official evaluation metric of MS MARCO passage dataset, which is MRR@10. MRR is an order-aware metric that gives a general measure of the quality of a ranked result list to queries. As previously mentioned, MRR only cares about the first relevant document to a query and ignores the other relevant results. Since more than 90\% of the MS MARCO queries have a single relevance judgement, the MRR@10 metric fits my end-to-end experiments on the MS MARCO dataset.

It is worth mentioning that graded-relevance-based metrics such as $CG@K$, $DCG@K$ and $NDCG@K$ are not the most adequate for the MS MARCO dataset which adopts a binary relevance paradigm. Consequently, I do not include them as evaluation metrics in the experiments.

% ====================================================================================================================

\section{Dataset}

Methods are experimented on 1) the MS MARCO dataset and 2) the MS MARCO Chameleons sets, which are subsets of the original MS MARCO known for their poor performance with rankers. I have also evaluated my work using three variants of relevance judgements sets as demonstrated in the next lines.

\subsection{MS MARCO Passage Dataset}

My experiments are conducted on the Microsoft MAchine Reading Comprehension (MS MARCO) passage collection v1\footnote{\url{https://microsoft.github.io/msmarco/}}.
Before the emergence of the passage ranking dataset, the first MS MARCO dataset was a question answering (QA) one, incorporating 100,000 real Bing questions with human-generated answers. The passage ranking collection \cite{msmarco} now comprises 8.8 million passages extracted from 3.5 million web documents retrieved by Bing, along with over 500k pairs of real anonymized search queries generated through Bing or Cortana; and judged-relevant passages for training purposes.

This dataset is introduced for the passage re-ranking task in IR \cite{bm25beyond}, and is targeted to provide a large-scale dataset for benchmarking neural IR methods \cite{Mitra2018AnIT}. It was also employed for the first time in Text REtrieval Conference (TREC)\footnote{\url{https://trec.nist.gov/}} 2019 in the “Deep Learning” track where passage and question collections were leveraged to set up an ad-hoc retrieval task. MS MARCO ranking dataset constitutes a great English resource that is broadly employed for training deep learning models for IR-related tasks, achieving considerable effectiveness on ``diverse zero-shot scenarios" \cite{msmarcomultilingual}. A multilingual version of the MS MARCO passage ranking dataset comprising 13 languages, known as mMARCO \cite{msmarcomultilingual}, was also created using machine translation.

Two main tasks are typically conducted using the MS MARCO passage dataset. I also cover these two tasks in the conducted experiments.
\begin{enumerate}
    \item Passage Re-Ranking: which requires re-ranking passages by relevance given a candidate top 1000 passages retrieved by BM25.
    \item Passage Full Ranking: which requires generating a  top 1000 candidate passages ordered by their relevance given the full corpus of passages and queries.
\end{enumerate}

The relevance annotations (i.e. qrels) provided by the human editors are shallow (at most 1 or 2 relevant passages per query). For less than 10\% of the queries, there are multiple (i.e. at most 2) judged relevant passages per query. The selected evaluation metrics also ignore the second relevant answer as previously highlighted. As a result, it is safe to assume that the provided qrels are of binary relevance (i.e relevant or not) in my case.

The ``MS MARCO Small Development Set” includes 6,980 queries dedicated to development and validation. This set, which constitutes 6.9\% of the full dev set, is usually used in experimentation. There is also a test set with private (not publicly available) relevance judgments for leaderboard purposes. In my work, I use both the training and the small development sets. I refer to the small development set as ``Dev'' set for conciseness.

\subsection{MS MARCO Chameleons}

I further test my techniques on the MS MARCO Chameleons\footnote{\url{https://github.com/Narabzad/Chameleons}} sets of obstinate queries \cite{hardqueries}.
These query sets are subsets of the original MS MARCO passage dataset's queries that are difficult for state-of-the-art rankers to satisfy. As a result, rankers show extremely poor performance when trying to find relevant matching for these queries.
The latter do not experience any performance improvement regardless of the underlying ranker, i.e. the overall improvement reported by a ranker always results from another subset of queries.
The rankers used as baselines to extract the sets of obstinate queries include BM25 as a stable traditional retriever, and five neural rankers which are DeepCT \cite{Dai2019ContextAwareST}, DocTTTTTQuery \cite{Nogueira2019FromDT}, RepBERT \cite{Zhan2020RepBERTCT}, ANCE \cite{xiong2020approximate} and TCT-ColBERT \cite{lin-etal-2021-batch}.

It is important to mention that hard queries do not have any special sentence structure compared to the easier ones. They share common characteristics with the easy queries such as the number of available relevant judgements and the query length. Nonetheless, two common patterns are discovered among these queries, which justify why they are resilient against any performance improvement attempt such as changing the neural ranker type or query reformulation. The first pattern is the existence of typographical errors or misspelled terms in the query text. The second pattern is the complex query structure that requires deep interpretation beyond the immediate term meaning \cite{hardqueries}.

Leveraging entity linking, I aim to overcome the two causes leading to the bad performance of these obstinate queries. Appending meaningful entities to the original text can consolidate the query structure by negating the effect of the misspelled words since the correct entity names are to be appended to the text, hence an exact matching will be possible if the appended words exist in a potentially relevant document. In addition, entities help provide further context to ambiguous query terms (such as the word ``eagles" in Example \ref{quote:one}), thus a deeper interpretation of the query is possible. Through my experiments, I demonstrate that collection expansion with linked entities helps rankers discover a higher percentage of matches for the hard queries of these sets. 

In general, the MS MARCO Chameleons dataset consists of three main sets: 
\begin{enumerate}
    \item Veiled Chameleon (or ``Hard'' set) comprises 3,119 hard queries that are common between the worst 50\% of the performing queries of at least four rankers among the previously mentioned ones. The following example illustrates a Hard query that is incorrectly matched by BM25. While the passage below seems relevant, it answers partially the query as it restricts the answer to the number of death in the US caused by venomous snakes.
    
    \begin{lquote}
    \textbf{Query:} \textit{how many people die from snake bites a year?} 

    \textbf{Relevant Passage according to BM25: }\textit{The exact percentage of a dry bite varies from venomous snake to venomous snake, but, for instance, around 50\% of Coral Snake bites are dry bites, delivering no venom. In fact, only 9-15 people per year in the U.S. die from snake bites out of about 8000 bites from venomous snakes per year.}
    \label{quote:hard_set_noentities}
    \end{lquote}

    \item Pygmy Chameleon (or ``Harder'' set) includes 2,473 hard queries that are common between the worst 50\% of queries of at least five rankers. Below is an example of a query from the Harder set for which five rankers fail to find a relevant match. The example also shows the `supposedly' relevant document that is retrieved by BM25 on the non-expanded MS MARCO passage dataset.
    
    \begin{lquote}
    \textbf{Query:} \textit{what is medical term for neck fusion?} 

    \textbf{Relevant Passage according to BM25: }\textit{ASA is the medical abbreviation for what medical term. What is the abbreviation for chronic obstructive pulmonary disease. What is the abbreviation for head, eyes, ears, neck, and throat. What is the abbreviation for complaint of. This abbreviation lbstands for what medical term.}
    \label{quote:harder_set_noentities}
    \end{lquote}

    \item Lesser Chameleon (or ``Hardest'' set) comprises 1,693 queries judged as the hardest by six rankers. The query in Example \ref{quote:one} is considered as one of the Hardest set that is usually wrong matched with all the six aforementioned rankers.
\end{enumerate}

\subsection{Relevance Judgments}

My experiments can be classified into two folds: sparse retrieval and dense retrieval.

Regarding the sparse-related experiments, I employ three types of relevance annotations. The intuition behind this choice is that I believe that the original qrel set is insufficient as ground truth for the evaluation of the constantly developed IR methods because of the limited human-based annotations. 
From another perspective, in a multi-stage ranking stack consisting of stages $S_{0}, S_{1}, ..., S{i}, ..., S_{n-1}, S_{n}$, a stage $S_{n-1}$ can be evaluated by its ability to retrieve the documents that would have been ranked highest at the stage $S_{n}$ using a method $m$, if $m$ is used to do the ranking at stage $S_{n}$ over the document set previously ranked by stage $S_{n-1}$.
In an attempt to compensate for the original qrels limitations, I generate two other qrel sets to evaluate the sparse methods in order to get a better indication of the entity linking effect.
In addition to the original qrels provided by MS MARCO assessors, I employ two other qrel sets based on the top results from the upper stages of my ranking stack, which allows to directly measure the ability of the first stage to satisfy the requirements of later stages. These qrels are generated from a run pool containing all of my generated runs for this purpose. The pool is constructed by generating four different runs: three runs using BM25 as a sparse retriever: (original run from the non-expanded corpus, run from the entity-equipped dataset, run from the hashed-entity-aware dataset) and one run using the ANCE dense retriever generated on the original non-expanded corpus. This pool is then re-ranked using MonoT5, and the top passage for each query is chosen to build the first qrel set. This qrel set is referred to as the MonoT5 qrels. The second set is formed by re-ranking the top 50 passages per query that were generated in the previous step using DuoT5 \cite{Pradeep2021TheED}. In short, it is generated by MonoT5+DuoT5 re-ranking of the run pool. I use the PyGaggle\footnote{\url{https://github.com/castorini/pygaggle}} implementation of the MonoT5 and DuoT5 re-ranking. In the remainder of the work, I refer to these qrel sets as ``original'', ``MonoT5'' and ``DuoT5'' qrels.

For the dense-retrieval-related experiments, I only evaluate my runs against the original set of qrels. Since there is no considerable change in the results with or without entities using the dense methods evaluated with the original qrels, I thought it is not necessary to conduct further experiments in that direction. 

% ====================================================================================================================
\section{Information Retrieval Method}

My primary objective is to test the effect of linked entities used to augment the corpus (queries and passages) on the retrieval quality. As a result, I perform two types of retrieval experiments:
\begin{itemize}
    \item Sparse retrieval experiments: where I focus on improving the recall@1000 of the first stage retrieval for later re-ranking stages. I evaluate the adopted methods using the original, MonoT5 and DuoT5 qrels as previously demonstrated.
    
    \item Dense retrieval experiments: where I evaluate the entity effect on an end-to-end state-of-the-art dense retriever using MRR@10 along with the original qrel set.
\end{itemize}

\subsection{Sparse Retrieval}
\label{sparse_technique}

As a sparse retriever, I employ BM25 as implemented by the open-source Anserini system \cite{Yang2017AnseriniET}, which provides state-of-the-art performance for sparse term-based retrievers.
Given a query $Q$ of $Q_{1}, Q_{2}, ..., Q_{n}$ terms and of $|Q|$ length, and a document $D$ of length $|D|$, the similarity function of Okapi BM25 between the query and the document is computed as \footnote{\url{http://ipl.cs.aueb.gr/stougiannis/bm25.html}}:
\begin{equation}
    BM25(Q, D) = \sum_{q=1}^{|Q|} IDF(Q_{q}) * \frac{TF(Q_{q}, D) (k_{1}+1)}{TF(Q_{q}, D)+ k_{1}.(1 - b + b. \frac{|D|}{dl\_avg}}
\end{equation}
where $dl\_avg$ is the average document length in the whole collection, $k_{1}$ and $b$ are BM25 hyperparameters, $TF$ is the term frequency, and $IDF$ is the inverse document frequency computed as:
\begin{equation}
    IDF(Q_{q}) = \log \mleft(\frac{N-DF(Q_{q})+0.5}{DF(Q_{q})+0.5} \mright)
\end{equation}
where $N$ is the number of the collection documents, $DF(Q_{q})$ is the frequency of the documents with $Q_{q}$ occurrences.
The Anserini\footnote{\url{https://github.com/castorini/anserini}}
implementation of BM25 has been widely adopted as the first stage retriever in many multi-stage ranking stacks~\cite{hofsttter2020local, hofsttter2020interpretable, nogueira2019multistage}. The BM25 tuned hyperparamaters ($k_{1}=0.82, b=0.68$) are optimized for recall@1000 on the MS MARCO dataset in Anserini.

In this type of experiment, I use the dense retriever ANCE ~\cite{xiong2020approximate} as a basis for my comparisons since it is a well-established contrastive representation learning mechanism for dense retrieval using an asynchronously updated ANN index. I have also leveraged ANCE run to build the combined run pool that was re-ranked to generate the MonoT5 and DuoT5 qrels as previously explained. In addition to ANCE, I also compare my best entity-related result against the BM25 with pseudo-relevance feedback (PRF) run. I employ PRF as implemented by the work of Büttcher et al. \cite{charles}.

\subsection{Dense Retrieval}

I use the dense retriever pipeline proposed by Zhan et al. \cite{staradore} that relies on two training strategies named: 1) ``Stable Training Algorithm for dense Retrieval (STAR)" and 2) ``Algorithm for Directly Optimizing Ranking pErformance (ADORE)". On one hand, STAR ameliorates the dense retrieval training stability using random negatives. It employs two types of negative samples: 1) static hard negatives for ranking improvement, and 2) random negatives for training stability. On the other hand, ADORE substitutes static hard negatives with dynamically sampled examples optimizing the ranking performance.
The STAR-ADORE combination achieves state-of-the-art performance on the dense retrieval task. In order to build STAR and ADORE architectures, the work of Zhan et al. \cite{staradore} investigates the trade-off between different sampling strategies such as: 1) Random negative sampling vs. Hard negative sampling, 2) Static hard negative sampling vs. Dynamic hard negative sampling. To better assess the differences between the studied sampling strategies, an additional context is provided. 

\subsubsection{Sampling Strategies}

Generally, dense retrieval models are trained by minimizing a specific loss function, usually the pairwise loss, on the training data where each training sample includes a query, a negative document, and a positive document. Due to the excessive cost of direct optimization over all the corpus samples, sampling strategies have emerged. A sampling strategy can be considered as assigning a different weight for each negative document. The first sampling strategy that is considered in this work is the random negative sampling that focuses on the minimization of the pairwise error sum. This approach typically leads to top-ranking performance loss because of the obstinate queries that dominate the training \cite{staradore}. In addition, this sampling strategy has an unbounded loss function. To compensate for the mentioned limitations, hard negative sampling is suggested to optimize the retrieval performance. The latter aims to only sample the top $K$ documents as negatives, instead of employing the total number of documents. 
Hard Negatives can be classified into two main categories: static and dynamic. Traditional dense retrievers usually employ static hard negatives which do not reflect the actual negatives constantly changing during the training of the dense retrieval model. As a result, dynamic hard negatives are introduced in \cite{staradore} to better fit the underlying problem. According to \cite{staradore}, dynamic negatives are defined as the non-relevant documents with the highest ranks that constantly change according to the dense retriever parameters at each training step.

\subsubsection{STAR}

STAR stands for ``Stable Training Algorithm for
dense Retrieval" \cite{staradore}. The algorithm leverages static hard negatives to boost the ranking performance.
The static hard negatives are generated by a warm-up model (the same warm-up model used in ANCE \cite{xiong2020approximate} experiments) which retrieves the highest-ranked documents for all the training queries. The latter are then fed to STAR model as static hard negatives that will be kept unchanged during the whole training process. In addition to the static negatives, STAR exploits random negatives to stabilize the training.
In an attempt to improve efficiency, STAR also employs a reusing strategy where non-relevant documents in the same batch are reused as approximate random negatives instead of explicitly adding random negative documents to the input.
% Figure \ref{fig:star_sampling} (a) shows the static hard negative sampling in one input batch of STAR, (b) emphasizes on the reusing strategy by demonstrating how the pairwise loss is computed for one query. 

% \begin{figure}%
%     \centering
%     \subfloat[\centering Input: one relevant doc and
% multiple static hard negatives are sampled for each query.]
%     {{\includegraphics[scale=0.5]{star_sampling_a.png} }}%
%     \qquad
%     \subfloat[\centering Reusing other document
% embeddings when computing pairwise loss.]
%     {{\includegraphics[scale=0.5]{star_sampling_b.png} }}%
    
%     \caption{Reusing strategy of STAR. The number of rows correspond to the batch size. (from \cite{staradore})}%
    
%     \label{fig:star_sampling}%
% \end{figure}

\subsubsection{ADORE}

The final stage of the dense retrieval pipeline proposed in \cite{staradore} is a ``query-side training Algorithm for Directly Optimizing Ranking pErformance (ADORE)".
As per the model's name, ADORE leverages dynamic hard negatives to optimize the top-ranking performance. Before the actual training, document embeddings are pre-computed with a pre-trained STAR used as a document encoder. Hence the index is built and stored. The generated static embeddings are then fed to ADORE for training. At each iteration, ADORE generates query encodings for each batch, and utilizes the document embeddings to fetch the top relevant results (i.e. dynamic hard negatives). Finally, the resultant negatives are used to train the dense retrieval model. ADORE is the first dense-retrieval-based model to introduce dynamic hard negatives to optimize the query encoder.

\begin{figure}
    \centering
    \includegraphics[scale=0.4]{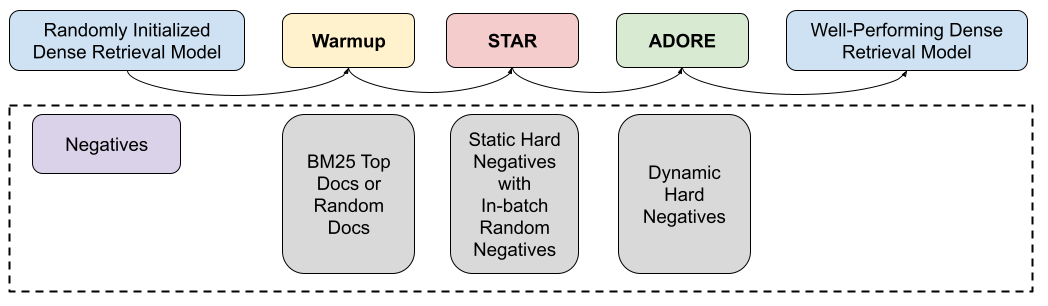}
    \caption{Dense Retriever pipeline using STAR and ADORE (adapted from \href{https://github.com/jingtaozhan/DRhard}{Github's documentation of the model implementation)}}
    \label{fig:star_adore}
\end{figure}

I use this implementation\footnote{\url{https://github.com/jingtaozhan/DRhard}} of the STAR-ADORE pipeline in my dense retrieval experiments. STAR is used for document encoder training, and the query encoder is trained with ADORE. This pipeline achieves state-of-the art results on both MS MACRO passage and documents dataset evaluating with MRR@10 and MRR@100 respectively. Figure \ref{fig:star_adore} shows the overall STAR-ADORE dense retrieval pipeline with the corresponding negative types generated at each stage.

% ====================================================================================================================
\section{Entity Linking}

I employ ELQ, an entity linking system for questions, to extract entity mentions from both queries and documents after extending its functionality to fit longer texts. ELQ is built as an extension to a zero-shot dense-retrieval-based entity linking model known as BLINK.
Before elaborating on ELQ architecture, it is important to review BLINK mechanism.

\subsection{BLINK}

BLINK \cite{blink} is a scalable two-stage zero-shot BERT-based entity linking algorithm with dense entity retrieval. Given an input document and an entity mention list, BLINK generates pairs of entity mentions and entity names where each entity name is an entry in a knowledge base (Wikipedia in this case). Unlike ELQ, BLINK\footnote{\url{https://github.com/facebookresearch/BLINK}} only performs entity disambiguation using pre-specified entity mention boundaries in the input. BLINK introduces a two-stage approach for zero-shot linking leveraging BERT \cite{bert} architectures.

The first stage of BLINK architecture relies on a bi-encoder consisting of two BERT transformers to independently encode the representations of mention context and entity descriptions into dense vectors. 
Before feeding the input to the corresponding transformer, the input text is pre-processed and modeled to a specific format according to its type (context with entity mentions or entity names with descriptions).
The modeling format of the context/entity mention is based on word tokens of the context surrounding the entity mention. The context input is represented as: 
\begin{quote}
    [CLS] text on the left [Ms] mention [Me] text on the right [SEP]
\end{quote}
[Ms] and [Me] are special tokens for entity tagging. The maximum allowed length of the input is 32.
The modeling format of the entity also relies on word tokens of the entity name (also known as ``entity title") and the corresponding description, where [ENT] is a special token to distinguish the entity name from its description.
\begin{quote}
    [CLS] entity name [ENT] description [SEP]
\end{quote}
The context representation and the entity representation generated from the two transformers are then used to determine the score to be used for network optimization. The score of entity candidates is given as vector dot products of the two transformer outputs. The network is then trained to maximize the score of the correct entity with respect to its counterparts of the same batch. In addition to the in-batch random negatives, the first stage also exploits hard negatives during training by generating the top 10 predicted entities for each training sample.

The retrieved candidates are then forwarded to the second stage for re-ranking. The latter takes place using a cross-encoder that concatenates the context and mention representation, and the entity representation together. The cross-encoder then encodes the concatenated representation in one transformer, and re-ranks the top 64 candidate entities produced in the previous stage.
The model is trained to maximize the score of the accurate entity given a set of candidate entities. Overall, the BERT-based end-to-end entity linking model achieves state-of-the-art results on many datasets: zero-shot entity linking dataset, WikilinksNED Unseen-Mentions dataset, and TACKBP-2010 benchmark \cite{blink} with no external entity knowledge or pre-defined heuristics.

\subsection{ELQ}

I use ELQ \cite{elq} as a fast end-to-end entity linking system. ELQ\footnote{\url{https://github.com/facebookresearch/BLINK/tree/main/elq}}, which stands for ``efficient one-pass end-to-end Entity Linking for Questions", is built on top of BLINK. Although no entity disambiguation system can fit all datasets, ELQ achieves state-of-the-art performance compared to other end-to-end tools \cite{shen2021entity}.
In contrast with BLINK, ELQ does not need pre-defined ``mention boundaries in the input" to extract the entity mentions. ELQ performs both entity NER and NED in one pass of BERT. In addition, it has a 2x faster end-to-end entity prediction time compared to neural baselines, making it suitable for downstream QA-related tasks. It is also designed to find entities in short noisy texts with a high level of accuracy. Hence, it is adequate for MS MARCO passage dataset, especially the queries and the obstinate ones of them.

The system determines each entity mention boundaries in a given query and the candidate Wikipedia entity using a BERT-based dual encoder. First, the entity encoder embeds every Wikipedia entity using its short description. Then, the query encoder calculates token-level query embeddings. These two embeddings are finally leveraged in mention boundary detection and disambiguation by computing their inner product. 
Figure 1 in \cite{elq} shows ELQ end-to-end pipeline. 
NER and NED components are jointly trained by optimizing their binary cross-entropy loss summation. In order to expedite training, a transfer learning technique is adopted by taking BLINK entity encoder that was trained on Wikipedia, freezing its weights, and fine-tuning the query encoder on a new dataset.

Contrary to a two-stage pipeline which first performs entity mention recognition, then disambiguation, ELQ joint approach, encompassing NER and NED, allows multiple possible candidate mentions to be considered for entity linking. This can be critical for queries known for their short and noisy texts as it may be difficult to extract mentions from them. It is also worth mentioning that under the same circumstances, ELQ outperforms BLINK by a significant margin using the following metrics: recall, precision and F1-score when evaluated on WebQSP and GraphQuestion datasets \cite{elq}.

% \begin{figure}[htp]
%     \centering
%     \includegraphics[scale=0.4]{elq.png}
%     \caption{ELQ entity linking
% mechanism. The query and the entity are separately encoded. (from \cite{elq})}
%     \label{fig:elq}
% \end{figure}

\subsubsection{ELQ for longer texts}

One major limitation is that ELQ description only fits MS MARCO queries. Hence, the current implementation cannot be used to extract entities from the passages.
Nonetheless, I want to perform entity linking on the MS MARCO passage ranking dataset for both the queries and the passages. Since ELQ is originally dedicated to short-length questions, I adopt an overlapping sliding window approach to extract entities from longer passages using a context window size of 128 tokens and an overlap stride of 42, so that the window overlaps by $1/3$ of the text length. 
The stride is chosen to represent 1/3 of the text on the left side to better reflect the passage theme and maximize context harmony within the identified entities.
This overlap ensures that for each sub-passage, the context is taken into account from both sides. The entity set of the whole passage is later deduplicated. I also retain the default parameter settings ($threshold=4.5$, $num\_cand\_mentions=10$, $num\_cand\_entities=10$) recommended by ELQ.

%===================================================================================================================

\section{Corpus Expansion}

After identifying entity mentions in the text and extracting the corresponding entity names using ELQ, I append these entities to the original BOW text.
I expand both the queries and the passages with a single instance of each retrieved entity name. Collection augmentation with entities is attempted using two forms: 1) explicit word form, and 2) MD5 hashed form.
The intuition behind the decision to experiment with MD5 hashed entities is to provide consistent representations for multi-word terms, hence avoiding partial or wrong matching between a query and a non-relevant passage. 

To elaborate on how query expansion and document expansion are performed using both the explicit and the hashed form, let's revisit Example \ref{quote:one}. The appended entities are underlined in red in the following examples.

\begin{enumerate}
    \item \textbf{Explicit Word Form: } 
    \begin{lquote}
    \textbf{Query Expansion:} \textit{ who are in the eagles \textcolor{red}{\underline{Eagles (band)}}}
    
    \textbf{Document Expansion:} \textit{ The scratch of an eagle in a dream means a sickness. A killed eagle in a dream means the death of a ruler. If a pregnant woman sees an eagle in her dream, it means seeing a midwife or a nurse. In a dream, an eagle also may be interpreted to represent a great ruler, a prophet or a righteous person. Eagle Dream Explanation  The eagle symbolizes a strong man, a warrior who can be trusted neither by a friend nor by a foe. Its baby is an intrepid son who mixes with rulers.
    \textcolor{red}{\underline{Prophet} \underline{Eagle} \underline{Midwife} \underline{Dream} \underline{Oy, to ne vecher} \underline{Nursing}}
    }
    \label{quote:expansion1}
    \end{lquote}
    
    \item \textbf{MD5 Hash Form: } 
    \begin{lquote}
    \textbf{Query Expansion:} \textit{ who are in the eagles \textcolor{red}{\underline{457e38cd8f6a6c4145a2038dc309f9e8}}}
    
    \textbf{Document Expansion:} \textit{ The scratch of an eagle in a dream means a sickness. A killed eagle in a dream means the death of a ruler. If a pregnant woman sees an eagle in her dream, it means seeing a midwife or a nurse. In a dream, an eagle also may be interpreted to represent a great ruler, a prophet or a righteous person. Eagle Dream Explanation  The eagle symbolizes a strong man, a warrior who can be trusted neither by a friend nor by a foe. Its baby is an intrepid son who mixes with rulers.
    \textcolor{red}{
    \underline{3efcc0e6934081e9f059d2d82b1152ba} \underline{7885830f9d3a8722f628e2985cd26daf}\\ \underline{b7ee7755f3f3812dc1f0feeca0b62806} \underline{2a2542f9e61a9a1d3b83ae31889ac954} \\ \underline{9aa96309fe5c059b13e87d942ab6d8d9} \underline{c1311fa3447790f02b8e9181846c2205}}
    }
    \label{quote:expansion2}
    \end{lquote}

\end{enumerate}
  
Examples \ref{quote:expansion1} and \ref{quote:expansion2} augment both the query and the passage that were previously demonstrated in Example \ref{quote:one} using explicit and hashed entities respectively. Appending entities to the original text enriches the context and clarifies ambiguous terms. Although this passage was previously considered relevant to the query using a standard BM25 in Example \ref{quote:one}, that is not the case after text expansion. It is important to underline that the expanded passage is not relevant to the expanded query (as you can probably deduce from the appended entities too), I am just using the same query and passage from the previous toy example for the sake of demonstrating the proposed text expansion method.

I have also attempted expansion with entities for the queries only and for the passages only. However, simultaneous query and passage augmentation achieves the best performance. In addition to expansion using one copy of each entity name, I have experimented with weighted expansion reflecting the number of entity mention occurrences in the text, and expansion with a constant factor. Overall, text expansion is performed on MS MARCO passage train and Dev sets. For cases where triple samples of queries, negative passages and positive passages are required as input (such as in the dense retrieval experiments), corpus expansion is conducted by expanding each of the query, negative passage and positive passage; then reconstructing the triple sample.

%===================================================================================================================

\section{Run Combination Strategies}

I adopt the methods explored in this section on sparse retrieval. Hence, my explanation is limited to this context excluding the STAR-ADORE dense retrieval pipeline. To further study the entity linking effect on sparse retrieval, I explore different combination strategies of diverse run types. Four different runs are generated: three runs using BM25 as a sparse retriever: (original run from the non-expanded corpus, run from the entity-equipped dataset, run from the hashed-entity-equipped dataset) and one run using the ANCE dense retriever generated from the original non-expanded corpus. Two main combination schemes are also considered: run fusion and run selection. To better assess the quality of each strategy, it is important to identify an upper-bound recall. As a result, I generate the ``Oracle" run which is defined as the run achieving the best possible performance when adopting an ideal fusion or selection method. An ideal fusion strategy will mix all the runs in the pool with corresponding weights, while an ideal selection scheme will pick the suitable run for each query from a run pool. 

\subsection{Oracle}

In order to estimate the maximum possible recall gain that can be achieved by entity linking, I generate hypothetical Oracle runs for each queries-qrels combination by selecting the run with the highest passage rank for each query. Four query sets are taken into account which are the Dev, Hard, Harder and Hardest sets. The qrel sets involve the Original, MonoT5 and DuoT5 query sets. By the end, I generate 12 Oracle runs for all queries-qrels combinations determining the ideal gain that may be obtained with the help of expansion methods with linked entities.
Given a query set; corpus passages; the corresponding qrels; a run pool of diverse runs that are previously generated using different IR methods on the same queries, passages and qrels; an Oracle run is generated by choosing the run containing the relevant passage to a query. If multiple runs have the same relevant passage to the underlying query, the run where the relevant passage is ranked the highest is chosen among all runs. This selection procedure is repeated for each query until building the Oracle run file with pairs of queries and relevant passages.
If the three pool runs under consideration (with no entities, with entities, with hashed-entities) do not include the passage required by the qrel set (the passage denoted as the correct answer in the ground truth) for a given query, the run selection is performed arbitrarily since the recall is always zero in any case.  For cases where there are multiple judged relevant passages per query (i.e. 2 qrels), I prioritize the judged passage with the highest rank across all runs.

\subsection{Reciprocal Rank Fusion (RFF)}

As previously highlighted, the Oracle runs are an unrealistic combination of the runs to generate an ideal recall.
In order to reduce the margin between the individual runs (no entities, with entities, with hashed entities) and the Oracle results, I attempt to realistically join the three runs together. The first approach I adopt is fusion. I experiment with Reciprocal Rank Fusion (RRF) \cite{rrf} for all the combinations of the three mentioned runs: pairwise RRF between the run without entities and the entity-equipped run, pairwise RRF between the run without entities and the hashed-entity-equipped run, pairwise RRF between the entity-aware and the hashed-entity-aware runs, RRF of the three runs. 

RRF combines the passage rankings from multiple runs by sorting the passages according to a simple scoring formula achieving better results than any individual run. RRF fuses ranks disregarding the arbitrary scores returned by ranking methods. Given $D$ documents to be ranked and $R$ rankings with permutations on $1..|D|$, RRF formula is computed as:
\begin{equation}
    RRFscore (d \in D) = \sum_{r \in R} \frac{1}{60+r(d)}
\end{equation}
The formulation takes into consideration the effect of lower-ranked documents even when they are not as important as the most relevant ones. The constant 60 is found to alleviate the impact of outlier rankings resultant from some IR methods.  

\subsection{BERT-based Classifier}
\label{chp3:bertClassifier}

Instead of the weighted fusion of runs, another idea is to intelligently select the best run by fine-tuning a contextualized pre-trained BERT-based model in a cross-encoder architecture followed by a linear classifier layer to identify the index type on which the retrieval should take place. I first expand the queries and the passages of the training set. Then, BM25 is run on the expanded and the non-expanded versions of the MS MARCO train dataset to produce the main three runs: original with no entities, with entities, with hashed entities. Leveraging the Oracle idea, I construct a label list where for each query, I select the run with the highest ranked relevant passage. The following labels are assigned for each run: 0 for the non-entity run, 1 for the entity-aware run and 2 for the hashed-entity-aware run. These labels are used to fine-tune BERT.

\begin{figure}[htp]
    \centering
    \includegraphics[scale=0.4]{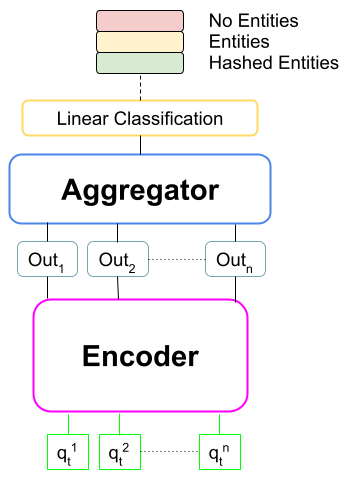}
    \caption{Overview of the Classifier Architecture. (adapted from \cite{arabzadeh2021predicting})}
    \label{fig:classifier}
\end{figure}

The classifier architecture follows the same paradigm previously explored in the work of Arabzadeh et al. \cite{arabzadeh2021predicting}. Figure \ref{fig:classifier} illustrates the classifier architecture where queries are used as inputs. The architecture relies on a contextualized pre-trained embedding representation of the queries, thus considering both the semantics and context. All tokens ( of the query and the special ones) are passed to the cross-encoder architecture. The model conducts full-cross self-attention over the given query and the corresponding label aiming for higher accuracy. Finally, the model utilizes a linear classification layer, and  binary cross entropy loss function for dimensionality reduction and probability computation for each class.

Through my experiments with RRF and BERT-based classification, I aim to perform a comparison between run fusion and run selection approaches to find the most suitable method with the best performance for joining runs generated by different IR methods.

%======================================================================
\chapter{Experiments and Results}
\label{Chapter4}
%======================================================================

\section{Experiments}
\subsection{Experimental Setup}
\begin{itemize}
    \item Sparse experiments are conducted using a cluster of Intel E5-2683 v4 Broadwell @ 2.1Ghz for entity inference on the MS MARCO passage train and Dev sets (both queries and passages), and for Anserini-related experiments such as run generation using the training and the Dev query sets, run evaluation using recall@1000, Oracle generation etc...
    
    \item I use 4 NVIDIA V100 Volta GPUs with 32G HBM2 for constructing the run pool with MonoT5 and DuoT5 re-ranking in order to generate the different qrel sets. 
    
    \item I employ 1 RTX 2080 GPU for training the classifier used for run selection using an input list of queries and corresponding class labels.
    
    \item I use 1 RTX A6000 GPU and 32 NVIDIA V100 Volta GPUs with 32G HBM2 to train each component of the dense STAR-ADORE pipeline.  The warm-up model is trained for 60K steps on the entity-equipped dataset using the 32 GPUS to generate the static hard negatives. The warm-up output is used by STAR to train the dense model. The latter is trained for 50K steps using the RTX A6000 GPU for 1.5 days. STAR checkpoint is finally used as a fixed document encoder, while ADORE is fine-tuned to optimize the query encoder using the same RTX GPU.
\end{itemize}

\subsection{Experiment Types}

I use ELQ extended version for longer texts to generate linked entities for the queries and passages of the MS MARCO passage train and Dev sets. Since a sliding window approach is employed for each text section, I end with duplicate entities for the same text piece. The final entity list of a text passage is then deduplicated.
After the entity generation, another hashed version of these entities is produced. I append both entity types (explicit and hashed) separately to the corresponding query or passage texts. The suggested text augmentation methods result in three types of datasets: the MS MARCO passage, the MS MARCO passage with linked entities, and the MS MARCO passage with linked hashed entities datasets. I experiment with corpus expansion using three approaches: 1) A single copy of the entity name. 2) A constant number of copies of each entity name such as 3 and 5. 3) Weighted expansion according to the number of entity mention occurrences. However, using a single entity term for each detected mention gives the best results, hence I adopt this expansion method in the rest of my experiments. In fact, I have found that the expansion with multiple copies of the same entity is inversely proportional to the recall performance, i.e factor 5 gives worse results than 3. As previously mentioned, two main types of retrieval experiments are conducted:

\subsubsection{Sparse Retrieval}
\begin{itemize}
    \item Anserini is used to generate the index for the three versions of the MS MARCO passage dataset. Since I am concerned with maximizing the performance of the first stage retrieval for later re-ranking, results are evaluated using recall@1000. I use BM25 tuned hyperparamaters ($k_{1}=0.82, b=0.68$) that are optimized for recall@1000 on the MS MARCO dataset in Anserini. Although, I have attempted tuning these parameters on the entity-equipped dataset with both versions: explicit and hashed, there were no considerable changes in the final hyperparamater values.
    \item I evaluate each of the three generated runs (original with no entities, with entities, with hashed entities) against the original qrels.
    \item As a comparative baseline, BM25 run with pseudo-relevance feedback (BM25 + PRF) is generated on the non-expanded dataset to assess the improvement gain introduced by entity linking in comparison with PRF. 
    \item An ANCE dense run is also generated for comparative evaluation purposes.
    \item In order to get a better perception of the entity linking effect on the sparse retrieval, two other types of qrels are formed. A run pool consisting of the three mentioned runs in addition to an ANCE dense run of the standard MS MARCO dataset is re-ranked with MonoT5 and DuoT5 using PyGaggle's implementation to generate the MonoT5 and DuoT5 qrels respectively.
    \item To define an upper bound of the maximum potential gain that may be achieved via entity liking, a hypothetical Oracle run is produced by choosing the best run among (no entities, entity-equipped, hashed-entity-equipped) runs. The Oracle's performance is also evaluated against the three qrel sets.
    \item TREC TOOLS\footnote{\url{https://github.com/joaopalotti/trectools}} are employed to perform reciprocal rank fusion on combinations of the three mentioned runs. I also evaluate the performance of all the RRF runs using the three qrel types.
    \item To perform run selection, the proposed classifier is trained on an input list of queries and corresponding labels where each label designates the run where the relevant document is found and ranked the highest. The classifier's input is generated from the BM25 run files performed on the MS MARCO training set with the three versions: original and expanded with entities, in their explicit and hashed forms. The trained classifier is then used to generate the best run file on the Dev set. The classifier's output run is also evaluated against the original, MonoT5 and DuoT5 qrels.
    \item In addition to the evaluation of the Dev set against three types of qrels, and in order to dive beyond the shallow evaluation of the Dev set, I was interested in the entity effect on the hard queries. The latter are known to have a bad performance not only with sparse retrievers like BM25, but also with neural rankers (i.e. dense retrievers). In order to examine the entity performance on these queries, I filter the queries, the three types of qrels and the previously generated runs (no entity, entity-equipped, hashed-entity-equipped, combinations of RRF, classifier Oracle, and ANCE) runs to evaluate the entity linking effect on the recall performance for the Hard, Harder and Hardest sets.
    
\subsubsection{Dense Retrieval}

For the dense experiments, only the entity-equipped MS MARCO passage dataset with the explicit entity format is utilized. The evaluation is also conducted on the original qrels only. I have limited my experiments for the dense retrieval due to three main reasons: 1) The time and computational resources required to train or fine-tune the selected dense model (i.e. STAR-ADORE pipeline) are huge. Faced with limited resources and time constraints, it was not possible to attempt as many dense-related experiments as needed. 2) Additionally, I decide to apply only the methods that show strong performance with sparse models. For example, sparse experiments have shown promising results for the explicit entities in comparison with their hashed form. As a result, I refrain from training the expensive dense model on the hashed-entity-aware dataset.
3) The results produced from the dense pipeline were not encouraging to continue additional exploration in that direction. Nonetheless, further investigation is still required in this context before reaching a definitive conclusion on entity linking impact on dense retrieval.

Before starting training the dense retrieval model proposed in \cite{staradore}, the MS MARCO dataset is pre-processed as per the required steps\footnote{\url{https://github.com/jingtaozhan/DRhard}}. 
The first stage of the Warmup-STAR-ADORE pipeline is the ANCE warm-up model. It relies on training a pre-trained BM25 warm-up checkpoint for 60k steps on the entity-equipped MS MARCO passage training set according to ANCE instructions\footnote{\url{https://github.com/microsoft/ANCE/blob/master/README.md}}. The generated warm-up model is then used to generate the static hard negatives required by the second stage of the pipeline: STAR. The latter leverages the generated static hard negatives to train the dense model for 50k steps. Finally, the generated STAR checkpoint is employed as a fixed document encoder, while ADORE is fine-tuned on the pre-processed data as a query encoder till convergence (~1 epoch). ADORE computes the query embeddings. The document embeddings are pre-computed by STAR. The evaluation is conducted for the whole pipeline on the Dev set against the original qrels using MRR@10.
    
\end{itemize}

%=========================================================================================================================
\section{Results and Evaluation}

Before jumping to the evaluation of the achieved results, Let's revisit the previous three examples of obstinate queries: Hard \ref{quote:hard_set_noentities}, Harder \ref{quote:harder_set_noentities}, and Hardest \ref{quote:one} after query and document expansion with linked entities.  Below are the same examples of the queries with the corresponding documents previously judged as relevant by BM25, and the new retrieved documents also by BM25 after corpus expansion. One can clearly see the impact of entity linking on query disambiguation, and the relevance quality of answer retrieval.

\begin{enumerate}
    
    \item \textbf{Hard query}
    \begin{lquote}
        \textbf{Query:} \textit{how many people die from snake bites a year? \textcolor{red}{\underline {Snakebite}}}
        
        \textbf{Previous Relevant Passage by BM25: }\textit{The exact percentage of a dry bite varies from venomous snake to venomous snake, but, for instance, around 50\% of Coral Snake bites are dry bites, delivering no venom. In fact, only 9-15 people per year in the U.S. die from snake bites out of about 8000 bites from venomous snakes per year. \textcolor{red}{\underline{United States} \underline{Venomous snake} \underline{Coral snake}}}
        
        \textbf{Current Relevant Passage by BM25: }\textit{Globally snake bite affects the lives of some 4.5 million people every year, and conservative estimates suggest that at least 100,000 people die from snake bite, and another 250,000 are permanently disabled. \textcolor{red}{\underline{Snakebite} \underline{Disability}}}
        \label{quote:hard_set_entities}
    \end{lquote}
    
    \item \textbf{Harder query}
    
    \begin{lquote}
        \textbf{Query:} \textit{what is medical term for neck fusion? \textcolor{red}{\underline{Spinal fusion} \underline{Medicine}}}
    
        \textbf{Previous Relevant Passage by BM25: }\textit{ASA is the medical abbreviation for what medical term. What is the abbreviation for chronic obstructive pulmonary disease. What is the abbreviation for head, eyes, ears, neck, and throat. What is the abbreviation for complaint of. This abbreviation lbstands for what medical term. \textcolor{red}{\underline{Chronic obstructive pulmonary disease} \underline{American Society of Anesthesiologists}}}
        
        \textbf{Current Relevant Passage by BM25: }\textit{During spinal fusion. Fusion from back of neck When spinal fusion is performed from the back of the neck (posterior cervical fusion), rods and screws are used to hold the vertebrae together. Fusion from front of neck In some cases, surgery on your neck (cervical) vertebrae occurs from the front (anterior) side of your neck. \textcolor{red}{\underline{Spinal fusion} \underline{Screw} \underline{Surgery} \underline{Rod end bearing}}}
        
        \label{quote:harder_set_entities}
    \end{lquote}
    
    \item \textbf{Hardest query}
    
    \begin{lquote}
    \textbf{Query:} \textit{ who are in the eagles \textcolor{red}{\underline{Eagles (band)}}}
    
    \textbf{Previous Relevant Passage by BM25:} \textit{ The scratch of an eagle in a dream means a sickness. A killed eagle in a dream means the death of a ruler. If a pregnant woman sees an eagle in her dream, it means seeing a midwife or a nurse. In a dream, an eagle also may be interpreted to represent a great ruler, a prophet or a righteous person. Eagle Dream Explanation  The eagle symbolizes a strong man, a warrior who can be trusted neither by a friend nor by a foe. Its baby is an intrepid son who mixes with rulers.
    \textcolor{red}{\underline{Prophet} \underline{Eagle} \underline{Midwife} \underline{Dream} \underline{Oy, to ne vecher} \underline{Nursing}}
    }
    
    \textbf{Current Relevant Passage by BM25:} \textit{ Who are the original members of The Eagles rock band? Glenn Frey, Don Henley, Bernie Leadon and Randy Meisner are the four original members who formed The Eagles rock band in Los Angeles, California in 1971.     \textcolor{red}{\underline{Glenn Frey} \underline{Don Henley} \underline{Randy Meisner} \underline{Bernie Leadon} \underline{Los Angeles} \underline{California} \underline{Eagles (band)}}}
    \label{quote:hardest_set_entities}
    \end{lquote}
    
\end{enumerate}

%==========================================================
\subsection{Sparse Retrieval}

\subsubsection{Evaluation against the Original Qrels}

\begin{figure}%
    \centering
    \subfloat[\centering The recall curves of the three main BM25 runs (without entities, with entities, with hashed entities) are represented in red, pink and green respectively. The hypothetical Oracle in blue represents the ideal gain achieved by entity linking by choosing the best run among the mentioned three for each query. The Dense yellow curve is generated from an ANCE run on the original MS MARCO passage Dev set.]
    {{\includegraphics[scale=0.26]{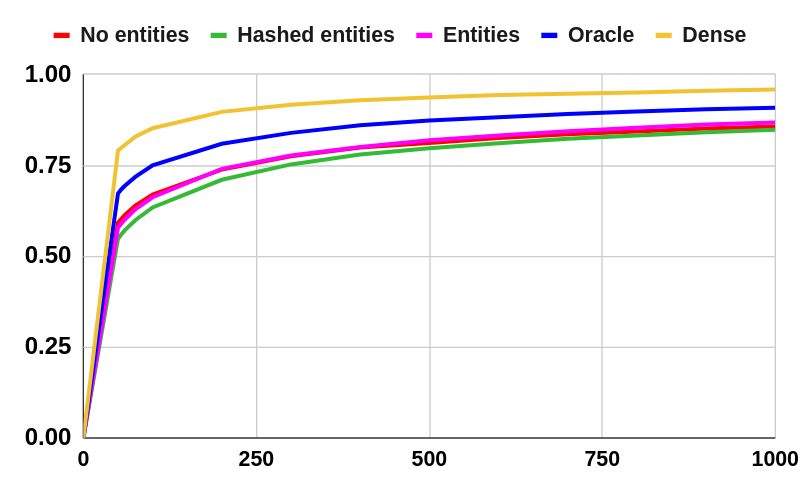} }} %
    \qquad
    \subfloat[\centering The graph shows all the generated RRF curves for the different run combinations on the MS MARCO passage Dev set. Ordered from best to worst, the RRF of the three main runs (blue) tops the list, then come the pairwise RRF curves (original/entities, original/hashed entities, entities/hashed entities) represented in red, green and orange respectively.]
    {{\includegraphics[scale=0.26]{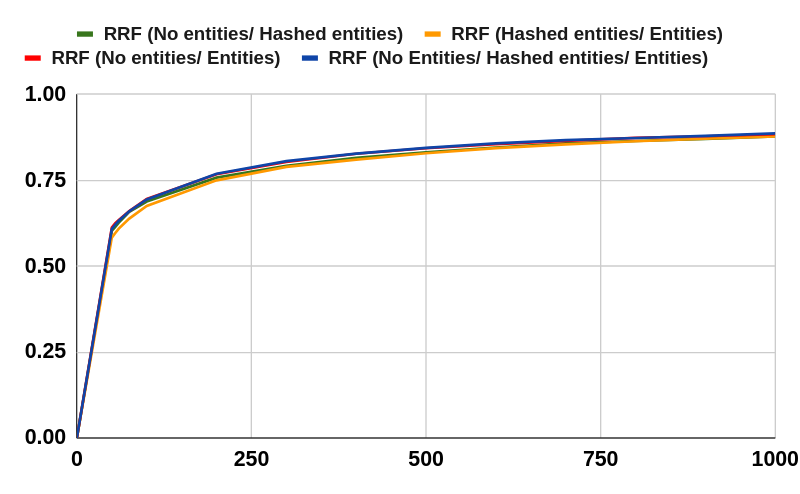} }}%
    \qquad
    \subfloat[\centering The best RRF curve is generated from fusing the three main runs. The corresponding curve is represented in red. The Classifier run is produced by selecting the best run containing the relevant document, and is represented in blue. As shown, the red curve outperforms the blue one for the Dev set.]
    {{\includegraphics[scale=0.26]{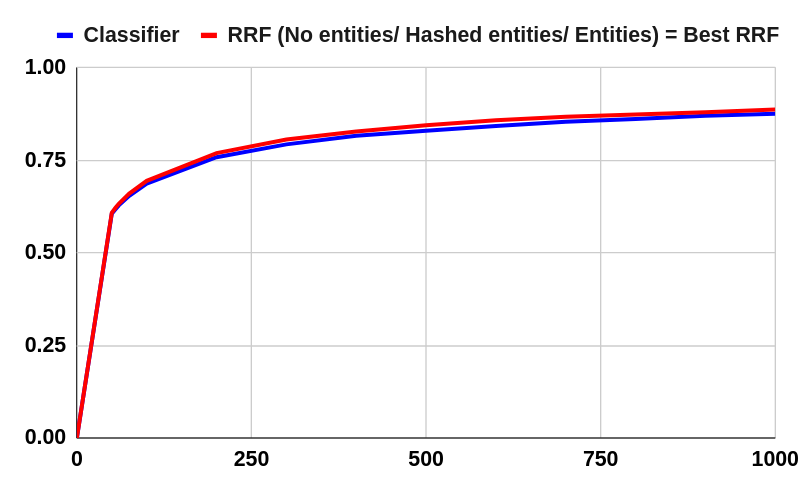} }}%
    \qquad
    \subfloat[\centering Although the ANCE (yellow) curve is the highest, entity linking introduces a significant improvement to the classical BM25 run (red). The Oracle (blue) shows the maximum possible gain achieved by entity linking. The best RRF of run combinations, shown in green, mediates the original BM25 (red) and the ANCE (yellow) runs. The same goes for the Oracle, suggesting the reduction of the recall gap between sparse and dense retrievers.]
    {{\includegraphics[scale=0.26]{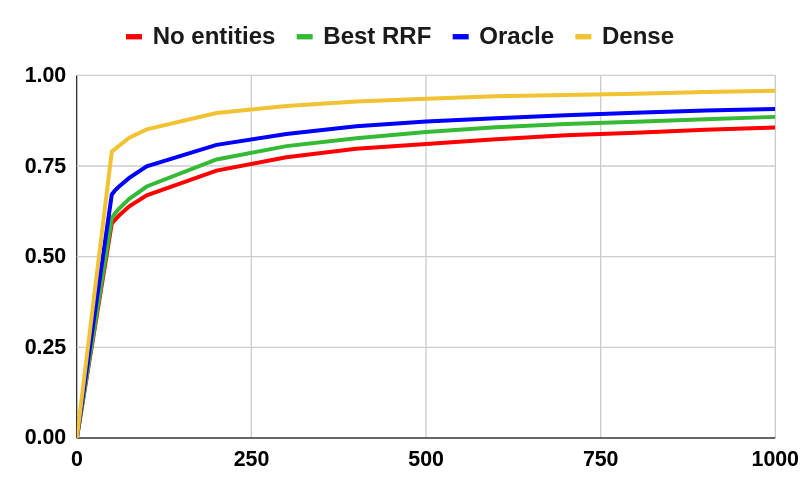} }}%
    \caption{Recall curves of the Dev query set with respect to the original qrels. The x-axis shows the cutoffs, and the y-axis is the corresponding recall value.}%
    \label{fig:dev_org}%
\end{figure}

\begin{figure}%
    \centering
    \subfloat[\centering The recall curves of the three main BM25 runs (without entities, with entities, with hashed entities) are represented in red, pink and green respectively. The hypothetical Oracle in blue represents the ideal gain achieved by entity linking by choosing the best run among the mentioned three for each query. The Dense yellow curve is generated from an ANCE run on the Hard set of the original MS MARCO passage.]
    {{\includegraphics[scale=0.26]{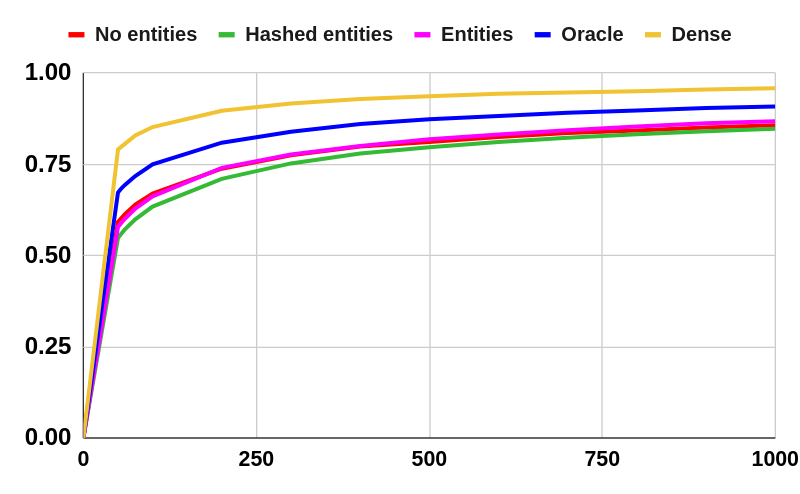} }}%
    \qquad
    \subfloat[\centering The graph shows all the generated RRF curves for the different run combinations on the MS MARCO passage Hard set. Ordered from best to worst, the RRF of the three main runs (blue) tops the list, then come the pairwise RRF curves (original/entities, original/hashed entities, entities/hashed entities) represented in red, green and orange respectively.]
    {{\includegraphics[scale=0.26]{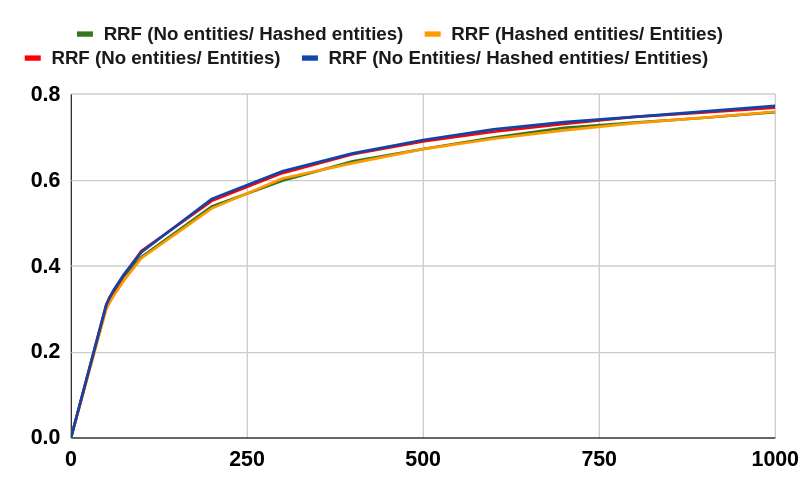} }}%
    \qquad
    \subfloat[\centering The best RRF curve is generated from fusing the three main runs. The corresponding curve is represented in red. The Classifier run is produced by selecting the best run containing the relevant document, and is represented in blue. As shown, the red curve outperforms the blue one for the Hard query set.]
    {{\includegraphics[scale=0.26]{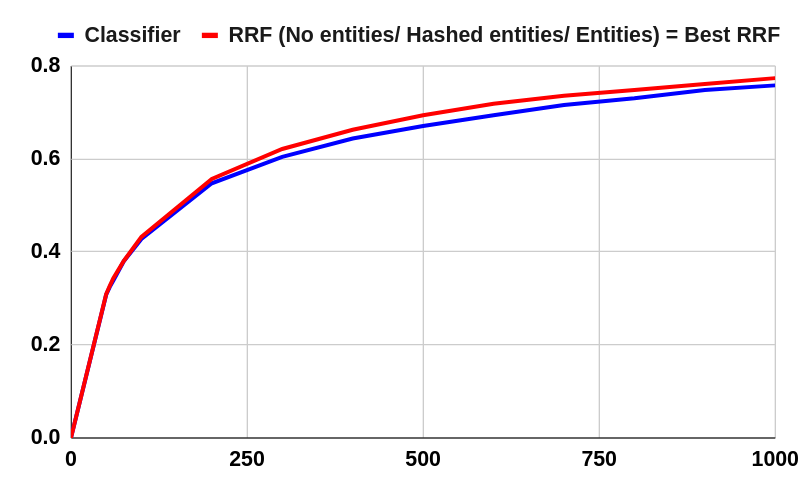} }}%
    \qquad
    \subfloat[\centering Although the ANCE (yellow) curve is the highest, entity linking introduces a significant improvement to the classical BM25 run (red). The Oracle (blue) shows the maximum possible gain achieved by entity linking. The best RRF of run combinations, shown in green, mediates the original BM25 (red) and the ANCE (yellow) runs. The same goes for the Oracle, suggesting the reduction of the recall gap between sparse and dense retrievers.]
    {{\includegraphics[scale=0.26]{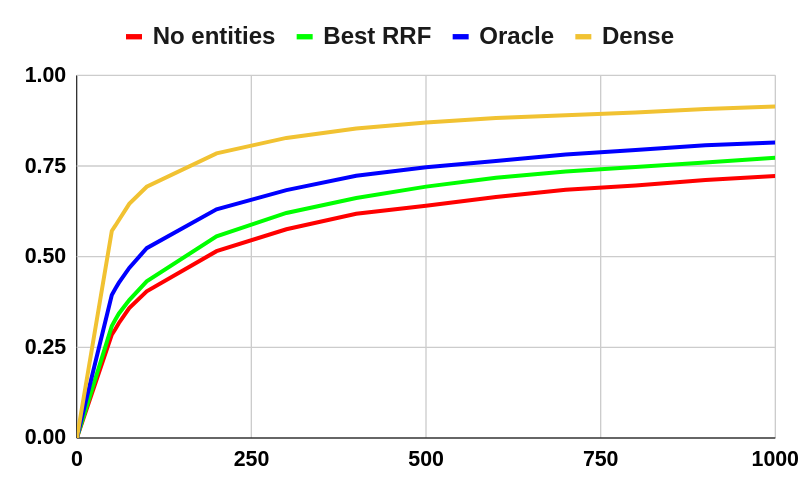} }}%
    \caption{Recall curves of the Hard query set with respect to the original qrels. The x-axis shows the cutoffs, and the y-axis is the corresponding recall value.}%
    \label{fig:veiled_org}%
\end{figure}

\begin{figure}%
    \centering
    \subfloat[\centering The recall curves of the three main BM25 runs (without entities, with entities, with hashed entities) are represented in red, pink and green respectively. The hypothetical Oracle in blue represents the ideal gain achieved by entity linking by choosing the best run among the mentioned three for each query. The Dense yellow curve is generated from an ANCE run on the Harder set of the original MS MARCO passage.]
    {{\includegraphics[scale=0.26]{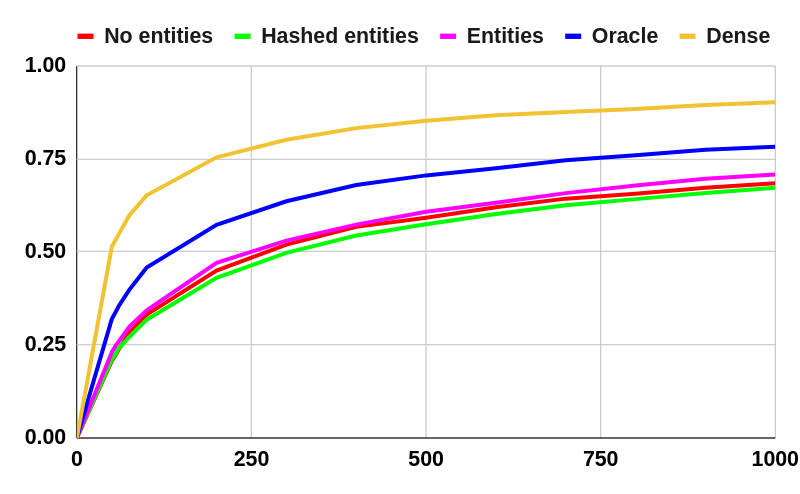} }}%
    \qquad
    \subfloat[\centering The graph shows all the generated RRF curves for the different run combinations on the MS MARCO passage Harder set. Ordered from best to worst, the RRF of the three main runs (blue) tops the list, then come the pairwise RRF curves (original/entities, original/hashed entities, entities/hashed entities) represented in red, green and orange respectively.]
    {{\includegraphics[scale=0.26]{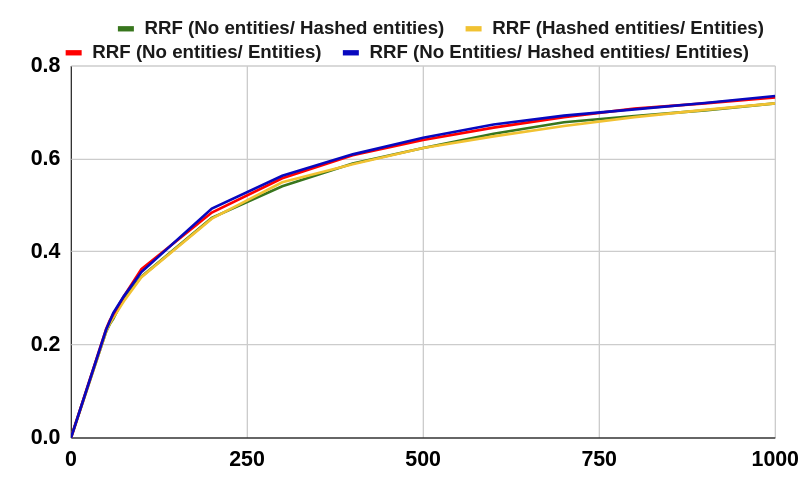} }}%
    \qquad
    \subfloat[\centering The best RRF curve is generated from fusing the three main runs. The corresponding curve is represented in red. The Classifier run is produced by selecting the best run containing the relevant document, and is represented in blue. As shown, the red curve outperforms the blue one for the Harder query set.]
    {{\includegraphics[scale=0.26]{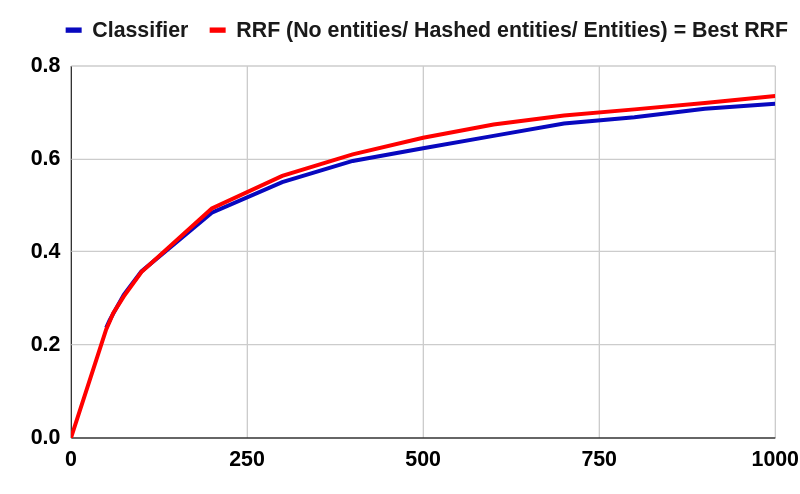} }}%
    \qquad
    \subfloat[\centering Although the ANCE (yellow) curve is the highest, entity linking introduces a significant improvement to the classical BM25 run (red). The Oracle (blue) shows the maximum possible gain achieved by entity linking. The best RRF of run combinations, shown in green, mediates the original BM25 (red) and the ANCE (yellow) runs. The same goes for the Oracle, suggesting the reduction of the recall gap between sparse and dense retrievers.]
    {{\includegraphics[scale=0.26]{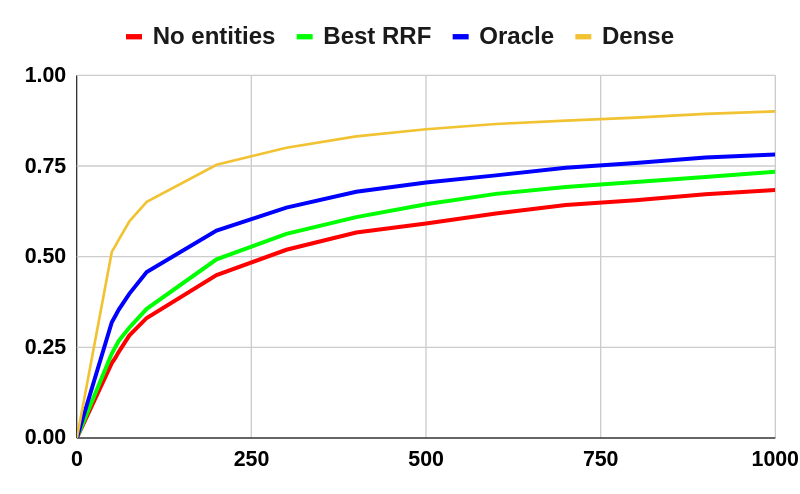} }}%
    \caption{Recall curves of the Harder query set with respect to the original qrels. The x-axis shows the cutoffs, and the y-axis is the corresponding recall value.}%
    \label{fig:pygmy_org}%
\end{figure}

\begin{figure}%
    \centering
    \subfloat[\centering The recall curves of the three main BM25 runs (without entities, with entities, with hashed entities) are represented in red, pink and green respectively. The hypothetical Oracle in blue represents the ideal gain achieved by entity linking by choosing the best run among the mentioned three for each query. The Dense yellow curve is generated from an ANCE run on the Hardest set of the original MS MARCO passage.]
    {{\includegraphics[scale=0.26]{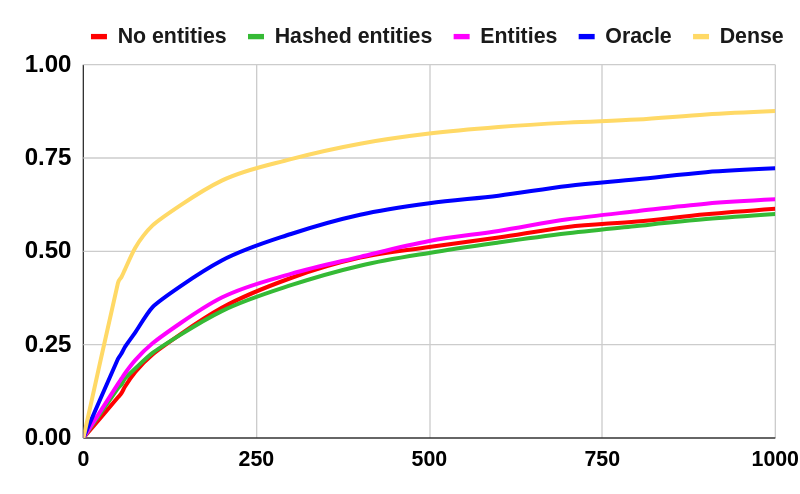} }}%
    \qquad
    \subfloat[\centering The graph shows all the generated RRF curves for the different run combinations on the MS MARCO passage Hardest set. Ordered from best to worst, the RRF of the three main runs (blue) tops the list, then come the pairwise RRF curves (original/entities, original/hashed entities, entities/hashed entities) represented in red, green and orange respectively.]
    {{\includegraphics[scale=0.26]{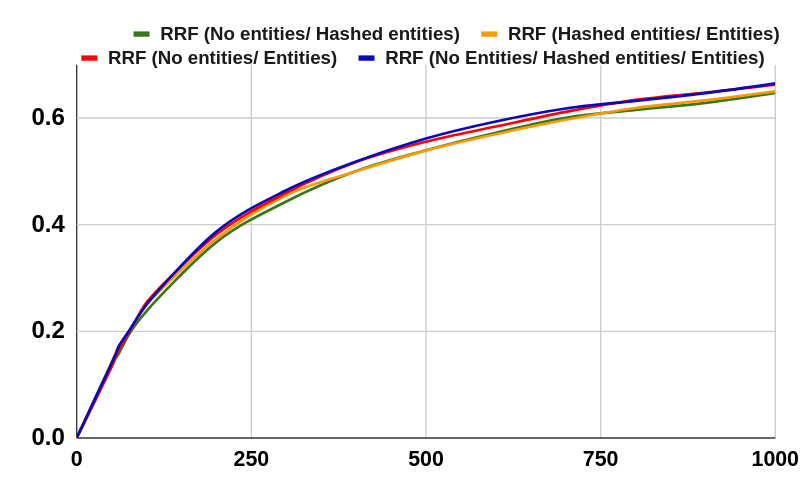} }}%
    \qquad
    \subfloat[\centering The best RRF curve is generated from fusing the three main runs. The corresponding curve is represented in red. The Classifier run is produced by selecting the best run containing the relevant document, and is represented in blue. As shown, the red curve outperforms the blue one for the Hardest query set.]
    {{\includegraphics[scale=0.26]{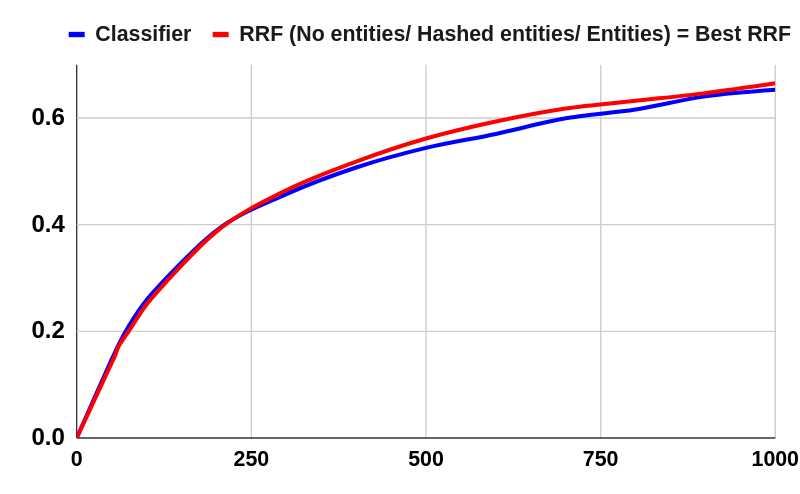} }}%
    \qquad
    \subfloat[\centering Although the ANCE (yellow) curve is the highest, entity linking introduces a significant improvement to the classical BM25 run (red). The Oracle (blue) shows the maximum possible gain achieved by entity linking. The best RRF of run combinations (green) and the Oracle (blue) mediate the original BM25 (red) and the ANCE (yellow) runs, suggesting the reduction of the recall gap between sparse and dense retrievers.]
    {{\includegraphics[scale=0.26]{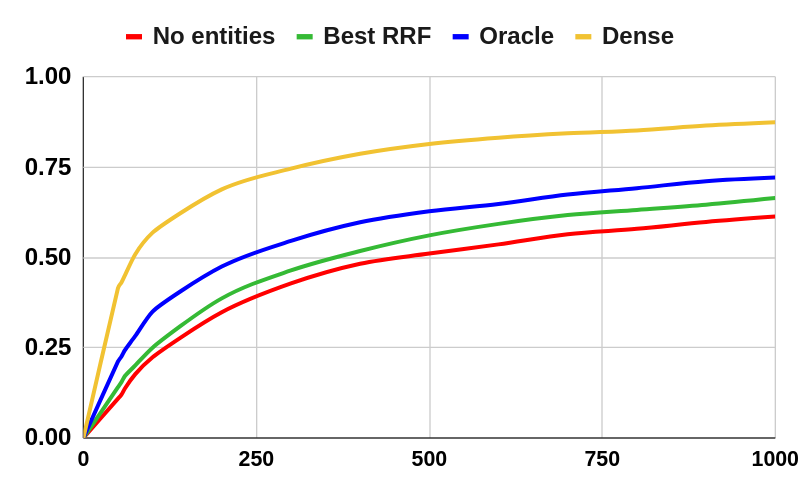} }}%
    \caption{Recall curves of the Hardest query set with respect to the original qrels. The x-axis shows the cutoffs, and the y-axis is the corresponding recall value.}%
    \label{fig:lesser_org}%
\end{figure}

Table \ref{tab:recall_org} and Figures \ref{fig:dev_org}, \ref{fig:veiled_org}, \ref{fig:pygmy_org} and \ref{fig:lesser_org} summarize my evaluation results for the Dev, Hard, Harder and Hardest query sets against the original qrels.
As shown in Table \ref{tab:recall_org}, entity-equipped runs, using the entity explicit format (row 3), give better recall performance compared to the original BM25 runs with no entities (row 1) across the MS MARCO Dev set and the three sets of obstinate queries.  The improvement gain is observed even without the adoption of further run fusion approaches. This result demonstrates that semantic expansion helps rankers disambiguate the hard queries. To further investigate the entity effect, I experiment with the performance with the hashed version. One can observe that the individual hashed-entity-equipped runs (row 2) have worse recall results than the original ones (row 1). Nonetheless, the pairwise reciprocal rank fusion between the original runs and those with the hashed entities (row 5) outperforms the three individual runs: original, with hashed entities, with entities  (i.e. the first 3 rows) for all types of queries. This could be justified as the runs expanded with hashed entities fetch complementary results that are not retrieved by BM25 using the non-expanded dataset. Nonetheless, further investigation is still required to hypothesize the bad performance of the individual hashed-entity-equipped runs.
The best recall results are achieved using the RRF of the three runs with a statistically significant performance improvement (p-value $< 0.01$) of 3.44\%, 6.97\%, 7.36\% and 8.38\% for the Dev, Hard, Harder and Hardest query sets respectively.  The statistical significance of the results was verified using paired t-test. 
As shown, the RRF of all possible run combinations (pairwise or triple-wise) beats the classifier (row 4) that shows a modest improvement over the entity-equipped run (row 3). 
I also notice that the hashed entity-equipped run contributes to the overall gain by only a small factor. This can be clearly seen when comparing the results of the pairwise RFF of the no-entity and the entity-aware runs (row 7), and the RRF of the three runs (row 8). The hypothetical Oracle runs exceed my best-achieved results (RRF) with percentages of 2.47\%, 5.44\%, 6.45\% and 8.57\% for the very same sets demonstrating that room for improvement remains available with the right run combination or selection strategy. 
The latter is worth exploring in a related future work.
In addition to the Oracle (row 10) and the ANCE (row 11) results that I use as a comparative reference, I also investigate the pseudo-relevance feedback (PRF) effect on the non-expanded MS MARCO (row 9). Although PRF causes a significant gain with a recall@1000 of 0.8759 compared to 0.8573 on the Dev set, I refrain from including the costly PRF in my entity-related experiments. It is interesting though to examine PRF effect on the entity-aware dataset with both types: explicit and hashed. One can also see that RRF of the three runs (row 8) still outperforms the BM25+PRF non-expanded run (row 9) across the Dev, Hard and Harder set. However, PRF results are still slightly higher for the Hardest set.

Figure \ref{fig:dev_org} shows the recall curves of my different runs on the Dev set evaluated against the original qrels, where the x-axis represents the different cutoffs, and the y-axis is the corresponding recall value at a given cutoff. Sub-figure \ref{fig:dev_org} (a) compares between the main runs without any combination or selection strategy. These runs include the original BM25 run generated on the non-expanded MS MARCO Dev set (red), the entity-equipped run (pink), the hashed-entity-equipped run (green), the Oracle run (blue) and the ANCE run (yellow). As previously highlighted, the Oracle signifies the maximum recall gain from linked entities by ideally selecting the suitable run among the first three mentioned ones. ANCE is used for comparison purposes to assess the quality of the obtained gain from semantic linking. As illustrated in (a), the entity-aware run (pink) introduces a performance gain in comparison with the original run (red). Surprisingly, the hashed-entity-run (green) gives the worst recall value, even worse than the original red curve. This result is true even after tuning BM25 hyperparameters on the hashed-entity-aware dataset. As expected, the Oracle run (blue) generates the ideal recall gain achieved from linked entities, while the ANCE run, my representative of the dense retrievers in these experiments, is still the best of all. Nonetheless, one can see that the recall margin between the BM25 and the ANCE runs (red and yellow curves) has shrunk with the help of linked entities (blue and yellow curves).
Sub-figure \ref{fig:dev_org} (b) demonstrates all of my RRF combinations of the three main runs (no entities, with entities, and with hashed entities). They consist of the three pairwise RRF curves, and the RRF of the three runs fused together. As shown from the graph, RRF of the three runs (blue curve) performs the best, then come the pairwise RRF between the original run and the entity-aware run (red), the pairwise RRF between the original run and the hashed-entity-aware run (green), and finally, the pairwise RRF between the entity-aware runs with both forms: explicit and hashed (orange) which gives relatively the worst recall performance. I can also observe that although the standalone hashed-entity-aware run performs poorly in (a) (even worse than the original run), fusing the hashed run with the other runs improves the overall performance since it retrieves complementary results in addition to those found by the original and the entity-equipped runs. In addition, it is noticed that the orange curve still introduces a recall improvement over each individual run despite being relatively the worst among RRF runs.
Sub-figure \ref{fig:dev_org} (c) presents a comparison between the best RRF achieved (which is the RRF of the three runs as deduced from (b)) represented by the red curve, and the classifier selection method (blue curve). The red curve comes in a higher position in comparison with the blue one. As a result, it is safe to conclude that the RRF outperforms the classifier's selection method for the Dev set. Yet, further experiments are still needed before judging that a run fusion is generally better than a run selection approach.
Sub-figure \ref{fig:dev_org} (d) illustrates the effectiveness differences between the recall curves of four main runs: the original BM25 with no entities (red), the best combination of no-entity and entity-aware BM25 runs that is achieved by RRF for a given query set (green), the hypothetical Oracle (blue) and the ANCE run (yellow). These curves cover the Dev query set. Like all the previous curves, the x-axis represents the different cutoffs, while the y-axis shows the corresponding recall results. As demonstrated by the yellow curves in (d), and also in the dense results of Table \ref{tab:recall_org}, ANCE retrieval still outperforms all BM25-dependent retrieval by a significant margin. Nonetheless, I observe that the effectiveness difference between BM25 and ANCE has considerably decreased with the help of semantic linking. The Oracle curve suggests that an additional performance improvement is still possible by taking advantage of linked entities, further reducing the recall gap between sparse and dense retrievers.

Figures \ref{fig:veiled_org}, \ref{fig:pygmy_org} and \ref{fig:lesser_org} show the recall curves on the Hard (Veiled), Harder (Pygmy) and Hardest (Lesser) query sets of the MS MARCO Chameleons dataset respectively, evaluated against the original qrels. The x-axis represents the different cutoffs, and the y-axis is the corresponding recall value at a given cutoff. The same pattern observed on the Dev set can also be found in the three sets of obstinate queries. The latter is true for all the Sub-figures (a), (b), (c) and (d) for each query set. Nonetheless, one can observe that for all the figures, the more difficult the underlying query set, the greater the recall improvement introduced by the suggested methods. For example, by looking at the recall@1000 value of the two curves: original BM25 (red) and entity-equipped BM25 (pink) in the Sub-figure (a) for all of \ref{fig:veiled_org}, \ref{fig:pygmy_org} and \ref{fig:lesser_org}, it is shown that the gain achieved by simply appending the entities to the original text is 3.22\%, 3.36\% and 4.12\% for the Hard, Harder and Hardest queries respectively, while the corresponding value for the Dev set is 1.27\%. In addition, the Oracle's recall@1000 in (a) for the three Figures of the hard sets highlights that the maximum potential recall gain for the very same sets is 12.78\%, 14.28 \% and 17.67\% (compare the red and blue curves). On the other hand, the recall@1000 improvement of the Oracle run over a traditional BM25 for the Dev set is only 6.00\%. Sub-figure(c) in the three Figures of the hard sets also elaborates on the same idea. The classifier's recall@1000 gain over the original BM25 is 4.81\%, 4.96\% and 6.39\% for the hard queries ordered in an increased level of difficulty, while the corresponding value for the Dev set is 2.11\%. In addition, my best RRF's recall@1000 gain is 3.44\%, 6.97\%, 7.36\% and 8.38\% for the Dev, Hard, Harder and Hardest query sets respectively. By taking a closer look at each Sub-figure, it is observed that 
Sub-figures \ref{fig:veiled_org} (a), \ref{fig:pygmy_org} (a) and \ref{fig:lesser_org} (a) underline that expansion with entities improves the first stage recall generated by BM25 as shown between the pink (run with entities) and red (run without entities) curves. The Oracle (blue) demonstrates that a recall enhancement potential is still ideally possible leveraging entity linking methods. The Dense retrieval is still unarguably the best for all the Hard sets. Nonetheless, the performance margin between the sparse model (represented by BM25) and the dense model (represented by ANCE) is narrowed. Take a look at the curve difference between the original BM25 and ANCE curves (red and yellow), and between the ideal BM25 with entity support and ANCE curves (blue and yellow). Again, the hashed-entity-aware run shows poor performance across all the hard query types.
Sub-figures \ref{fig:veiled_org} (b), \ref{fig:pygmy_org} (b) and \ref{fig:lesser_org} (b) accentuate that the RRF of the three main BM25 runs: original and entity-equipped (explicit and hashed) gives the best recall performance with the original qrels among all the other combinations of run fusion. The pairwise RRF between the two runs with entities offers the least recall performance gain compared to the other RRF runs. Nonetheless, it is worth mentioning that all the RRF runs still perform better than the three standalone main runs. The latter is true for the Hard, Harder and Hardest query sets.
Sub-figures \ref{fig:veiled_org} (c), \ref{fig:pygmy_org} (c) and \ref{fig:lesser_org} (c) show that the best RRF run also beats the classifier's method of run selection for the sets of obstinate queries. However, the run generated by the classifier still outperforms each of the three independent runs.
Sub-figures \ref{fig:veiled_org} (d), \ref{fig:pygmy_org} (d) and \ref{fig:lesser_org} (d) highlights that the semantic linking is effective in reducing the performance gap between sparse retrievers (red) and dense retrievers (yellow). The best RRF (green) represents the maximum result I was able to actually achieve leveraging entity linking, while the Oracle (blue) is the ideal recall gain that can be achieved.

Further information about the generation of the recall curves for Figures \ref{fig:dev_org}, \ref{fig:veiled_org}, \ref{fig:pygmy_org} and \ref{fig:lesser_org} can be found in Tables \ref{tab:recall_dev_org}, \ref{tab:recall_veiled_org}, \ref{tab:recall_pygmy_org} and \ref{tab:recall_lesser_org} respectively in Appendix \ref{AppendixA}.

\begin{table}[!t] 
\centering
\caption{Recall@1000 of the 4 query sets (Dev, Hard, harder and Hardest) with respect to the original qrels. PRF stands for Pseudo-Relevance Feedback.
}

\begin{tabular}{l|cccc} 
% \hline          %inserts double horizontal lines
% \cline{2-9}
&\multicolumn{3}{c} \textbf{\textbf{Query Set Type}}
 \\ [0.5ex]
\hline
\textbf{Run type}&\textbf{ Dev}&\textbf{Hard}&\textbf{Harder}&\textbf{Hardest}
\\[0.5ex] % inserts table 
%heading
\hline 
No entities&0.8573&0.7234&0.6849&0.6136 \\
Hashed entities&0.8479&0.7146&0.6727&0.5995\\
Entities&0.8682&0.7467&0.7079&0.6389 \\
Classifier&0.8754&0.7582&0.7189&0.6528 \\
No entities/ Hashed entities  RRF&0.8780&0.7591&0.7195&0.6471 \\
Hashed entities/ Entities RRF&0.8784&0.7599&0.7196&0.6498 \\
No entities/ Entities RRF&0.8844&0.7695&0.7323	&0.6625 \\
No entities/ Entities/ Hashed RRF&\textbf{0.8868$^{}$}&\textbf{0.7738$^{}$}&\textbf{0.7353$^{}$}&\textbf{0.6650$^{}$}\\
\hline
No entities + PRF &0.8759&0.7622&0.7272&0.6674\\
\hline
Oracle&0.9087&0.8159&0.7827&0.7220\\
\hline
Dense&0.9587&0.9152&0.9022& 0.8753 \\
\hline
\end{tabular} 
\label{tab:recall_org}
\end{table}

%%%%%%%%%%%%%%%%%%%%%%%%%%%%%%%%%%%%%%%%%%%%%%%%%%%%%%%%%%
%========Dev queries with Original qrels================

%========Hard queries with Original qrels================

%========Harder queries with Original qrels================

%========Hardest queries with Original qrels================

%%%%%%%%%%%%%%%%%%%%%%%%%%%%%%%%%%%%%%%%%%%%%%%%%%%%%%%%%%%%====================================================================================================================================================================================

\subsubsection{Evaluation against MonoT5 Qrels}

\begin{figure}%
    \centering
    \subfloat[\centering The recall curves of the three main BM25 runs (without entities, with entities, with hashed entities) are represented in red, pink and green respectively. The hypothetical Oracle in blue represents the ideal gain achieved by entity linking by choosing the best run among the mentioned three for each query. The Dense yellow curve is generated from an ANCE run on the original MS MARCO passage Dev set.]
    {{\includegraphics[scale=0.26]{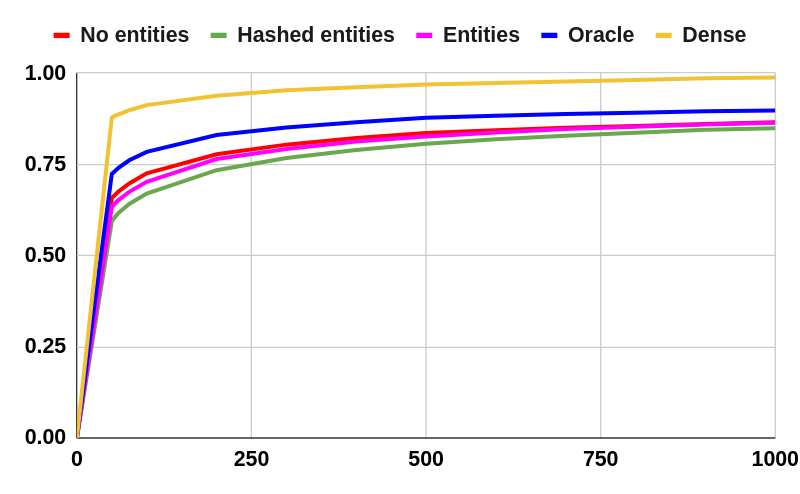} }}%
    \qquad
    \subfloat[\centering The graph shows all the generated RRF curves for the different run combinations on the MS MARCO passage Dev set. Ordered from best to worst, the
    pairwise RRF between the original and the entity-aware run (red) tops the list, then come the RRF of the three main runs (blue), and the pairwise RRF curves (original/hashed entities, entities/hashed entities) represented in green and orange respectively.]
    {{\includegraphics[scale=0.26]{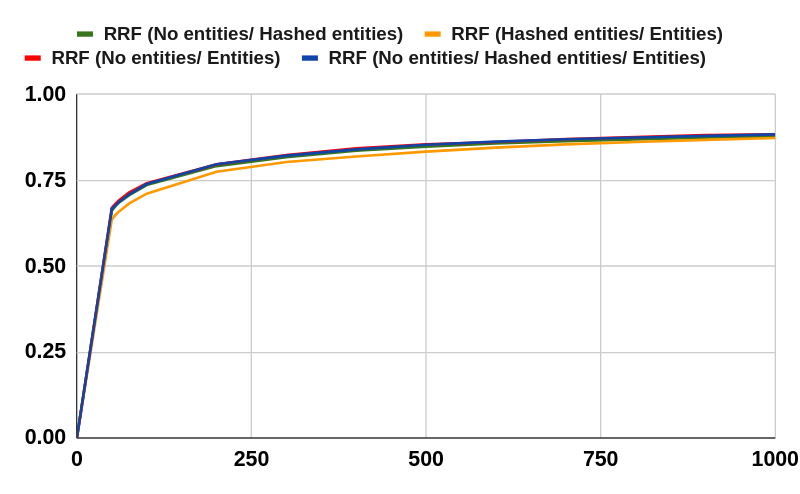} }}%
    \qquad
    \subfloat[\centering The best RRF curve is generated from fusing the BM25 original run and the entity-aware run in the explicit entity format. The corresponding curve is represented in red. The Classifier run is produced by selecting the best run containing the relevant document, and is represented in blue. As shown, the red curve outperforms the blue one for the Dev set.]
    {{\includegraphics[scale=0.26]{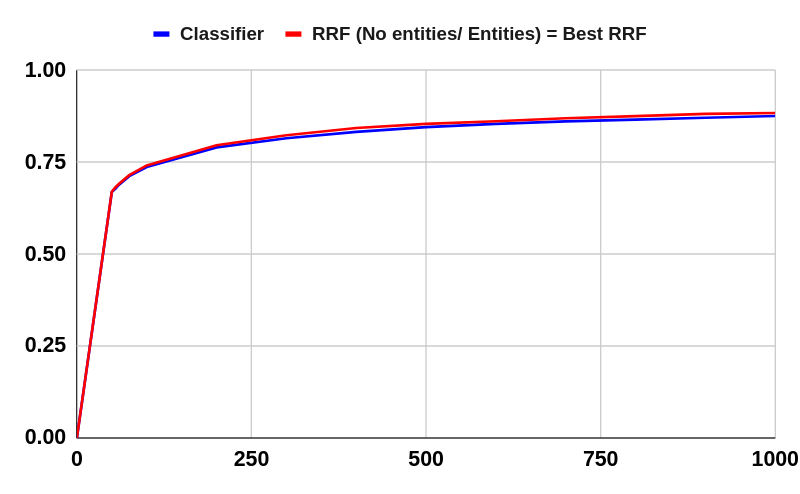} }}%
    \qquad
    \subfloat[\centering Although the ANCE (yellow) curve is the highest, entity linking introduces a significant improvement to the classical BM25 run (red). The Oracle (blue) shows the maximum possible gain achieved by entity linking. The best RRF of run combinations, shown in green, mediates the original BM25 (red) and the ANCE (yellow) runs. The same goes for the Oracle, suggesting the reduction of the recall gap between sparse and dense retrievers.]
    {{\includegraphics[scale=0.26]{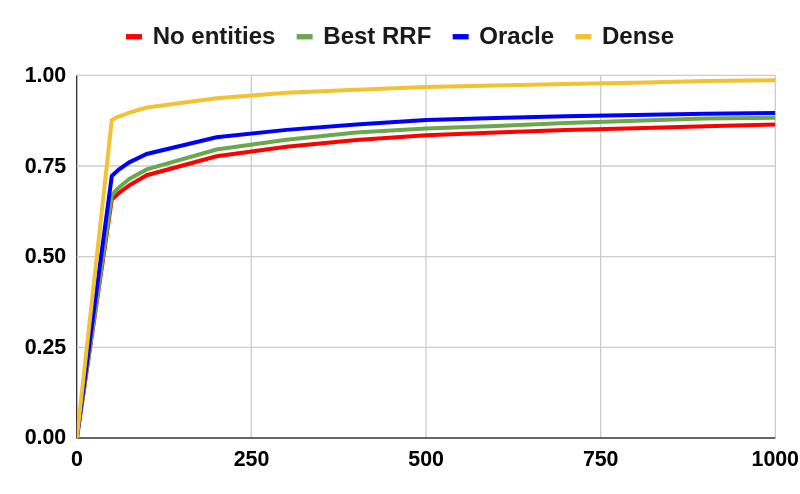} }}%
    \caption{Recall curves of the Dev query set with respect to MonoT5 qrels. The x-axis shows the cutoffs, and the y-axis is the corresponding recall value.}%
    \label{fig:dev_monoT5}%
\end{figure}

\begin{figure}%
    \centering
    \subfloat[\centering The recall curves of the three main BM25 runs (without entities, with entities, with hashed entities) are represented in red, pink and green respectively. The hypothetical Oracle in blue represents the ideal gain achieved by entity linking by choosing the best run among the mentioned three for each query. The Dense yellow curve is generated from an ANCE run on the original MS MARCO passage Hard set.]
    {{\includegraphics[scale=0.26]{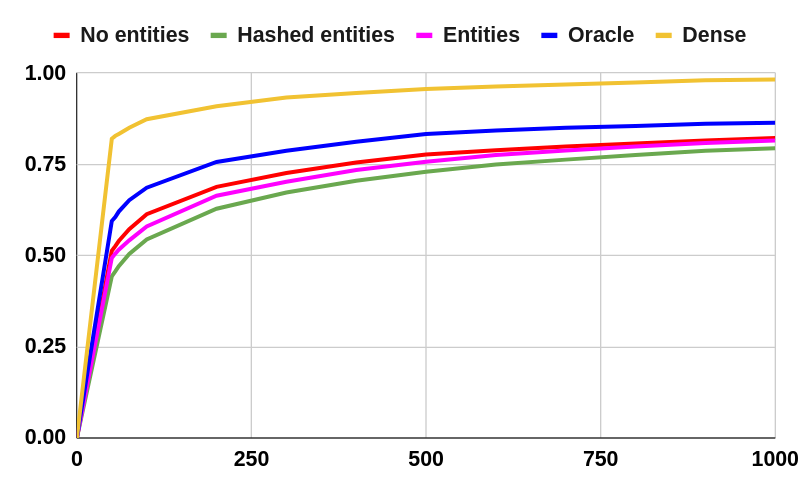} }}%
    \qquad
    \subfloat[\centering The graph shows all the generated RRF curves for the different run combinations on the MS MARCO passage Hard set. Ordered from best to worst, the
    pairwise RRF between the original and the entity-aware run (red) tops the list, then come the RRF of the three main runs (blue), and the pairwise RRF curves (original/hashed entities, entities/hashed entities) represented in green and orange respectively.]
    {{\includegraphics[scale=0.26]{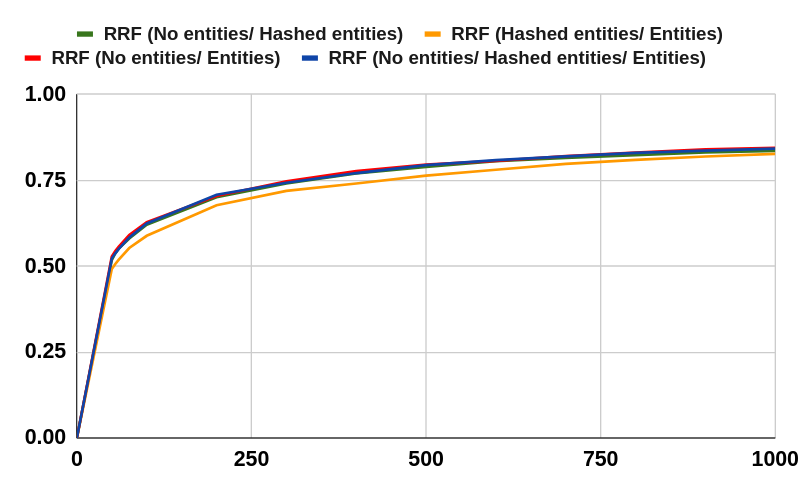} }}%
    \qquad
    \subfloat[\centering The best RRF curve is generated from fusing the BM25 original run and the entity-aware run in the explicit entity format. The corresponding curve is represented in red. The Classifier run is produced by selecting the best run containing the relevant document, and is represented in blue. As shown, the red curve outperforms the blue one for the Hard set.]
    {{\includegraphics[scale=0.26]{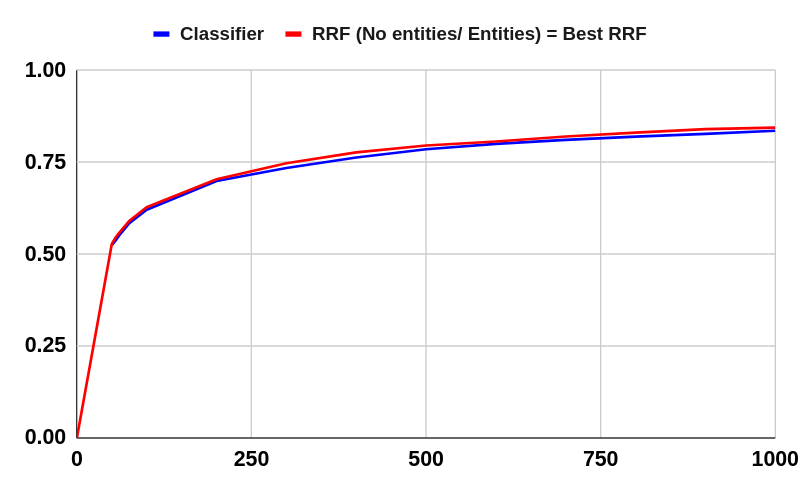} }}%
    \qquad
    \subfloat[\centering Although the ANCE (yellow) curve is the highest, entity linking introduces a significant improvement to the classical BM25 run (red). The Oracle (blue) shows the maximum possible gain achieved by entity linking. The best RRF of run combinations, shown in green, mediates the original BM25 (red) and the ANCE (yellow) runs. The same goes for the Oracle, suggesting the reduction of the recall gap between sparse and dense retrievers.]
    {{\includegraphics[scale=0.26]{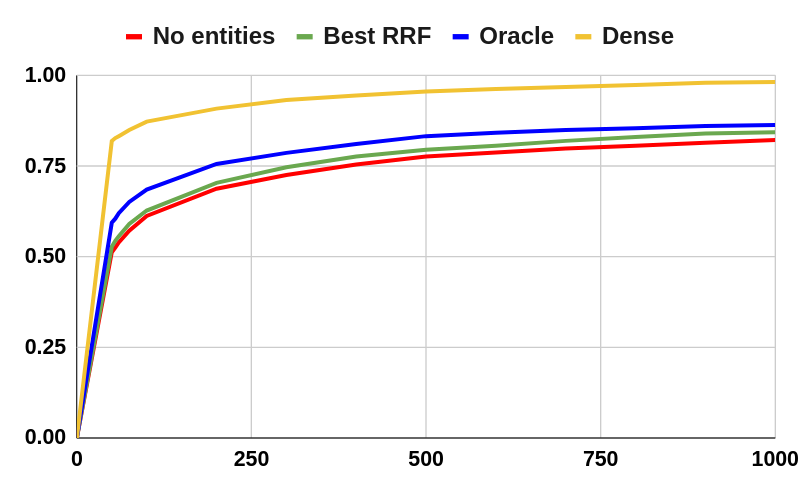} }}%
    \caption{Recall curves of the Hard query set with respect to MonoT5 qrels. The x-axis shows the cutoffs, and the y-axis is the corresponding recall value.}%
    \label{fig:veiled_monoT5}%
\end{figure}

\begin{figure}%
    \centering
    \subfloat[\centering The recall curves of the three main BM25 runs (without entities, with entities, with hashed entities) are represented in red, pink and green respectively. The hypothetical Oracle in blue represents the ideal gain achieved by entity linking by choosing the best run among the mentioned three for each query. The Dense yellow curve is generated from an ANCE run on the original MS MARCO passage Harder set.]
    {{\includegraphics[scale=0.26]{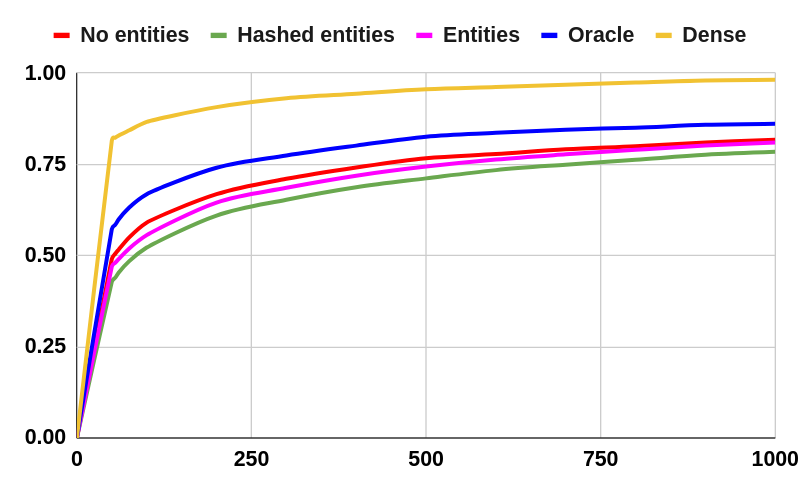} }}%
    \qquad
    \subfloat[\centering The graph shows all the generated RRF curves for the different run combinations on the MS MARCO passage Harder set. Ordered from best to worst, the pairwise RRF between the original and the entity-aware run (red) tops the list, then come the RRF of the three main runs (blue), and the pairwise RRF curves (original/hashed entities, entities/hashed entities) represented in green and orange respectively.]
    {{\includegraphics[scale=0.26]{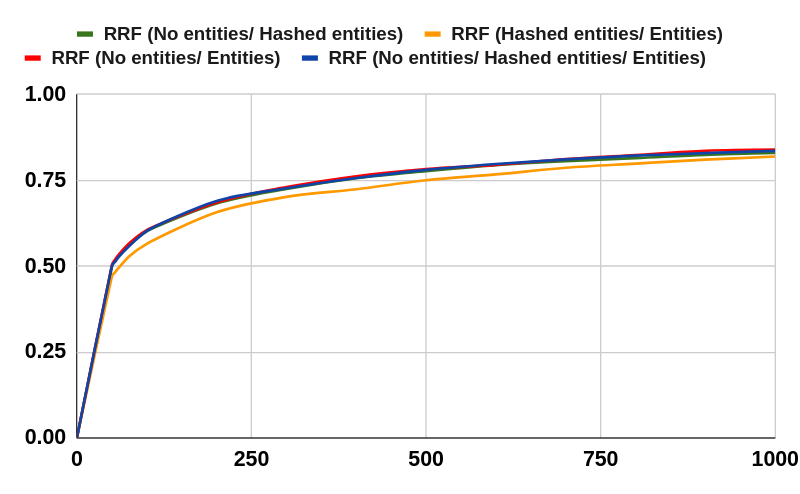} }}%
    \qquad
    \subfloat[\centering The best RRF curve is generated from fusing the BM25 original run and the entity-aware run in its explicit format. The corresponding curve is represented in red. The Classifier run is produced by selecting the best run containing the relevant document, and is represented in blue. As shown, the red curve outperforms the blue one for the Harder set.]
    {{\includegraphics[scale=0.26]{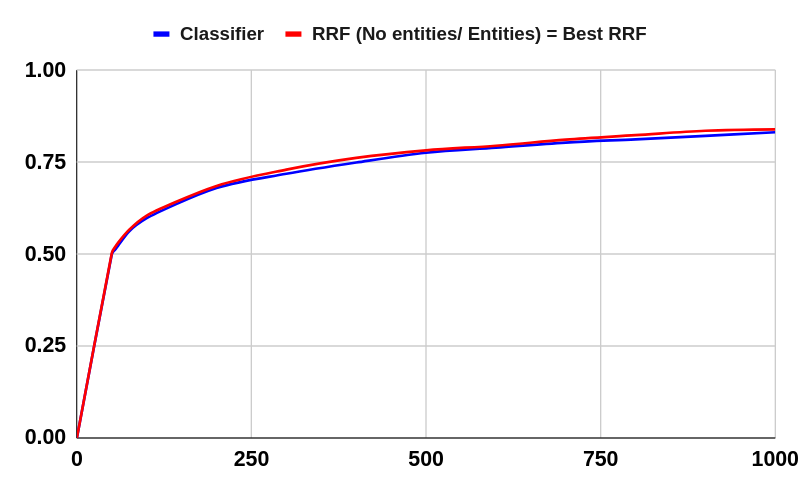} }}%
    \qquad
    \subfloat[\centering Although the ANCE (yellow) curve is the highest, entity linking introduces a significant improvement to the classical BM25 run (red). The Oracle (blue) shows the maximum possible gain achieved by entity linking. The best RRF of run combinations, shown in green, mediates the original BM25 (red) and the ANCE (yellow) runs. The same goes for the Oracle, suggesting the reduction of the recall gap between sparse and dense retrievers.]
    {{\includegraphics[scale=0.26]{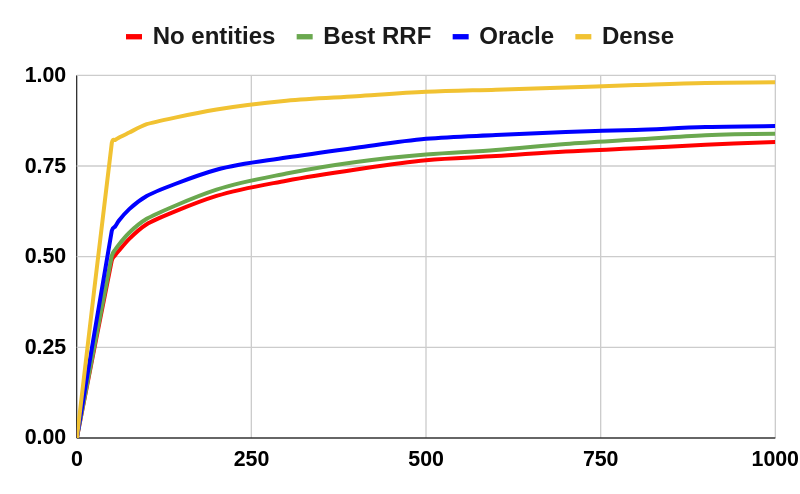} }}%
    \caption{Recall curves of the Harder query set with respect to MonoT5 qrels. The x-axis shows the cutoffs, and the y-axis is the corresponding recall value.}%
    \label{fig:pygmy_monoT5}%
\end{figure}

\begin{figure}%
    \centering
    \subfloat[\centering The recall curves of the three main BM25 runs (without entities, with entities, with hashed entities) are represented in red, pink and green respectively. The hypothetical Oracle in blue represents the ideal gain achieved by entity linking by choosing the best run among the mentioned three for each query. The Dense yellow curve is generated from an ANCE run on the original MS MARCO passage Hardest set.]
    {{\includegraphics[scale=0.26]{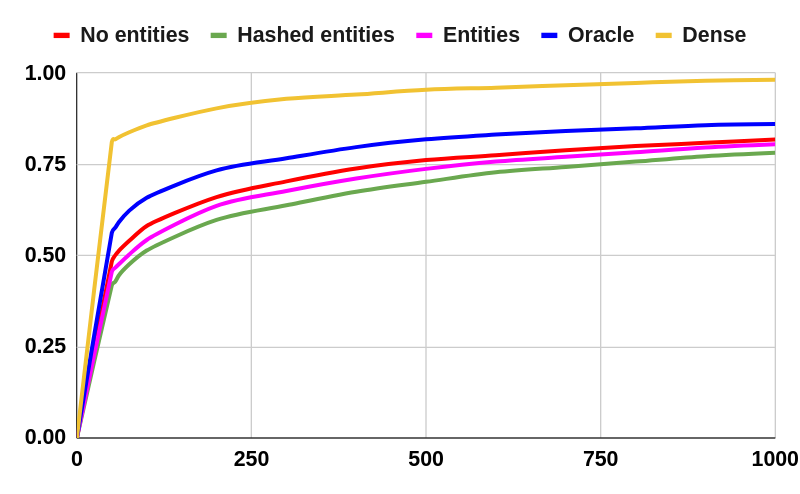} }}%
    \qquad
    \subfloat[\centering The graph shows all the generated RRF curves for the different run combinations on the MS MARCO passage Hardest set. Ordered from best to worst, the pairwise RRF between the original and the entity-aware run (red) tops the list, then come the RRF of the three main runs (blue), and the pairwise RRF curves (original/hashed entities, entities/hashed entities) represented in green and orange respectively.]
    {{\includegraphics[scale=0.26]{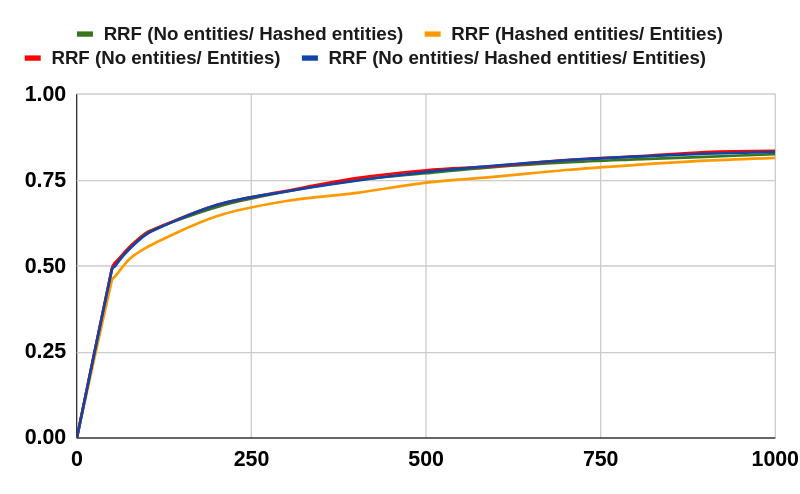} }}%
    \qquad
    \subfloat[\centering The best RRF curve is generated from fusing the BM25 original run and the entity-aware run in its explicit format. The corresponding curve is represented in red. The Classifier run is produced by selecting the best run containing the relevant document, and is represented in blue. As shown, the red curve outperforms the blue one for the Hardest set.]
    {{\includegraphics[scale=0.26]{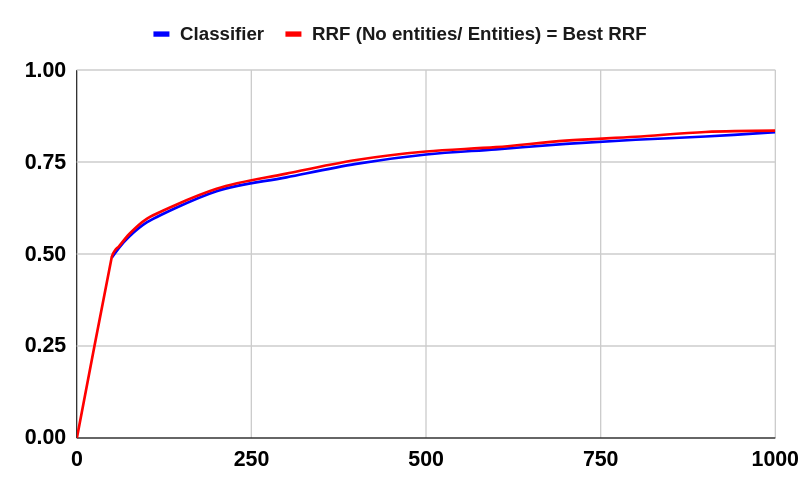} }}%
    \qquad
    \subfloat[\centering Although the ANCE (yellow) curve is the highest, entity linking introduces a significant improvement to the classical BM25 run (red). The Oracle (blue) shows the maximum possible gain achieved by entity linking. The best RRF of run combinations, shown in green, mediates the original BM25 (red) and the ANCE (yellow) runs. The same goes for the Oracle, suggesting the reduction of the recall gap between sparse and dense retrievers.]
    {{\includegraphics[scale=0.26]{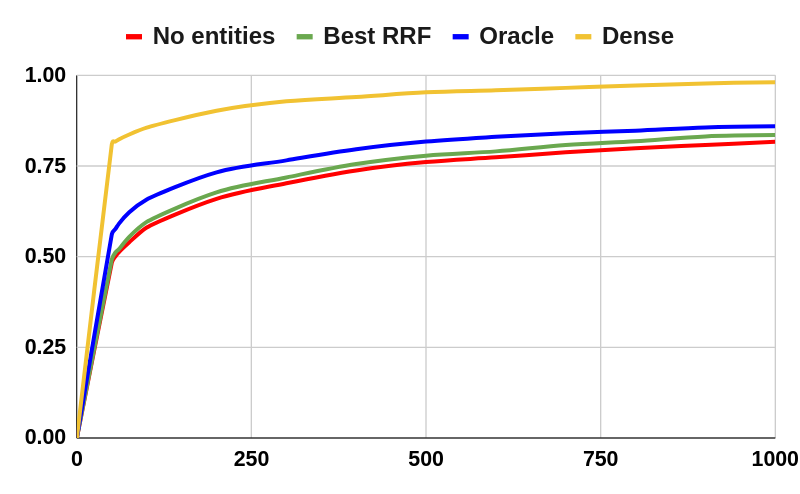} }}%
    \caption{Recall curves of the Hardest query set with respect to MonoT5 qrels. The x-axis shows the cutoffs, and the y-axis is the corresponding recall value.}%
    \label{fig:lesser_monoT5}%
\end{figure}

Table \ref{tab:recall_monoT5} and Figures \ref{fig:dev_monoT5}, \ref{fig:veiled_monoT5}, \ref{fig:pygmy_monoT5} and \ref{fig:lesser_monoT5} summarize my evaluation results for the Dev, Hard, Harder and Hardest query sets against the MonoT5 qrels.
In order to derive a consistent conclusion about entity linking, and to make up for any shortcomings of the original relevance judgments provided by MS MARCO assessors, I evaluate the results against another set of qrels: MonoT5 qrels. As previously described, MonoT5 qrels are generated by re-ranking with the MonoT5 model a run pool consisting of the three main BM25 runs: without entities, with entities and with hashed entities; and the ANCE dense run. The top answer retrieved for each query is then used as the relevance judgment.
From Table \ref{tab:recall_monoT5}, it is noticed that the evaluation against MonoT5 qrels experiences some changes when compared to the original qrels. Both runs equipped with entities: explicit and hashed formats have worse recall@1000 than the original BM25 run without entities. Consistently with the original qrels evaluation, the hashed run has the lowest performance. In addition, one can see that both the RRF and the classifier selection approaches are still effective since they outperform the standalone runs. Despite the bad performance of the entity-aware runs, their integration with the original BM25 run still achieves the best output. It is possible to attribute this behaviour to the fact that entity-based augmentation helps the rankers find complementary results. Also aligned with the previous findings, RRF still beats the classifier's selection method. Nonetheless, it is observed that the inclusion of the hashed entities in the fused run harms the overall performance when evaluated against MonoT5 qrels. As a result, the pairwise RRF between the no-entities and the entity-equipped runs has the highest performance among all the independent and combined runs. Compared to the non-expanded dataset, my best results realize performance gains of 2.20\%, 2.65\%, 2.73\% and 2.24\% for the Dev set and the three hard sets: Hard, Harder and Hardest respectively. The Oracle results also indicate additional increase potentials of 1.49\%, 2.36\%, 2.60\% and 2.89\% for the same sets. As shown, both the actual (best RRF) and the ideal (Oracle) recall@1000 gains have shrunk compared to the original-qrels-based evaluation, suggesting that the entities are not as effective with MonoT5 qrels as they were with the original qrels.

Figure \ref{fig:dev_monoT5} describes the recall curves of the different runs on the Dev set evaluated against MonoT5 qrels, where the x-axis represents the different cutoffs, and the y-axis is the corresponding recall value at a given cutoff. Similar to the original qrels, Sub-figure \ref{fig:dev_monoT5} (a) demonstrates the three main curves: original BM25 with no entities (red), BM25 run generated from the entity-aware dataset (pink), and the hashed-entity-equipped run (green). The hypothetical Oracle generated from ideally selecting the best run for each query out of the previous three runs (represented by the blue curve), and the ANCE dense run generated on the original non-expanded MS MACRO dataset serve as comparative baselines to bound the potential recall improvement gained by entity linking methods. In contrast with the performance with the original judgments, the red curve tops both the pink and green ones, emphasizing that the original BM25 retrieves the largest number of relevant passages for the Dev query set when using MonoT5 qrels as a ground truth. Nonetheless, the Oracle curve is higher than the original BM25 one as it mediates the sparse and dense curves. The latter confirms the effectiveness of entity linking in finding new matches that were not previously retrieved with the standard methods. The bad performance of the standalone entity-aware runs can be justified by saying that they add mixed signals that may act as noise during the BM25 retrieval. This hypothesis needs further investigation though before being judged as a fact. As a result, I observe that the original run outperforms the two runs with entities. However, exploiting entity-aware runs as a complementary method increases the pool of relevant documents (as can be seen from the blue curve), therefore improving the first stage recall. Sub-figure \ref{fig:dev_monoT5} (b) shows the different RRF combinations of the three main runs. Aligned with findings in (a) suggesting that the hashed entities hurt the performance, the best RRF is the pairwise combination of the original and the entity-equipped run in the explicit entity form (red) introducing a recall@1000 gain of 2.20\% . Similar to Sub-figure \ref{fig:dev_org} (b), the pairwise RRF of the entity-aware runs (orange) is the worst. Comes in second place the RRF of the three runs (blue) with a recall@1000 gain of 2.11\%, and in the third position is the RRF of the original and hashed-entity-aware runs (green) with a gain of 1.37\% over the traditional BM25 run. It is worth mentioning that even the worst RRF (orange curve) still outperforms the original BM25 (red curve in (a)) by a factor of 0.91\% for the recall@1000. In general, the improvement margin between the different RRF runs is very small. This can also be deduced from how the RRF curves coincide with each other. Sub-figure \ref{fig:dev_monoT5} (c) underlines that the best RRF run (red) still outperforms the classifier run (blue) by a small factor of 0.92\% for the recall@1000. Contrary to the corresponding graph in \ref{fig:dev_org} (c) with the original qrels, the curves of the best RRF and the classifier runs nearly coincide for the MonoT5 assessments. Sub-figure \ref{fig:dev_monoT5} (d) summarizes my best results by highlighting the recall margin between 1) my curves: best RRF (green) and Oracle (blue); and 2) the original sparse and dense curves: BM25 (red) and ANCE (yellow). The best RRF improves the recall@1000 by a factor of 2.20\% over the original run; while the Oracle shows a possible ideal recall improvement by a 3.72\%. Although entity linking did narrow the recall gap, the ANCE run still offers a recall@1000 gain of 14.15\% over the original run indicating the dense run dominates the run pool re-ranked by MonoT5 to generate this type of qrels.

Figures \ref{fig:veiled_monoT5}, \ref{fig:pygmy_monoT5} and \ref{fig:lesser_monoT5} show the recall curves of the Hard (Veiled), Harder (Pygmy) and Hardest (Lesser) query sets of the MS MARCO Chameleons dataset respectively, evaluated against MonoT5 qrels. The x-axis represents the different cutoffs, and the y-axis is the corresponding recall value at a given cutoff. The same pattern of the Dev set can be observed for the three sets of obstinate queries in each of the Sub-figures (a), (b), (c) and (d). 
In addition, the evaluation paradigm using the original qrels for the three hard sets also applies to the MonoT5 qrels' case in most of the cases. Nonetheless, contrary to the original judgments, I find that the recall improvement introduced by the entity linking methods is not always directly proportional to the difficulty degree of the underlying query set. For example, the ideal recall@1000 gain of the Oracle is bound to 5.07\%, 5.39 \% and 5.19 for the Hard (Sub-figure \ref{fig:veiled_monoT5} (a)), Harder (Sub-figure \ref{fig:pygmy_monoT5} (a)) and Hardest (Sub-figure \ref{fig:lesser_monoT5} (a)) query sets respectively. By taking a look at the Sub-figure (c) for the very same sets, it is found that the recall@1000 increase generated by the best RRF is 2.65\%, 2.73\% and 2.24\%, while the classifier's gain is 1.64\%, 1.79\% and 1.74\% for the Hard, Harder and Hardest sets respectively. While my methods achieve the largest recall gain on the Hardest query set using the original qrels, this does not apply to the MonoT5 qrels case where entity linking techniques are shown to be most effective for the Harder set.
Sub-figures \ref{fig:veiled_monoT5} (a), \ref{fig:pygmy_monoT5} (a) and \ref{fig:lesser_monoT5} (a)
follow the scheme of Sub-figure \ref{fig:dev_monoT5} (a) of the Dev set where both of my entity-equipped runs: explicit (pink) and hashed (green) show poor performance. However, the Oracle (blue) demonstrates that a recall improvement is still possible leveraging entity linking methods, hence compensating for the semantic shortcomings between the original BM25 and ANCE curves (red and yellow). The ANCE (yellow) tops all the curves for all the Hard sets. 
Sub-figures \ref{fig:veiled_monoT5} (b), \ref{fig:pygmy_monoT5} (b) and \ref{fig:lesser_monoT5} (b) show that the pairwise RRF of the original and the entity-equipped BM25 runs outperforms all other RRF runs. I notice that there is a very slight difference between the top RRF and the second best one achieved by fusing the three main runs. The two curves (red and blue) nearly coincide. While the recall@1000 gains of the best RRF are 2.65\%, 2.73\% and 2.24\% on the hard, Harder and Hardest sets respectively, the second best RRF achieves 2.42\%, 2.27\% and 1.80\% for the same sets.
In accordance with Sub-figures \ref{fig:veiled_org} (b), \ref{fig:pygmy_org} (b) and \ref{fig:lesser_org} (b), one can also see that the pairwise RRF between the two runs with entities has the worst recall gain among all the RRF runs. However, the aforementioned run still outperforms the original BM25 run without entities.
The results achieved by Sub-figures \ref{fig:veiled_monoT5} (c), \ref{fig:pygmy_monoT5} (c) and \ref{fig:lesser_monoT5} (c) are consistent with every previous Sub-figure (c) for all query-qrel combinations. The best RRF shows a slight recall enhancement over the classifier with factors of 0.99\%, 0.93\% and 0.49\% for the Hard, Harder and Hardest sets respectively. Despite that, selecting the best run with the help of a classifier introduces a recall increase over the classical BM25 run.
Sub-figures \ref{fig:veiled_monoT5} (d), \ref{fig:pygmy_monoT5} (d) and \ref{fig:lesser_monoT5} (d) underline that entity linking is useful in narrowing the recall margin between sparse retrievers (red) and dense retrievers (yellow). The best RRF (green) represents the maximum result achieved by leveraging entity linking, while the Oracle (blue) is the ideal recall gain that can be achieved.

Further information about the generation of the recall curves for Figures \ref{fig:dev_monoT5}, \ref{fig:veiled_monoT5}, \ref{fig:pygmy_monoT5} and \ref{fig:lesser_monoT5} can be found in Tables \ref{tab:recall_dev_monoT5}, \ref{tab:recall_veiled_monoT5}, \ref{tab:recall_pygmy_monoT5} and \ref{tab:recall_lesser_monoT5} respectively in Appendix \ref{AppendixA}.

%%%%%%%%%%%%%%%%%%%%%%%%%%%%%%%%%%%%%%%%%%%%%%%%%%%%%%%%%%
%=====================MonoT5 qrels========================

\begin{table}[!t] 
\centering
\caption{Recall@1000 of the 4 query sets (Dev, Hard, harder and Hardest) with respect to the MonoT5 qrels. %$^{*}$ denotes a statistically significant result with a p-value < 0.05.
}
% \vspace{-1em}
\begin{tabular}{l|cccc} 
\hline  
% \cline{2-9}
&
\multicolumn{3}{c} \textbf{\textbf{Query Set Type}} \\ [0.5ex]
\hline
\textbf{Run type}&\textbf{ Dev}&\textbf{Hard}&\textbf{Harder}&\textbf{Hardest}
 
\\[0.5ex] % inserts table 
%heading
\hline 
No entities& 0.8655&0.8227&0.8176&0.8181 \\

Hashed entities &0.8493&0.7945&0.7845&0.7820\\

Entities &0.8639&0.8156&0.8099&0.8051\\

Classifier &0.8764&0.8362&0.8322&0.8323 \\

No entities/ Hashed entities  RRF &0.8774 &0.8362&0.8302&0.8263 \\

Hashed entities/ Entities RRF &0.8734&0.8272&0.8197&0.8157 \\

No entities/ Entities RRF
&\textbf{0.8845}&\textbf{0.8445} &\textbf{0.8399}&\textbf{0.8364}\\

No entities/ Entities/ Hashed RRF
&0.8838&0.8426&0.8362&0.8328\\
\hline
Oracle &0.8977&0.8644&0.8617&0.8606\\
\hline
Dense &0.9880&0.9830&0.9822&0.9823\\
\hline
\end{tabular} 
\label{tab:recall_monoT5}
\end{table}

%%%%%%%%%%%%%%%%%%%%%%%%%%%%%%%%%%%%%%%%%%%%%%%%%%%%%%%%%%%%%%%
%========Dev queries with MonoT5 qrels================

%========Hard queries with MonoT5 qrels================

%========Harder queries with MonoT5 qrels================

%========Hardest queries with MonoT5 qrels================

%%%%%%%%%%%%%%%%%%%%%%%%%%%%%%%%%%%%%%%%%%%%%%%%%%%%%%%%%%%%====================================================================================================================================================================================

\subsubsection{Evaluation against DuoT5 Qrels}

\begin{figure}%
    \centering
    \subfloat[\centering The recall curves of the three main BM25 runs (without entities, with entities, with hashed entities) are represented in red, pink and green respectively. The hypothetical Oracle in blue represents the ideal gain achieved by entity linking by choosing the best run among the mentioned three for each query. The Dense yellow curve is generated from an ANCE run on the original MS MARCO passage Dev set.]
    {{\includegraphics[scale=0.26]{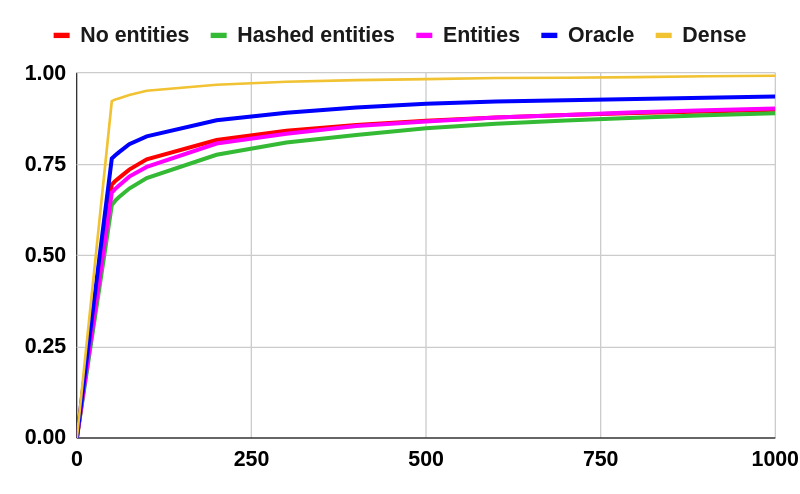} }}%
    \qquad
    \subfloat[\centering The graph shows all the generated RRF curves for the different run combinations on the MS MARCO passage Dev set. Ordered from best to worst, the RRF of the three main runs (blue) tops the list, then come the pairwise RRF curves (original/entities, original/hashed entities, entities/hashed entities) represented in red, green and orange respectively. The differences between the curves are very slight, and they nearly coincide.]
    {{\includegraphics[scale=0.26]{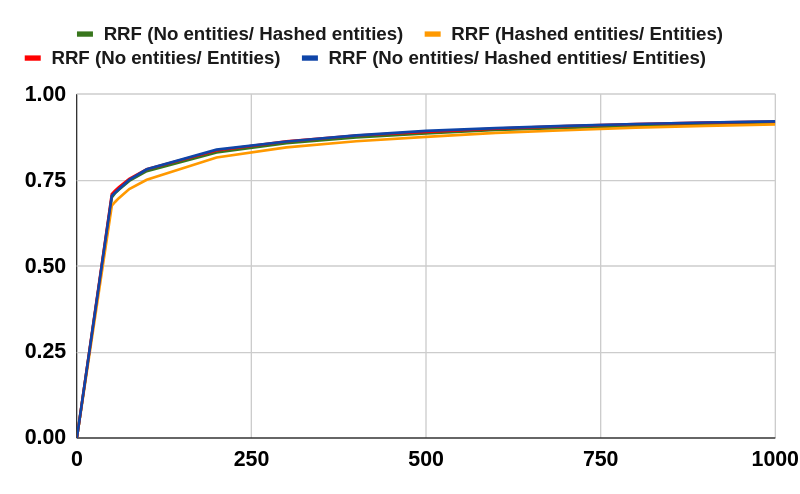} }}%
    \qquad
    \subfloat[\centering The best RRF curve is generated from fusing the three main runs. The corresponding curve is represented in red. The Classifier run is produced by selecting the best run containing the relevant document, and is represented in blue. As shown, the red curve outperforms the blue one for the Dev set.]
    {{\includegraphics[scale=0.26]{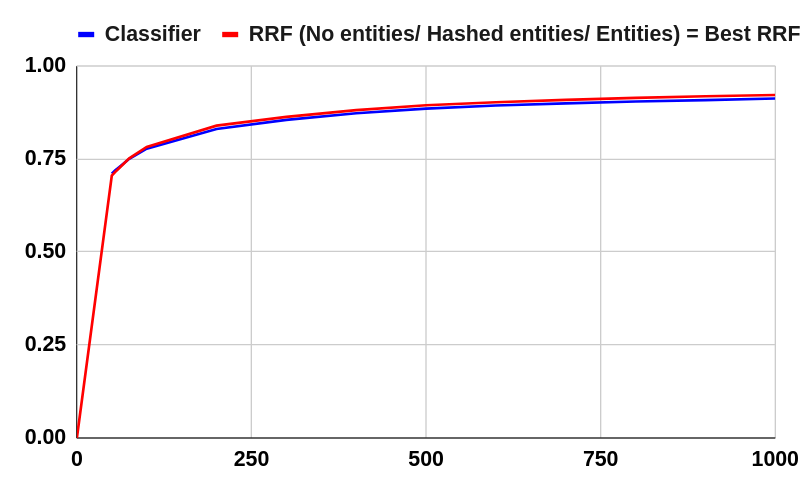} }}%
    \qquad
    \subfloat[\centering Although the ANCE (yellow) curve is the highest, entity linking introduces a significant improvement to the classical BM25 run (red). The Oracle (blue) shows the maximum possible gain achieved by entity linking. The best RRF of run combinations, shown in green, mediates the original BM25 (red) and the ANCE (yellow) runs. The same goes for the Oracle, suggesting the reduction of the recall gap between sparse and dense retrievers.]
    {{\includegraphics[scale=0.26]{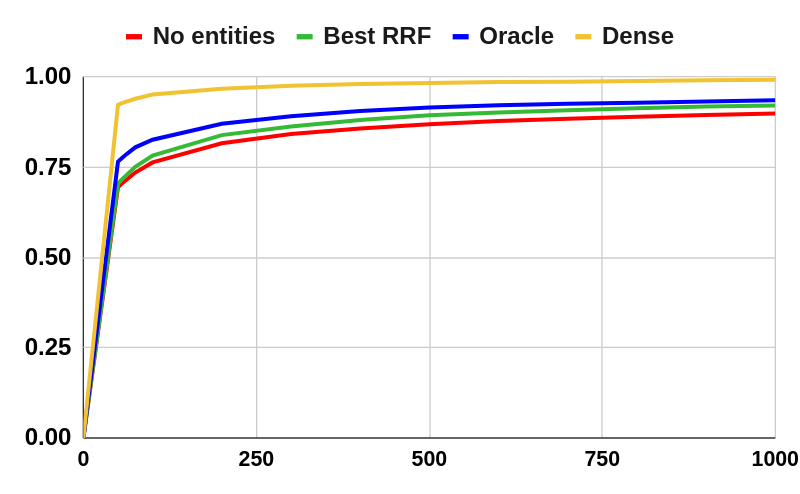} }}%
    \caption{Recall curves of the Dev query set with respect to DuoT5 qrels. The x-axis shows the cutoffs, and the y-axis is the corresponding recall value.}%
    \label{fig:dev_duoT5}%
\end{figure}

\begin{figure}%
\vspace{-2cm}
    \centering
    \subfloat[\centering The recall curves of the three main BM25 runs (without entities, with entities, with hashed entities) are represented in red, pink and green respectively. The hypothetical Oracle in blue represents the ideal gain achieved by entity linking by choosing the best run among the mentioned three for each query. The Dense yellow curve is generated from an ANCE run on the original MS MARCO passage Hard set.]
    {{\includegraphics[scale=0.26]{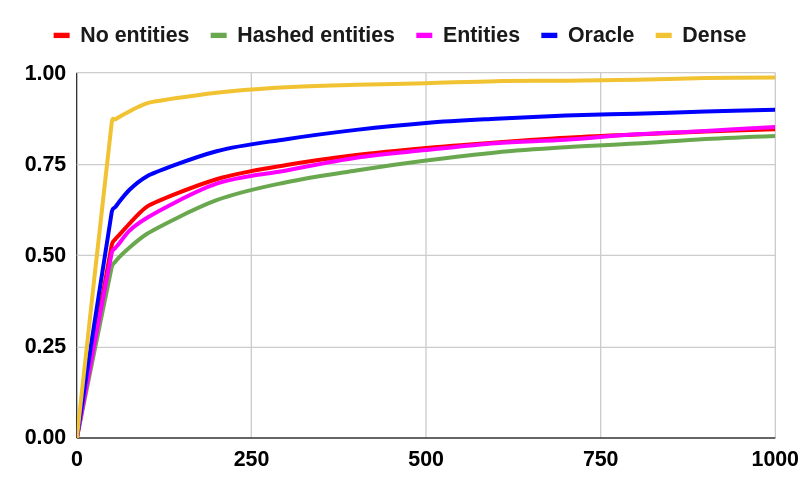} }}%
    \qquad
    \subfloat[\centering The graph shows all the generated RRF curves for the run combinations on the Hard set. Ordered from best to worst, The two red and blue curves coincide at the recall@1000 as they top the list. The red and blue curves represent the pairwise RRF between the original and the entity-aware run, and the RRF of the three main runs respectively. Then come the pairwise RRF curves (original/hashed entities, entities/hashed entities) represented in green and orange respectively. The differences between the curves are very slight.]
    {{\includegraphics[scale=0.26]{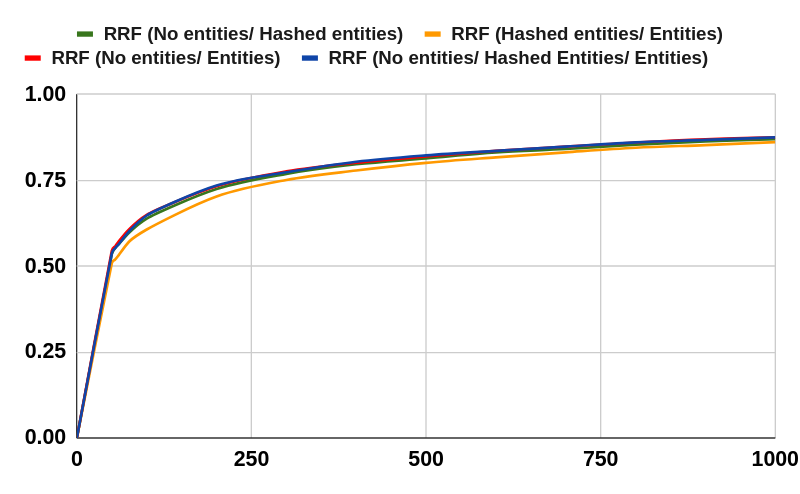} }}%
    \qquad
    \subfloat[\centering Since the two curves: RRF of the three runs and the pairwise RRF (no entities/entities) overlap with each other on several recall occasions. It is hard to determine the best among the two. Nonetheless, both of them outperform the classifier run. For the sake of the graph,  I represent the best RRF curve as the RRF of the three main runs (red). The Classifier run is produced by selecting the best run containing the relevant document, and is represented in blue. As shown, the red curve outperforms the blue one for the Hard set.]
    {{\includegraphics[scale=0.26]{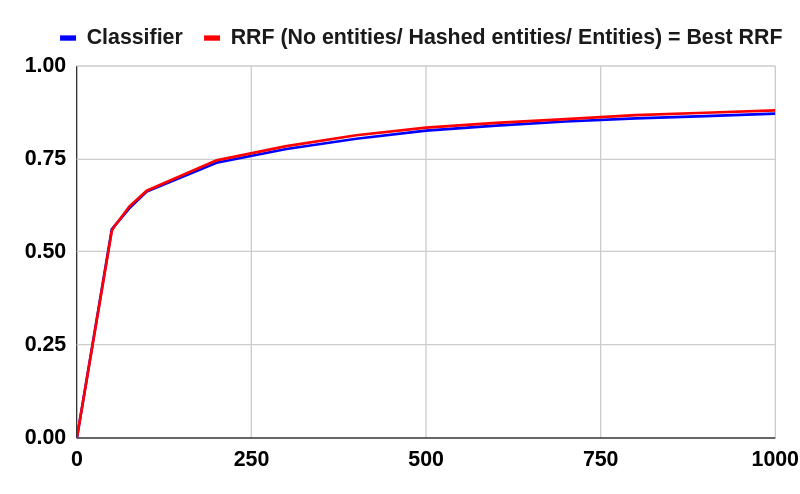} }}%
    \qquad
    \subfloat[\centering Although the ANCE (yellow) curve is the highest, entity linking introduces a significant improvement to the classical BM25 run (red). The Oracle (blue) shows the maximum possible gain achieved by entity linking. The best RRF of run combinations, shown in green, mediates the original BM25 (red) and the ANCE (yellow) runs. The same goes for the Oracle, suggesting the reduction of the recall gap between sparse and dense retrievers.]
    {{\includegraphics[scale=0.26]{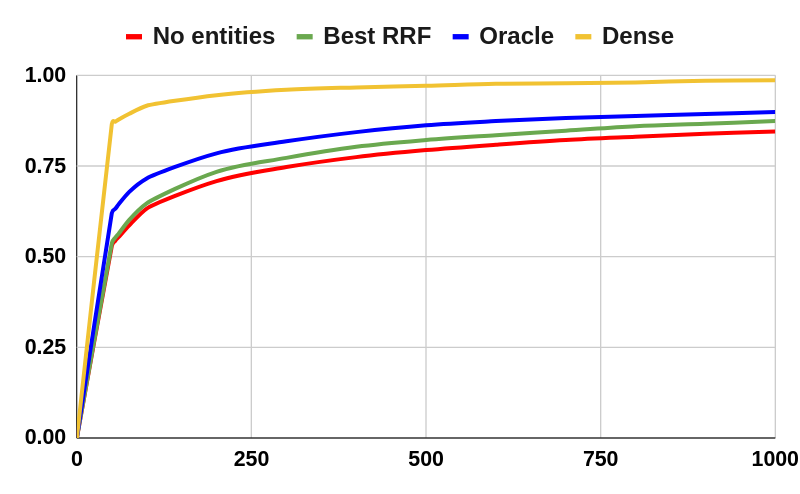} }}%
    \caption{Recall curves of the Hard query set with respect to DuoT5 qrels. The x-axis shows the cutoffs, and the y-axis is the corresponding recall value.}%
    \label{fig:veiled_duoT5}%
\end{figure}

\begin{figure}%
\vspace{-2cm}
    \centering
    \subfloat[\centering The recall curves of the three main BM25 runs (without entities, with entities, with hashed entities) are represented in red, pink and green respectively. The hypothetical Oracle in blue represents the ideal gain achieved by entity linking by choosing the best run among the mentioned three for each query. The Dense yellow curve is generated from an ANCE run on the original MS MARCO passage Harder set.]
    {{\includegraphics[scale=0.26]{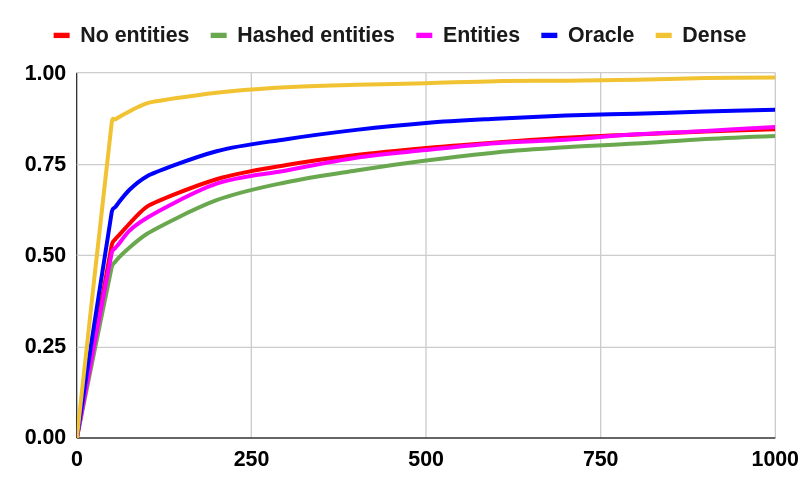} }}%
    \qquad
    \subfloat[\centering The graph shows all the generated RRF curves for the run combinations on the Harder set. Ordered from best to worst, The two red and blue curves coincide at the recall@1000 as they top the list. The red and blue curves represent the pairwise RRF between the original and the entity-aware run, and the RRF of the three main runs respectively. Then come the pairwise RRF curves (original/hashed entities, entities/hashed entities) represented in green and orange respectively. The differences between the curves are very slight.]
    {{\includegraphics[scale=0.26]{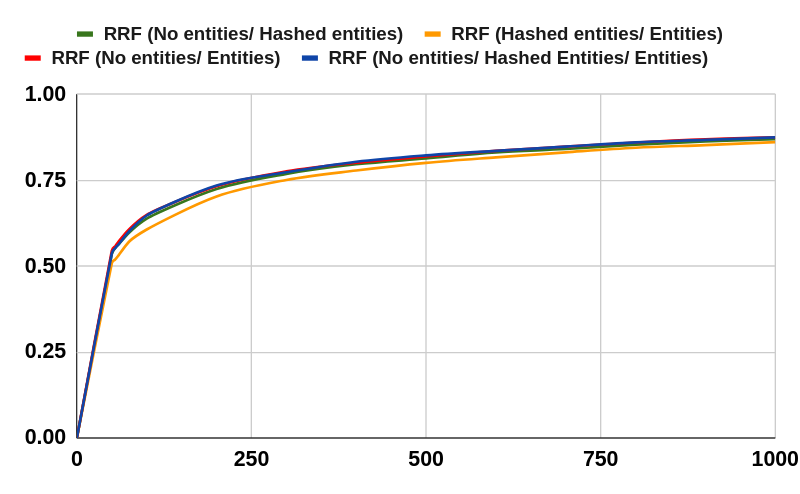} }}%
    \qquad
    \subfloat[\centering Since the two curves: RRF of the three runs and the pairwise RRF (no entities/entities) overlap with each other on several recall occasions. It is hard to determine the best among the two. Nonetheless, both of them outperform the classifier run. For the sake of the graph,  I represent the best RRF curve as the RRF of the three main runs (red). The Classifier run is produced by selecting the best run containing the relevant document, and is represented in blue. As shown, the red curve outperforms the blue one for the Harder set.]
    {{\includegraphics[scale=0.26]{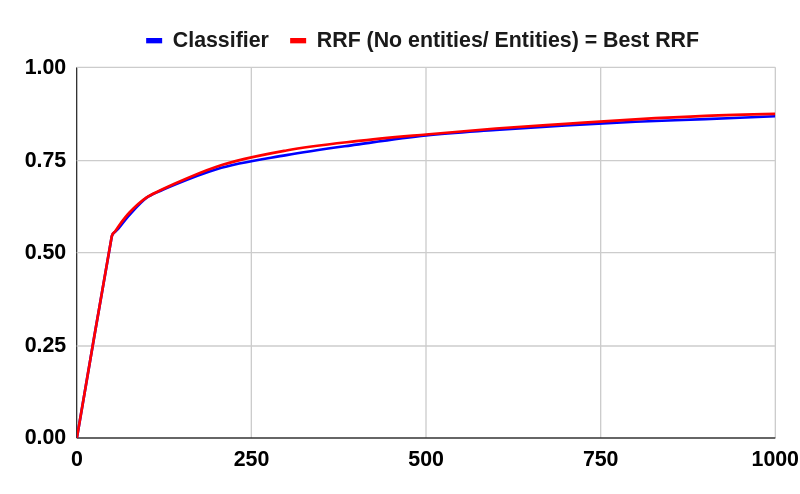} }}%
    \qquad
    \subfloat[\centering Although the ANCE (yellow) curve is the highest, entity linking introduces a significant improvement to the classical BM25 run (red). The Oracle (blue) shows the maximum possible gain achieved by entity linking. The best RRF of run combinations, shown in green, mediates the original BM25 (red) and the ANCE (yellow) runs. The same goes for the Oracle, suggesting the reduction of the recall gap between sparse and dense retrievers.]
    {{\includegraphics[scale=0.26]{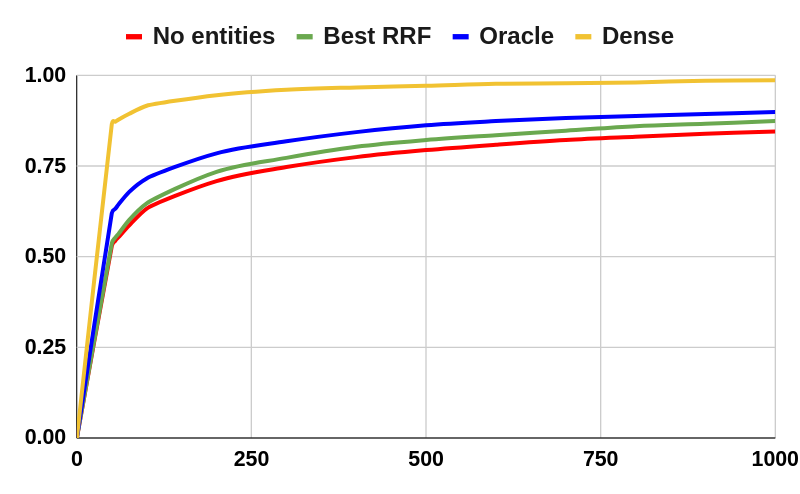} }}%
    \caption{Recall curves of the Harder query set with respect to DuoT5 qrels. The x-axis shows the cutoffs, and the y-axis is the corresponding recall value.}%
    \label{fig:pygmy_duoT5}%
\end{figure}

\begin{figure}%
    \centering
    \subfloat[\centering The recall curves of the three main BM25 runs (without entities, with entities, with hashed entities) are represented in red, pink and green respectively. The hypothetical Oracle in blue represents the ideal gain achieved by entity linking by choosing the best run among the mentioned three for each query. The Dense yellow curve is generated from an ANCE run on the original MS MARCO passage Hardest set.]
    {{\includegraphics[scale=0.26]{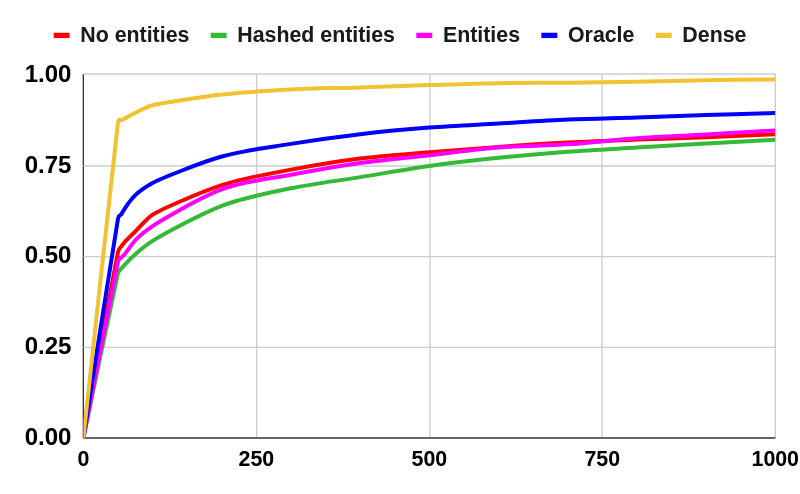} }}%
    \qquad
    \subfloat[\centering The graph shows all the generated RRF curves for the run combinations on the Hardest set. Ordered from best to worst, the pairwise RRF between the original and the entity-aware run (red) tops the list, then come the RRF of the three main runs (blue), and the pairwise RRF curves (original/hashed entities, entities/hashed entities) represented in green and orange respectively. The differences between the curves are very slight.]
    {{\includegraphics[scale=0.26]{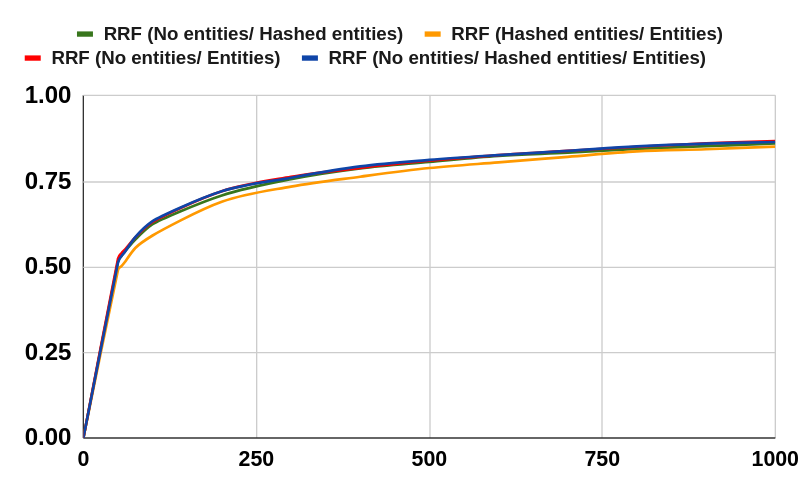} }}%
    \qquad
    \subfloat[\centering The best RRF curve is generated from fusing the BM25 original run and the entity-aware run in its explicit format. The corresponding curve is represented in red. The Classifier run is produced by selecting the best run containing the relevant document, and is represented in blue. As shown, the red curve outperforms the blue one for the Hardest set.]
    {{\includegraphics[scale=0.26]{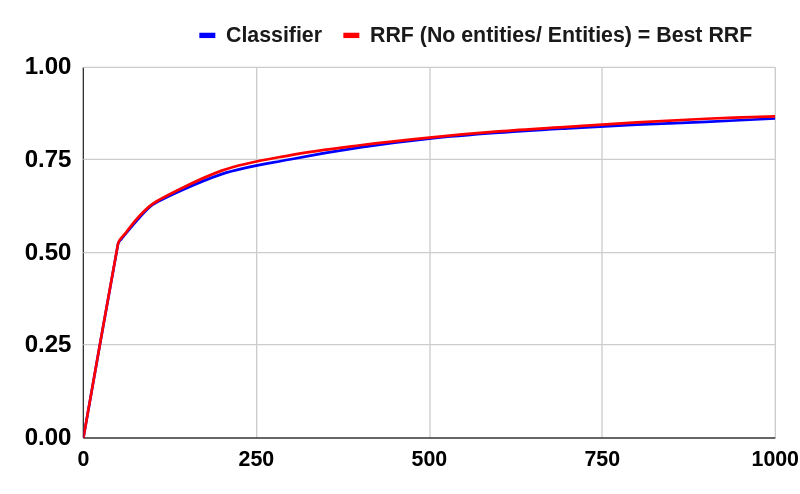} }}%
    \qquad
    \subfloat[\centering Although the ANCE (yellow) curve is the highest, entity linking introduces a significant improvement to the classical BM25 run (red). The Oracle (blue) shows the maximum possible gain achieved by entity linking. The best RRF of run combinations, shown in green, mediates the original BM25 (red) and the ANCE (yellow) runs. The same goes for the Oracle, suggesting the reduction of the recall gap between sparse and dense retrievers.]
    {{\includegraphics[scale=0.26]{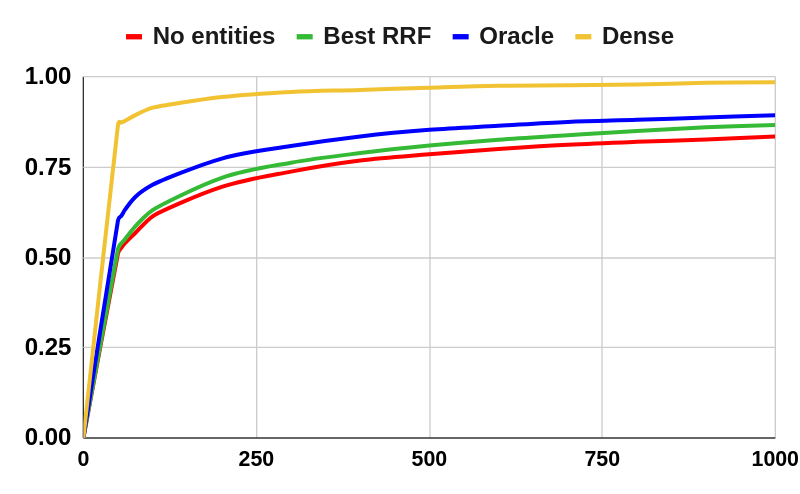} }}%
    \caption{Recall curves of the Hardest query set with respect to DuoT5 qrels. The x-axis shows the cutoffs, and the y-axis is the corresponding recall value.}%
    \label{fig:lesser_duoT5}%
   
\end{figure}

Table \ref{tab:recall_duoT5} and Figures \ref{fig:dev_duoT5}, \ref{fig:veiled_duoT5}, \ref{fig:pygmy_duoT5} and \ref{fig:lesser_duoT5} describe the evaluation results for the Dev, Hard, Harder and Hardest query sets against DuoT5 qrels.
As previously explained, I believe that the original qrel set is insufficient as ground truth for the evaluation of the constantly developed IR methods. In an attempt to compensate for its limitations, my methods are evaluated using different types of relevance assessments.
DuoT5 qrels are the third type of relevance judgments employed to evaluate the effect of query and document expansion with linked entities. The qrels are generated by re-ranking with the MonoT5 model a run pool consisting of the three main BM25 runs: without entities, with entities and with hashed entities; and the ANCE dense run. The top 50 passages for each query are then re-ranked using the DuoT5 model. Finally, the top answer retrieved for each query is used as the correct answer.
With respect to DuoT5 qrels, Table \ref{tab:recall_duoT5} shows that the best-fused run varies across the query sets, in contrast with the two previous qrel types. The best recall for the Dev set is 0.9219 with a performance increase of 2.47\% achieved by the RRF of the three main runs, and a potential improvement of 1.54\% (computed from the difference between the Oracle and the best RRF). For the Hardest query set, it is shown that adding the hashed-entity-aware run to the fusion pool weakens the retrieval. The best recall is 0.8677 produced by the pairwise fusion between the no-entities and the entity-aware runs with a gain of 3.82\% and an improvement window of 3.07\%. It is also possible to notice that for both the Hard and the Harder sets, the hashed entities have a neutral effect on the overall recall@1000 gain. Both the fusion of the three runs and the pairwise fusion of the original and the entity-equipped runs achieve the best results. The latter contribute with a performance gain of 3.35\% and possible additional increases of 2.66\% and 2.86\% for the Hard and the Harder sets respectively. It is also observed that the classifier outperforms each of the three individual runs (no entities, with entities, with hashed entities) by factors of 1.42\%, 2.34\%, 2.63\% and 3.11\% over the classical BM25 run for the Dev, Hard, Harder and Hardest sets respectively. The ANCE performance recalls the large gap between sparse (i.e. BM25 without entities) and dense retrievers with margins of 10.38\%, 15.95\%, 16.72\% and 18.02\% for the Dev, Hard, Harder and Hardest query sets respectively. This huge gap is expected since DuoT5 qrels are generated from a pool containing the ANCE run. Hence, the qrels are dominated by dense-retriever-generated matches.

Figure \ref{fig:dev_duoT5} shows the recall curves of the generated runs on the Dev set evaluated against DuoT5 qrels, where the x-axis represents the different cutoffs, and the y-axis is the corresponding recall value at a given cutoff.
This Figure is the same as the ones generated from the original and MonoT5 qrels. I omit the description of each sub-figure for conciseness. The reader can always refer back to the description in \ref{fig:dev_org} and \ref{fig:dev_monoT5}. The focus in the next few lines is on the differences in comparison with the original and MonoT5 qrels. Overall, Figure \ref{fig:dev_duoT5} (with each of its four sub-figures) follows the same pattern of Figure \ref{fig:dev_org} of the original qrels evaluation. In Sub-figure \ref{fig:dev_duoT5} (a), it is observed that the margin between the original run (red) and the entity-equipped run (pink) is very narrow. While the red curve tops the pink one at the small recall value, the pink curve representing the entity-aware run outperforms the red curve as the recall value increases. The performance improvement introduced by the entity-aware run in its explicit form is minor with a factor of 0.43\% over the classic BM25 run (red). The ideal recall@1000 gain represented by the Oracle (blue) is within a margin of 4.05\%. ANCE (yellow) is still presenting a large effectiveness gap of 10.38\% with respect to the classic run (red). Similar to the previous findings, the standalone hashed-equipped run performs poorly.
Sub-figure \ref{fig:dev_duoT5} (b) presents the different RRF combinations. Ordered from best to worst, one can mention the RRF of the three main runs (blue), the pairwise RRF of the original and the entity-aware run (red), the pairwise RRF between the original and the hashed-entity-aware run (green), and the pairwise RRF of the entity-equipped runs: explicit and hashed (orange). Although the hashed-run has a bad independent performance (as previously seen in (a)), its inclusion in the run fusion achieves the best recall results. The recall@1000 gains of the same four runs over the classic BM25 are 2.47\%, 2.42\%, 1.98\% and 1.50\% in their respective order. As shown, even the worst RRF offers an improvement over the non-expanded dataset. 
Sub-figure \ref{fig:dev_duoT5} (c) underlines that run fusion (red) is more effective than run selection (blue) adopting the proposed classifier's architecture.
Sub-figure \ref{fig:dev_duoT5} (d) concludes that using entity linking methods (represented by the green and blue curves) improves the retrieval recall of the sparse methods (red) in an attempt to get a step closer to the dense performance (yellow) with minimum computational resources.

Figures \ref{fig:veiled_duoT5}, \ref{fig:pygmy_duoT5} and \ref{fig:lesser_duoT5} show the recall curves of the Hard (Veiled), Harder (Pygmy) and Hardest (Lesser) query sets of the MS MARCO Chameleons dataset respectively, evaluated against DuoT5 qrels. The x-axis represents the different cutoffs, and the y-axis is the corresponding recall value at a given cutoff. Although I have combined the explanation of the three sets of hard queries together for the evaluations with the original and MonoT5 qrels, this approach cannot be adopted for all the sub-figures in this case since the performance varies across the hard sets using DuoT5 qrels.
All of the Sub-figures \ref{fig:veiled_duoT5} (a), \ref{fig:pygmy_duoT5} (a) and \ref{fig:lesser_duoT5}(a) of the Hard, Harder and Hardest sets follow the same performance as the Dev set (Sub-figure \ref{fig:dev_duoT5} (a)). The Oracle (blue) mediates the no-entity BM25 run (red) and the no-entity ANCE run (yellow) with recall@1000 gains of 6.09\%, 6.31\% and 7.00\% in their respective mentioned order. The runs augmented with hashed entities still underperform the original runs for all types of hard queries.  The improvement introduced by the explicit entity augmentation is subtle: The red and pink curves nearly coincide. The recall@1000 gains from the entity-aware runs are 0.47\%, 0.67\% and 1.27\% for the Hard, Harder and Hardest query sets respectively. Consistent with the previous behaviour with the original qrels, and by taking a look on the entity-equipped and Oracle gains, one can also infer that the recall gain obtained from the entity linking methods is directly proportional to the difficulty of the underlying query set. The harder the queries are, the larger the recall gain achieved with semantic incorporation is.
For Sub-figure (b), both the Hard \ref{fig:veiled_duoT5} and Harder \ref{fig:pygmy_duoT5} sets show the same paradigm. It is found that the RRF of the three main runs (blue) and the pairwise RRF of the original and entity-aware BM25 runs (red) coincide at the recall@1000 value. The two curves overlap at several locations, i.e. the red curve may show a higher recall at a cutoff $x$, but the blue one may perform better at a cutoff $x+1$. One curve may act as the best at a certain value, but the worst at another position. However, no consistent pattern can be deduced with respect to the cutoff value. This behaviour is worth exploration in a future extension to this work. It is worth mentioning though that the difference is so minor that can be negligible. Additionally, the two curves show the exact same recall@1000 value which is 0.8804 and 0.8751 for the Hard and Harder sets respectively. Further information about the exact value at each cutoff is provided in Tables \ref{tab:recall_veiled_duoT5} and \ref{tab:recall_pygmy_duoT5} for the Hard and Harder sets. For the Hardest set in Sub-figure \ref{fig:lesser_duoT5} (b), I observe that considering the hashed entities in the fusion lessens the overall recall gain. The best RRF is the pairwise fusion between the original and the entity-equipped BM25 runs (red). This behaviour is similar to the one witnessed with MonoT5 qrels in Sub-figure (b) of Figures \ref{fig:dev_monoT5}, \ref{fig:veiled_monoT5}, \ref{fig:pygmy_monoT5} and \ref{fig:lesser_monoT5}. The RRF of the entity-equipped runs: explicit and hashed (orange) still gives the most modest recall improvement over the classic BM25 with gains of 1.84\%, 1.72\% and 1.90\% for the Hard, Harder and Hardest query sets. The green curve representing the pairwise RRF of the original and hashed-entity-aware BM25 runs comes in third place across all the query sets in the same order with gains of 2.63\%, 2.78\% and 3.04\%. The best RRF introduces recall@1000 gains of 3.35\%, 3.35\% and 3.82\% for the Hard, Harder and Hardest sets respectively. The main conclusion provided by the Sub-figure (c) is consistent across all the query sets:  \ref{fig:veiled_duoT5}, \ref{fig:pygmy_duoT5} and \ref{fig:lesser_duoT5}. It is also uniform with all of my previous findings. The best RRF (red) run beats the classifier run (blue) by minor factors of 0.99\%, 0.70\% and 0.68\% for the Hard, Harder and Hardest sets respectively.
Similarly, Sub-figure (d) across all the hard queries underlines that augmenting the corpus with linked entities is effective for improving the recall of sparse retrievers like BM25 with respect to DuoT5 qrels. While the best RRF (green) demonstrates a considerable improvement over the original run (red), the Oracle confirms promising results for adopting entity linking methods with advanced run fusion or selection methods.

Further information about the generation of the recall curves for Figures \ref{fig:dev_duoT5}, \ref{fig:veiled_duoT5}, \ref{fig:pygmy_duoT5} and \ref{fig:lesser_duoT5} can be found in Tables \ref{tab:recall_dev_duoT5}, \ref{tab:recall_veiled_duoT5}, \ref{tab:recall_pygmy_duoT5} and \ref{tab:recall_lesser_duoT5} respectively in Appendix \ref{AppendixA}.

%=====================DuoT5 qrels========================

\begin{table}[!t] 
% \vspace{1cm}
\centering
\caption{Recall@1000 of the 4 query sets (Dev, Hard, harder and Hardest) with respect to the DuoT5 qrels. %$^{*}$ denotes a statistically significant result with a p-value < 0.05.
}
\begin{tabular}{l|cccc} 
\hline 
% \cline{2-9}
&
\multicolumn{3}{c} \textbf{\textbf{Query Set Type}} \\ [0.5ex]
\hline
\textbf{Run type}&\textbf{ Dev}&\textbf{Hard}&\textbf{Harder}&\textbf{Hardest} \\[0.5ex] % inserts table 
%heading
\hline 
No entities&0.8997&0.8519&0.8467&0.8358\\ 

Hashed entities&0.8904&0.8352&0.8281&0.8204\\

Entities&0.9036&0.8557&0.8524&0.8464\\ 

Classifier &0.9125&0.8718&0.8690&0.8618\\

No entities/ Hashed entities  RRF &0.9175&0.8743&0.8702&0.8612\\

Hashed entities/ Entities RRF&0.9132&0.8676&0.8613&0.8517\\

No entities/ Entities RRF
&0.9215&\textbf{0.8804}&\textbf{0.8751}&\textbf{0.8677}\\

No entities/ Entities/ Hashed RRF
&\textbf{0.9219}&\textbf{0.8804}&\textbf{0.8751}&0.8653\\
 
\hline
Oracle &0.9361&0.9038&0.9001&0.8943\\ 
\hline
Dense &0.9931&0.9878&0.9883&0.9864\\ 
\hline

\end{tabular} 
\label{tab:recall_duoT5}
\end{table}

%%%%%%%%%%%%%%%%%%%%%%%%%%%%%%%%%%%%%%%%%%%%%%%%%%%%%%%%%%%%%%%
%========Dev queries with DuoT5 qrels================

%========Hard queries with DuoT5 qrels================

%========Harder queries with DuoT5 qrels================

%========Hardest queries with DuoT5 qrels================

%%%%%%%%%%%%%%%%%%%%%%%%%%%%%%%%%%%%%%%%%%%%%%%%%%%%%%%%%%%%%%%%%%%%%%%%%%%%%%%%%%%%%%%%%%%%%%%%%%%%%%%%%%%%%%%%%%%%%%%%

\subsection{Dense Retrieval}

I have started my experiments with the objective of leveraging the advancement of end-to-end entity linking tools to disambiguate queries and documents, hence improving the retrieval quality and increasing the recall of sparse retrievers like BM25. Since the latter is usually employed in the first ranking stage, I have evaluated my results using the recall@1000 metric. Both the queries and passages of the MS MARCO dataset are expanded with semantic terms in an attempt to shrink the effectiveness gap between sparse and dense retrievers. The success of corpus augmentation in improving the sparse-retrieval-generated recall, as was demonstrated in the previous sparse evaluation, arouses the curiosity to experiment with entity linking with the dense retrievers as well.
Without deviating from the primary objective, I attempt dense retrieval on the expanded dataset with linked entities in their explicit form. The evaluation is conducted using the original qrels, and the corresponding results are reported in MRR@10. I refrain from experimenting with the hashed entities since the previous experiments have shown their underperformance when individually employed.

\begin{table}[!ht]
   \caption{MRR@10 for each component (Warm-up, STAR, ADORE) in the STAR-ADORE dense pipeline. `No entities' refers to the dense retrieval using the non-expanded MS MARCO passage dataset. These MRR@10 values generated using the original dataset are successfully replicated before applying the model to the entity-equipped dataset referred to as `Entities' in the table.}
    \centering
    \begin{tabular}{c|c|c|c}
    \hline
    \textbf{Metric/ Model} & \textbf{Warm-up} & \textbf{STAR} & \textbf{ADORE} \\ [0.5ex]
    \hline
    \textbf{No entities} & 0.311 & 0.340 & 0.347 \\ 
    \textbf{Entities} & 0.313 & 0.338 & 0.347\\ 
    \hline
    \end{tabular}
    \label{tab:mrr_dense}
\end{table}

Table \ref{tab:mrr_dense} gives an overview of my results of the STAR-ADORE dense pipeline using the MS MARCO expanded dataset: the train set is used for training/fine-tuning purposes, and the Dev is used for the evaluation against the original qrels. The first row provides the MRR@10 for each stage in the pipeline (Warm-up, STAR and ADORE) respectively. These values are first replicated on the original dataset to confirm their validity before corpus augmentation. The second row shows the corresponding values on the entity-equipped dataset. For each model, I pick the checkpoint that evaluates the highest on the Dev set, and employ it in the next stage of the pipeline. For the warm-up model trained on the entity-aware MS MARCO training set, the highest MRR@10 value achieved on the Dev set is 0.313. Although it is slightly higher than the MRR@10 resultant from the original data, the improvement is too small (0.64\%) that it can be negligible. A little noise in the data is able to cause such a spike. This hypothesis is later confirmed in the STAR stage where my best MRR@10 on the entity-equipped data is slightly worse than the MRR@10 of the original dataset by a factor of 0.59\%. This result is also obtained by the best STAR checkpoint. As previously explained, the best warm-up checkpoint is used to generate the static hard negatives needed by STAR algorithm to train the dense model with the objective of optimizing the document encoder. The final stage is training the dense model with ADORE in order to optimize the query encoder, leveraging the document embeddings generated by the best STAR from the previous stage. ADORE's MRR@10 refers to the MRR@10 of the whole dense pipeline evaluated on the Dev set using the original qrels. From Table \ref{tab:mrr_dense}, one can see that ADORE's MRR@10 with or without entities is exactly the same. The entity effect is neutral on the retrieval performance. This result is also achieved by the best ADORE checkpoint.

The main conclusion drawn from the dense experiment is that linked entities are not as useful as it was believed they would be for the dense models. Due to the ineffectivenss of the linked entities with the STAR-ADORE model, I did not explore method variations as with sparse retrieval experiments. Nonetheless, it is worth experimenting with query expansion and document expansion individually. Different evaluation schemes using other qrel types such as MonoT5 and DuoT5 qrels may provide additional context. I have also limited my dense experiments to the Dev set, but it is important to examine the dense retriever's behaviour on the three obstinate query sets: Hard, Harder and Hardest. The STAR-ADORE model was selected for being a state-of-the-art dense model. Due to the time constraints and limited computational resources, it was not possible to experiment with other dense retrievers. The latter may provide further coverage of the entity linking effect in the IR systems. It is also worth exploring corpus expansion with hashed entities (or even with entity descriptions) in addition to the explicit entity names in the context of dense retrieval.

%======================================================================
\chapter{Conclusion}
\label{Chapter5}
%======================================================================

Despite their substantial performance advances, dense retrievers require costly GPU resources, and suffer from a significant latency overhead. In contrast, sparse retrievers offer higher efficiency benefits at the expense of semantic comprehension. In an attempt to bridge the gap, I propose leveraging recent advances in entity linking to expand IR collections, hence reducing vocabulary discrepancies. My primary objective focuses on boosting the retrieval recall in the first retrieval stage of cascaded ranking architectures using BM25 as a solid sparse retriever. 
I apply my methods to the MS MARCO passage ranking dataset, and on the three MS MARCO Chameleons sets of obstinate queries: Hard, Harder and Hardest.
I experiment with variations of expansion methods exploiting linked entities. These variations consider the type of the expanded text (i.e. query expansion, document expansion and both combined), the number of entity copies (i.e. single term expansion, weighted expansion, and expansion by a constant factor of entities), and entity format (i.e. explicit and hashed). 

My results indicate that augmenting both the queries and the passages by a single entity instance in the explicit word form for each mention is the most effective approach. I generate different runs on both the original and the expanded dataset such as: BM25 run without entities, BM25 run with entities, BM25 run with hashed entities, the hypothetical Oracle of the three previous runs, BM25 run with pseudo-relevance feedback without entities and ANCE run that is used as a comparative basis. I evaluate my sparse retrieval results using recall@1000 with the help of three types of relevance judgements: the original qrel set provided by MS MARCO assessors, the MonoT5 qrels generated from a large run pool re-ranked by MonoT5 model, and the DuoT5 qrels generated by re-ranking the previously re-ranked MonoT5 runs.
I also attempt two main types of run conjunction strategies: RRF of different combinations of the three main runs, and a classifier-based selection of the best run for each query using a contextualized pre-trained BERT-based model. 
My best results are achieved by RRF between different runs across the query sets. Although hashed entities manifest poor individual performance, they seem useful to retrieve complementary results when used along with a strong run.
Through comparative evaluation of the experimented query-qrel combinations, it is proven that my approach enhances the performance of traditional sparse retrieval, with additional potential for improvement through the ideal Oracle runs. 

The recall improvement achieved with entity linking on sparse retrievers encourages further exploration. As a result, I train a state-of-the-art dense retrieval model optimized with STAR and ADORE algorithms on the expanded MS MARCO dataset (queries and passages). I evaluate my results with MRR@10 using the original qrels. Contrary to the expectations, entity augmentation shows a neutral effect on the performance of dense retrievers. 
This result concludes my experiments at the intersection of entity linking and document retrieval.

My findings across the sparse and dense experiments suggest further study and exploration of entity linking for IR advancement. In this work, I limit my expansion to entity names retrieved from KBs for each mention identified in the original text. However, it is interesting to attempt expanding the corpus, not only with entity names, but also with entity descriptions. The latter is usually formed of one or two sentences explaining the corresponding entity name. Some terms from the description can be relevant to the query text, hence improving the exact matching procedure of BOW retrievers. The expansion with entity description can also be tested on queries only, passages only or both. 

Another idea is to attempt packing both hashed and non-hashed entity names simultaneously, and append them to the original text before the indexing. it would be interesting to experiment with the effect of both entity formats on the retrieval. As previously underlined, the idea of MD5 hashing of the entity names was suggested to account for multi-word expressions, so that it is possible to encode each multi-word terminology into a single representation, hence diminishing mismatching possibilities. Nonetheless, it is found that enriching the text with hashed entity names hurts the overall BM25 performance, even after tuning the $k$ and $b_{1}$ parameters on the expanded corpus with hashed entities. This behaviour with BM25 necessitates further investigation. Maybe MD5 hashing is not the best approach to represent multi-word terms, or maybe BM25 model does not align well with the hashing idea. Could the performance have changed if I attempt dense retrieval on the hashed-entity-equipped dataset? How about hashing the whole text, not just the entities? Additional experiments along those lines might shed some light on why hashed entities fail to rise to the expectations. Despite that, it is no secret that the inclusion of the hashed-entity-aware run in the fusion pool gives my best-achieved performance using the original qrels for all query sets, and using the DuoT5 qrels for the Dev, Hard and Harder query sets. This implies that the idea is not a total failure, but further adjustment is still required.

One imminent extension to the current work might also be the enrichment of the run pool on which I apply RRF or the classifier's selection method. The current run pool incorporates only the following three runs: the original BM25 run generated on the non-expanded dataset, the BM25 run generated on the entity-aware dataset (or simply the entity run), and the BM25 run produced on the hashed-entity-equipped dataset (also known as the hashed entity run). RRF fuses combinations of these runs or all of them together to outperform the recall gain achieved by each standalone run. On the other hand, my classifier tries to choose the best run among the three, for each query with the same objective. In addition to the previous runs, it is useful to augment the pool with variations of sparse (beside BM25) runs applied on both the original and the expanded dataset. Dense runs might also be an option in case of time and computational resource availability. In this work, I limit the PRF application with BM25 on the original data due to time constraints. However, it is important to explore PRF on the entity-equipped data in both versions: explicit and hashed. PRF can also be adopted with other retrievers other than BM25. The PRF-based runs might be leveraged as a potential source for RRF/ classifier's pool augmentation as well. The enriched run pool would also provide the classifier with more options to select from.

Aligned with the previous idea, it is interesting to experiment with entity linking gains with neural sparse retrievers like DeepTR, DeepImpact or DeepCT to comparatively evaluate entity linking effect on diverse sparse methods. The same can be told about dense retrievers. Besides the STAR-ADORE pipeline, it is worth studying the effect of entity expansion with variations of dense models such as RepCONC \cite{repconc}, ANCE, RepBERT, ColBERT, etc. This might provide further explanation of the neutral effect I previously achieved with the STAR-ADORE-based dense model, whether the bad performance is inherent for dense retrieval, or the dense retriever choice was not the best. Since I have constrained the investigation in the dense-related experiments, there is a larger scope to attempt the same ideas explored with the sparse methods on the dense ones such as the evaluation with MonoT5 and DuoT qrels, the application of the dense model on the hashed-entity-equipped dataset, the selection or fusion of variations of dense runs generated on data with or without entities, the Oracle generation for the dense runs to quantify the potential recall upper bound, and the study of the entity linking effect with dense retrieval on the three sets of hard queries.

From another perspective, the Oracle's results generated for the sparse-based runs confirm that there is still a room for recall improvement with better run combination methods. For sparse experiments, the margin between my best actual run (RRF) and the ideal possible gain (Oracle) is of a significant factor across all the query sets: Dev, hard, Harder and Hardest. I have only experimented with RRF and classifier-based selection methods. Nonetheless, it is worth exploring other fusion or selection approaches to narrow the gap between the actual and the ideal recall gain, and to maximize the benefits of linked entities. The suggested classifier's idea can also be enhanced by modifying the model architecture, tuning the parameters or both.

Furthermore, it is possible to try expanding the run pool used in the generation of the MonoT5 and DuoT5 qrels. As previously described, the run pool used for this purpose includes the original BM25 run, the entity-based BM25 run, the hashed-entity-based BM25 run and the original ANCE run. As shown, my run pool is not diverse enough to account for all the correct answers. Including other dense runs generated on the dataset (with and without entities) before the re-ranking procedures can refine the resultant ground truth. Testing the behaviour against the new qrels is a good idea for a related future work.

It is also of interest to try a mixture of expansion methods together and evaluate the impact of such a scheme on the Dev and Hard queries. For example, combining my entity linking expansion methods with doc2query or docTTTTTquery approaches to solidify semantic understanding of the BOW sparse retrieval might serve as an adequate extension to the current research.

Finally, all of my experiments are conducted on the MS MARCO passage dataset. In order to draw a definitive conclusion about the entity linking effect on the retrieval, it is critical to adopt my methods on longer texts such as the MS MARCO document dataset. Documents are challenging because of their length, and the unavailability of off-the-shelf semantic tools that can recognize and disambiguate entity mentions in them. Extending the current entity linking end-to-end tools to accommodate longer texts, or even building new versions of these models will be required to augment the text before jumping into the retrieval experiments.

% \input{chapter-observations.tex}

%----------------------------------------------------------------------
% END MATERIAL
% Bibliography, Appendices, Index, etc.
%----------------------------------------------------------------------

% Bibliography

% The following statement selects the style to use for references.  
% It controls the sort order of the entries in the bibliography and also the formatting for the in-text labels.
\bibliographystyle{plain}
% This specifies the location of the file containing the bibliographic information.  
% It assumes you're using BibTeX to manage your references (if not, why not?).
\cleardoublepage % This is needed if the "book" document class is used, to place the anchor in the correct page, because the bibliography will start on its own page.
% Use \clearpage instead if the document class uses the "oneside" argument
\phantomsection  % With hyperref package, enables hyperlinking from the table of contents to bibliography             
% The following statement causes the title "References" to be used for the bibliography section:
\renewcommand*{\bibname}{References}

% Add the References to the Table of Contents
\addcontentsline{toc}{chapter}{\textbf{References}}

\bibliography{uw-ethesis}
% Tip: You can create multiple .bib files to organize your references. 
% Just list them all in the \bibliogaphy command, separated by commas (no spaces).

% The following statement causes the specified references to be added to the bibliography even if they were not cited in the text. 
% The asterisk is a wildcard that causes all entries in the bibliographic database to be included (optional).
% \nocite{*}
%----------------------------------------------------------------------

% Appendices

% The \appendix statement indicates the beginning of the appendices.
\appendix
% Add an un-numbered title page before the appendices and a line in the Table of Contents
\chapter*{APPENDICES}
\addcontentsline{toc}{chapter}{APPENDICES}
% Appendices are just more chapters, with different labeling (letters instead of numbers).
\chapter[Additional Results]{Additional Results}
\label{AppendixA}
\vspace{-1.1cm}
The recall curves are plotted using the recall values at each cutoff in $[0, 1000]$. The following tables demonstrate the corresponding values for each query-qrel combinations.

\begin{table}[!ht]
\caption{Recall Values for the Dev set Curves of Figure \ref{fig:dev_org} evaluated against the Original Qrels. No entities is the original BM25 run generated from the non-expanded dataset. Hashed designates the hashed-entity run, RRF1 is the RRF (No entities/ Hashed entities), RRF2 is the	RRF (Hashed entities/ Entities), RRF3 is the RRF (No entities/ Entities), RRF4 is the RRF (No Entities/ Hashed entities/ Entities).}
  
% \hspace*{-2.4cm}
    \centering
    \begin{sideways}
    \scriptsize
    \begin{tabular}{|c|c|c|c|c|c|c|c|c|c|c|}
    \hline
    
    \textbf{Recall} & \textbf{No entities} & \textbf{Hashed} & \textbf{Entities} & \textbf{Classifier} & \textbf{RRF1} & \textbf{RRF2} & \textbf{RRF3} &\textbf{RRF4} & \textbf{Oracle} & \textbf{Dense} \\ [0.5ex]
     
    \hline
        0 & 0 & 0 & 0 & 0 & 0 & 0 & 0 & 0 & 0 & 0 \\ \hline
        50 & 0.5923 & 0.5482 & 0.5787 & 0.6056 & 0.6029 & 0.5834 & 0.6125 & 0.6081 & 0.6743 & 0.6727 \\ \hline
        55 & 0.6032 & 0.5604 & 0.5902 & 0.6171 & 0.6148 & 0.5968 & 0.6255 & 0.6211 & 0.6862 & 0.6845 \\ \hline
        60 & 0.6139 & 0.5718 & 0.601 & 0.6282 & 0.6272 & 0.6097 & 0.6346 & 0.6325 & 0.6955 & 0.6939 \\ \hline
        75 & 0.6397 & 0.5992 & 0.6291 & 0.6542 & 0.6583 & 0.6387 & 0.6605 & 0.6602 & 0.7198 & 0.7182 \\ \hline
        100 & 0.6701 & 0.6344 & 0.6623 & 0.6869 & 0.6886 & 0.6756 & 0.6957 & 0.6945 & 0.7522 & 0.7502 \\ \hline
        200 & 0.7383 & 0.7104 & 0.7406 & 0.7583 & 0.7581 & 0.7501 & 0.768 & 0.7693 & 0.8122 & 0.8095 \\ \hline
        300 & 0.7751 & 0.7527 & 0.777 & 0.7929 & 0.7922 & 0.7894 & 0.8037 & 0.806 & 0.8418 & 0.8395 \\ \hline
        400 & 0.799 & 0.7797 & 0.8005 & 0.8159 & 0.8159 & 0.8104 & 0.8277 & 0.8278 & 0.8627 & 0.8606 \\ \hline
        500 & 0.8116 & 0.7969 & 0.8191 & 0.8295 & 0.8319 & 0.8291 & 0.8436 & 0.8448 & 0.8755 & 0.8737 \\ \hline
        600 & 0.8251 & 0.8112 & 0.8322 & 0.8424 & 0.8459 & 0.844 & 0.8552 & 0.8579 & 0.8844 & 0.8828 \\ \hline
        700 & 0.8359 & 0.8234 & 0.8439 & 0.8541 & 0.8582 & 0.8551 & 0.8651 & 0.8673 & 0.8932 & 0.8916 \\ \hline
        800 & 0.8431 & 0.8326 & 0.8537 & 0.8614 & 0.8648 & 0.8641 & 0.8737 & 0.8735 & 0.8996 & 0.898 \\ \hline
        900 & 0.8513 & 0.8409 & 0.8627 & 0.8703 & 0.8711 & 0.8714 & 0.8787 & 0.8799 & 0.9064 & 0.9047 \\ \hline
        1000 & 0.8573 & 0.8479 & 0.8682 & 0.8754 & 0.878 & 0.8784 & 0.8844 & 0.8868 & 0.9104 & 0.9087 \\ \hline
    \end{tabular}
    \end{sideways}
    \label{tab:recall_dev_org}
    
\end{table}

\begin{table}[!ht]
    \caption{Recall Values for the Hard set Curves of Figure \ref{fig:veiled_org} evaluated against the Original Qrels. No entities is the original BM25 run generated from the non-expanded dataset. Hashed designates the hashed-entity run, RRF1 is the RRF (No entities/ Hashed entities), RRF2 is the	RRF (Hashed entities/ Entities), RRF3 is the RRF (No entities/ Entities), RRF4 is the RRF (No Entities/ Hashed entities/ Entities).} \vspace{0.5cm}
% \hspace*{-2.4cm}
    \centering
    \begin{sideways}
    \scriptsize
    \begin{tabular}{|c|c|c|c|c|c|c|c|c|c|c|}
    \hline
    
    \textbf{Recall} & \textbf{No entities} & \textbf{Hashed} & \textbf{Entities} & \textbf{Classifier} & \textbf{RRF1} & \textbf{RRF2} & \textbf{RRF3} &\textbf{RRF4} & \textbf{Oracle} & \textbf{Dense} \\ [0.5ex]
    \hline
        0 & 0 & 0 & 0 & 0 & 0 & 0 & 0 & 0 & 0 & 0 \\ \hline
        50 & 0.2854 & 0.2747 & 0.2977 & 0.3080 & 0.3052 & 0.2991 & 0.3104 & 0.309 & 0.3948 & 0.5718 \\ \hline
        55 & 0.3002 & 0.2877 & 0.3129 & 0.3244 & 0.3203 & 0.3151 & 0.3288 & 0.3266 & 0.4122 & 0.5861 \\ \hline
        60 & 0.3169 & 0.3049 & 0.3262 & 0.3380 & 0.3328 & 0.3295 & 0.3402 & 0.3431 & 0.4282 & 0.6013 \\ \hline
        75 & 0.3586 & 0.3364 & 0.3646 & 0.3800 & 0.3765 & 0.3663 & 0.3789 & 0.3806 & 0.4695 & 0.6466 \\ \hline
        100 & 0.4049 & 0.3853 & 0.4114 & 0.4280 & 0.4215 & 0.4189 & 0.4356 & 0.4325 & 0.5242 & 0.6937 \\ \hline
        200 & 0.5158 & 0.495 & 0.5294 & 0.5476 & 0.5394 & 0.5348 & 0.5523 & 0.557 & 0.6316 & 0.786 \\ \hline
        300 & 0.5764 & 0.5554 & 0.5858 & 0.6043 & 0.5994 & 0.6043 & 0.617 & 0.6214 & 0.6846 & 0.8287 \\ \hline
        400 & 0.6191 & 0.5979 & 0.6247 & 0.6441 & 0.6444 & 0.6403 & 0.6613 & 0.6629 & 0.724 & 0.8548 \\ \hline
        500 & 0.6411 & 0.6266 & 0.6562 & 0.6710 & 0.6732 & 0.673 & 0.6907 & 0.6941 & 0.7477 & 0.8712 \\ \hline
        600 & 0.6653 & 0.6513 & 0.6782 & 0.6938 & 0.7 & 0.6971 & 0.7133 & 0.7189 & 0.7648 & 0.8838 \\ \hline
        700 & 0.6853 & 0.672 & 0.7008 & 0.7162 & 0.7223 & 0.7166 & 0.7316 & 0.7358 & 0.7826 & 0.8914 \\ \hline
        800 & 0.6974 & 0.6876 & 0.7201 & 0.7303 & 0.7347 & 0.7333 & 0.7479 & 0.7481 & 0.7953 & 0.8987 \\ \hline
        900 & 0.7124 & 0.7022 & 0.7367 & 0.7482 & 0.746 & 0.7464 & 0.7581 & 0.761 & 0.8086 & 0.9084 \\ \hline
        1000 & 0.7234 & 0.7146 & 0.7467 & 0.7582 & 0.7591 & 0.7599 & 0.7695 & 0.7738 & 0.8159 & 0.9152 \\ \hline
    \end{tabular}
    \label{tab:recall_veiled_org}
    \end{sideways}
\end{table}

\begin{table}[!ht]
      \caption{Recall Values for the Harder set Curves of Figure \ref{fig:pygmy_org} evaluated against the Original Qrels . No entities is the original BM25 run generated from the non-expanded dataset. Hashed designates the hashed-entity run, RRF1 is the RRF (No entities/ Hashed entities), RRF2 is the	RRF (Hashed entities/ Entities), RRF3 is the RRF (No entities/ Entities), RRF4 is the RRF (No Entities/ Hashed entities/ Entities).} \vspace{0.5cm}
% \hspace*{-2.4cm}
    \centering
    \begin{sideways}
    \scriptsize
    \begin{tabular}{|c|c|c|c|c|c|c|c|c|c|c|}
    \hline
    
    \textbf{Recall} & \textbf{No entities} & \textbf{Hashed} & \textbf{Entities} & \textbf{Classifier} & \textbf{RRF1} & \textbf{RRF2} & \textbf{RRF3} &\textbf{RRF4} & \textbf{Oracle} & \textbf{Dense} \\ [0.5ex]
    \hline
        0 & 0 & 0 & 0 & 0 & 0 & 0 & 0 & 0 & 0 & 0 \\ \hline
        50 & 0.2068 & 0.2108 & 0.23 & 0.2368 & 0.2288 & 0.2294 & 0.2345 & 0.2332 & 0.3195 & 0.5138 \\ \hline
        55 & 0.2208 & 0.2239 & 0.246 & 0.2532 & 0.2442 & 0.2455 & 0.2525 & 0.2505 & 0.3366 & 0.5306 \\ \hline
        60 & 0.2378 & 0.2403 & 0.2597 & 0.2673 & 0.2563 & 0.2586 & 0.2656 & 0.2682 & 0.3546 & 0.548 \\ \hline
        75 & 0.2828 & 0.2714 & 0.2989 & 0.3078 & 0.3007 & 0.2939 & 0.3045 & 0.3045 & 0.398 & 0.5979 \\ \hline
        100 & 0.3314 & 0.3179 & 0.3433 & 0.3581 & 0.3469 & 0.3455 & 0.3629 & 0.3567 & 0.458 & 0.6523 \\ \hline
        200 & 0.4499 & 0.4302 & 0.4705 & 0.4842 & 0.4732 & 0.4716 & 0.4848 & 0.4934 & 0.5727 & 0.7544 \\ \hline
        300 & 0.5197 & 0.4975 & 0.5305 & 0.5500 & 0.5411 & 0.5498 & 0.5585 & 0.5637 & 0.6363 & 0.8016 \\ \hline
        400 & 0.5674 & 0.5443 & 0.5735 & 0.5957 & 0.5903 & 0.5886 & 0.608 & 0.6099 & 0.6801 & 0.8329 \\ \hline
        500 & 0.592 & 0.5747 & 0.608 & 0.6231 & 0.6236 & 0.6235 & 0.6407 & 0.6456 & 0.7058 & 0.8527 \\ \hline
        600 & 0.6201 & 0.602 & 0.6325 & 0.6496 & 0.6542 & 0.6484 & 0.6674 & 0.6741 & 0.7252 & 0.8672 \\ \hline
        700 & 0.6433 & 0.6257 & 0.6581 & 0.6764 & 0.679 & 0.6708 & 0.6898 & 0.6933 & 0.7462 & 0.8764 \\ \hline
        800 & 0.6567 & 0.642 & 0.6787 & 0.6897 & 0.6923 & 0.6899 & 0.7081 & 0.7066 & 0.7597 & 0.8845 \\ \hline
        900 & 0.6728 & 0.6586 & 0.6969 & 0.7079 & 0.7046 & 0.7054 & 0.7196 & 0.7204 & 0.7745 & 0.8949 \\ \hline
        1000 & 0.6849 & 0.6727 & 0.7079 & 0.7189 & 0.7195 & 0.7196 & 0.7323 & 0.7353 & 0.7827 & 0.9022 \\ \hline
    \end{tabular}
    \label{tab:recall_pygmy_org}
    \end{sideways}
\end{table}

\begin{table}[!ht]
      \caption{Recall Values for the Hardest set Curves of Figure \ref{fig:lesser_org} evaluated against the Original Qrels . No entities is the original BM25 run generated from the non-expanded dataset. Hashed designates the hashed-entity run, RRF1 is the RRF (No entities/ Hashed entities), RRF2 is the	RRF (Hashed entities/ Entities), RRF3 is the RRF (No entities/ Entities), RRF4 is the RRF (No Entities/ Hashed entities/ Entities).} \vspace{0.5cm}
% \hspace*{-2.4cm}
    \centering
    \begin{sideways}
    \scriptsize
    \begin{tabular}{|c|c|c|c|c|c|c|c|c|c|c|}
    \hline
    
    \textbf{Recall} & \textbf{No entities} & \textbf{Hashed} & \textbf{Entities} & \textbf{Classifier} & \textbf{RRF1} & \textbf{RRF2} & \textbf{RRF3} &\textbf{RRF4} & \textbf{Oracle} & \textbf{Dense} \\ [0.5ex]
    \hline
        0 & 0 & 0 & 0 & 0 & 0 & 0 & 0 & 0 & 0 & 0 \\ \hline
        50 & 0.1089 & 0.1332 & 0.1449 & 0.1475 & 0.1335 & 0.1418 & 0.1343 & 0.1406 & 0.2119 & 0.4155 \\ \hline
        55 & 0.12 & 0.143 & 0.1588 & 0.1611 & 0.1479 & 0.1575 & 0.1505 & 0.1548 & 0.2259 & 0.4311 \\ \hline
        60 & 0.1371 & 0.1589 & 0.1726 & 0.1748 & 0.1576 & 0.1705 & 0.1641 & 0.1718 & 0.2434 & 0.4508 \\ \hline
        75 & 0.1769 & 0.1858 & 0.2083 & 0.2108 & 0.195 & 0.2002 & 0.1985 & 0.2011 & 0.2828 & 0.5094 \\ \hline
        100 & 0.2237 & 0.2278 & 0.2534 & 0.2588 & 0.238 & 0.2497 & 0.2543 & 0.2502 & 0.3504 & 0.5691 \\ \hline
        200 & 0.3484 & 0.3396 & 0.3759 & 0.3890 & 0.3671 & 0.3735 & 0.3814 & 0.3869 & 0.4751 & 0.6891 \\ \hline
        300 & 0.4283 & 0.409 & 0.4394 & 0.4571 & 0.4434 & 0.4551 & 0.4597 & 0.4644 & 0.5462 & 0.7465 \\ \hline
        400 & 0.4829 & 0.4613 & 0.485 & 0.5071 & 0.5004 & 0.4999 & 0.5174 & 0.5182 & 0.5978 & 0.7878 \\ \hline
        500 & 0.5112 & 0.4955 & 0.5274 & 0.5439 & 0.5394 & 0.5388 & 0.5555 & 0.5615 & 0.6285 & 0.815 \\ \hline
        600 & 0.5366 & 0.5232 & 0.5544 & 0.5701 & 0.5723 & 0.5696 & 0.5839 & 0.5934 & 0.6487 & 0.8324 \\ \hline
        700 & 0.5649 & 0.5479 & 0.5854 & 0.5994 & 0.6003 & 0.5966 & 0.6115 & 0.6178 & 0.6748 & 0.8444 \\ \hline
        800 & 0.5797 & 0.5677 & 0.6072 & 0.6160 & 0.6152 & 0.6188 & 0.6339 & 0.6322 & 0.6925 & 0.8526 \\ \hline
        900 & 0.5988 & 0.586 & 0.6272 & 0.6405 & 0.6284 & 0.6331 & 0.6478 & 0.6466 & 0.7118 & 0.8661 \\ \hline
        1000 & 0.6136 & 0.5995 & 0.6389 & 0.6528 & 0.6471 & 0.6498 & 0.6625 & 0.665 & 0.722 & 0.8753 \\ \hline
    \end{tabular}
    \label{tab:recall_lesser_org}
    \end{sideways}
\end{table}

\begin{table}[!ht]
      \caption{Recall Values for the Dev set Curves of Figure \ref{fig:dev_monoT5} evaluated against MonoT5 Qrels. No entities is the original BM25 run generated from the non-expanded dataset. Hashed designates the hashed-entity run, RRF1 is the RRF (No entities/ Hashed entities), RRF2 is the	RRF (Hashed entities/ Entities), RRF3 is the RRF (No entities/ Entities), RRF4 is the RRF (No Entities/ Hashed entities/ Entities).} \vspace{0.5cm}
% \hspace*{-2.4cm}
    \centering
    \begin{sideways}
    \scriptsize
    \begin{tabular}{|c|c|c|c|c|c|c|c|c|c|c|}
    \hline
    
    \textbf{Recall} & \textbf{No entities} & \textbf{Hashed} & \textbf{Entities} & \textbf{Classifier} & \textbf{RRF1} & \textbf{RRF2} & \textbf{RRF3} &\textbf{RRF4} & \textbf{Oracle} & \textbf{Dense} \\ [0.5ex]
    \hline
        0 & 0 & 0 & 0 & 0 & 0 & 0 & 0 & 0 & 0 & 0 \\ \hline
        50 & 0.6585 & 0.5938 & 0.6335 & 0.6692 & 0.6622 & 0.6364 & 0.6702 & 0.6668 & 0.7235 & 0.8779 \\ \hline
        55 & 0.667 & 0.6069 & 0.6437 & 0.6781 & 0.6742 & 0.6491 & 0.6818 & 0.6771 & 0.7324 & 0.8838 \\ \hline
        60 & 0.6762 & 0.6178 & 0.6527 & 0.6887 & 0.6852 & 0.659 & 0.6913 & 0.6862 & 0.7413 & 0.8874 \\ \hline
        75 & 0.6977 & 0.6414 & 0.6748 & 0.7125 & 0.7073 & 0.683 & 0.7152 & 0.7095 & 0.7612 & 0.8986 \\ \hline
        100 & 0.7256 & 0.6701 & 0.7023 & 0.7374 & 0.7364 & 0.7113 & 0.7417 & 0.7395 & 0.7844 & 0.9125 \\ \hline
        200 & 0.7778 & 0.7342 & 0.7648 & 0.7908 & 0.7917 & 0.7751 & 0.7966 & 0.7971 & 0.8305 & 0.9381 \\ \hline
        300 & 0.8042 & 0.7675 & 0.7924 & 0.8156 & 0.8178 & 0.8037 & 0.8238 & 0.8223 & 0.8509 & 0.9532 \\ \hline
        400 & 0.8226 & 0.7895 & 0.8125 & 0.8332 & 0.8365 & 0.8196 & 0.8437 & 0.8403 & 0.8659 & 0.9617 \\ \hline
        500 & 0.8362 & 0.8069 & 0.8266 & 0.8461 & 0.8486 & 0.8338 & 0.8549 & 0.8539 & 0.8779 & 0.9692 \\ \hline
        600 & 0.8436 & 0.8192 & 0.8377 & 0.8547 & 0.8579 & 0.8453 & 0.8619 & 0.8628 & 0.8837 & 0.9735 \\ \hline
        700 & 0.8503 & 0.8288 & 0.8468 & 0.8617 & 0.8646 & 0.855 & 0.8701 & 0.8698 & 0.8883 & 0.9775 \\ \hline
        800 & 0.8554 & 0.837 & 0.853 & 0.8663 & 0.8696 & 0.8619 & 0.8759 & 0.8749 & 0.8918 & 0.9812 \\ \hline
        900 & 0.861 & 0.8453 & 0.8593 & 0.8716 & 0.8745 & 0.8682 & 0.8822 & 0.8799 & 0.8956 & 0.9862 \\ \hline
        1000 & 0.8655 & 0.8493 & 0.8639 & 0.8764 & 0.8774 & 0.8734 & 0.8845 & 0.8838 & 0.8977 & 0.988 \\ \hline
    \end{tabular}
    \label{tab:recall_dev_monoT5}
    \end{sideways}
\end{table}

\begin{table}[!ht]
    \caption{Recall Values for the Hard set Curves of Figure \ref{fig:veiled_monoT5} evaluated against MonoT5 Qrels. No entities is the original BM25 run generated from the non-expanded dataset. Hashed designates the hashed-entity run, RRF1 is the RRF (No entities/ Hashed entities), RRF2 is the	RRF (Hashed entities/ Entities), RRF3 is the RRF (No entities/ Entities), RRF4 is the RRF (No Entities/ Hashed entities/ Entities).} \vspace{0.5cm}
% \hspace*{-2.4cm}
    \centering
    \begin{sideways}
    \scriptsize
    \begin{tabular}{|c|c|c|c|c|c|c|c|c|c|c|}
    \hline
    
    \textbf{Recall} & \textbf{No entities} & \textbf{Hashed} & \textbf{Entities} & \textbf{Classifier} & \textbf{RRF1} & \textbf{RRF2} & \textbf{RRF3} &\textbf{RRF4} & \textbf{Oracle} & \textbf{Dense} \\ [0.5ex]
    \hline
        0 & 0 & 0 & 0 & 0 & 0 & 0 & 0 & 0 & 0 & 0 \\ \hline
        50 & 0.513 & 0.4431 & 0.4934 & 0.5245 & 0.5181 & 0.4921 & 0.5281 & 0.5226 & 0.5947 & 0.8201 \\ \hline
        55 & 0.5261 & 0.4566 & 0.5056 & 0.5364 & 0.5374 & 0.5066 & 0.5444 & 0.5377 & 0.6056 & 0.8282 \\ \hline
        60 & 0.5399 & 0.471 & 0.5162 & 0.5499 & 0.5518 & 0.5194 & 0.5569 & 0.5505 & 0.6207 & 0.8336 \\ \hline
        75 & 0.5726 & 0.5043 & 0.5418 & 0.5848 & 0.581 & 0.5531 & 0.5909 & 0.5838 & 0.6518 & 0.8506 \\ \hline
        100 & 0.6133 & 0.5444 & 0.5797 & 0.6210 & 0.621 & 0.5887 & 0.6281 & 0.6246 & 0.6861 & 0.8737 \\ \hline
        200 & 0.6884 & 0.6284 & 0.664 & 0.6996 & 0.7012 & 0.6775 & 0.7044 & 0.7082 & 0.7567 & 0.9096 \\ \hline
        300 & 0.7262 & 0.673 & 0.7025 & 0.7349 & 0.7413 & 0.7195 & 0.7477 & 0.7432 & 0.7874 & 0.9336 \\ \hline
        400 & 0.7554 & 0.7054 & 0.7345 & 0.7634 & 0.7704 & 0.7409 & 0.7772 & 0.7714 & 0.8121 & 0.9461 \\ \hline
        500 & 0.7772 & 0.7297 & 0.757 & 0.7858 & 0.789 & 0.764 & 0.7961 & 0.7945 & 0.8336 & 0.957 \\ \hline
        600 & 0.7887 & 0.7496 & 0.7756 & 0.8003 & 0.8057 & 0.781 & 0.807 & 0.8089 & 0.8432 & 0.9638 \\ \hline
        700 & 0.7993 & 0.7631 & 0.7878 & 0.8115 & 0.8156 & 0.7983 & 0.8205 & 0.8198 & 0.8503 & 0.9692 \\ \hline
        800 & 0.807 & 0.7756 & 0.7993 & 0.8201 & 0.8237 & 0.8096 & 0.831 & 0.8301 & 0.8551 & 0.9747 \\ \hline
        900 & 0.8156 & 0.7874 & 0.8089 & 0.8278 & 0.8317 & 0.8195 & 0.8407 & 0.8368 & 0.8618 & 0.9808 \\ \hline
        1000 & 0.8227 & 0.7945 & 0.8156 & 0.8362 & 0.8362 & 0.8272 & 0.8445 & 0.8426 & 0.8644 & 0.983 \\ \hline
    \end{tabular}
    \label{tab:recall_veiled_monoT5}
    \end{sideways}
\end{table}

\begin{table}[!ht]
      \caption{Recall Values for the Harder set Curves of Figure \ref{fig:pygmy_monoT5} evaluated against MonoT5 Qrels. No entities is the original BM25 run generated from the non-expanded dataset. Hashed designates the hashed-entity run, RRF1 is the RRF (No entities/ Hashed entities), RRF2 is the	RRF (Hashed entities/ Entities), RRF3 is the RRF (No entities/ Entities), RRF4 is the RRF (No Entities/ Hashed entities/ Entities).} \vspace{0.5cm}
% \hspace*{-2.4cm}
    \centering
    \begin{sideways}
    \scriptsize
    \begin{tabular}{|c|c|c|c|c|c|c|c|c|c|c|}
    \hline
    
    \textbf{Recall} & \textbf{No entities} & \textbf{Hashed} & \textbf{Entities} & \textbf{Classifier} & \textbf{RRF1} & \textbf{RRF2} & \textbf{RRF3} &\textbf{RRF4} & \textbf{Oracle} & \textbf{Dense} \\ [0.5ex]
    \hline
        0 & 0 & 0 & 0 & 0 & 0 & 0 & 0 & 0 & 0 & 0 \\ \hline
        50 & 0.4905 & 0.4278 & 0.4695 & 0.4998 & 0.4974 & 0.4695 & 0.503 & 0.4994 & 0.5722 & 0.8148 \\ \hline
        55 & 0.5051 & 0.4395 & 0.4804 & 0.5127 & 0.5156 & 0.4832 & 0.5204 & 0.514 & 0.5847 & 0.8233 \\ \hline
        60 & 0.5168 & 0.4537 & 0.4909 & 0.5261 & 0.5301 & 0.4958 & 0.5338 & 0.5273 & 0.5997 & 0.8294 \\ \hline
        75 & 0.5499 & 0.4848 & 0.5196 & 0.5625 & 0.5617 & 0.5293 & 0.5669 & 0.5592 & 0.632 & 0.8435 \\ \hline
        100 & 0.5904 & 0.522 & 0.556 & 0.5981 & 0.6017 & 0.5653 & 0.6053 & 0.6025 & 0.6684 & 0.8666 \\ \hline
        200 & 0.6684 & 0.6098 & 0.6454 & 0.6801 & 0.6826 & 0.6571 & 0.6858 & 0.6899 & 0.7408 & 0.907 \\ \hline
        300 & 0.7105 & 0.6531 & 0.6858 & 0.7190 & 0.725 & 0.702 & 0.7303 & 0.7271 & 0.7744 & 0.9313 \\ \hline
        400 & 0.7416 & 0.6874 & 0.719 & 0.7497 & 0.7566 & 0.7242 & 0.7622 & 0.757 & 0.8015 & 0.9438 \\ \hline
        500 & 0.7671 & 0.7117 & 0.744 & 0.7764 & 0.7772 & 0.7505 & 0.7829 & 0.7808 & 0.8261 & 0.9559 \\ \hline
        600 & 0.7784 & 0.7343 & 0.7634 & 0.7901 & 0.7946 & 0.7675 & 0.7954 & 0.7978 & 0.8366 & 0.962 \\ \hline
        700 & 0.7913 & 0.7493 & 0.7776 & 0.8039 & 0.8063 & 0.7869 & 0.812 & 0.8112 & 0.8451 & 0.9681 \\ \hline
        800 & 0.7998 & 0.7626 & 0.7901 & 0.8128 & 0.8148 & 0.799 & 0.8241 & 0.8221 & 0.8508 & 0.9745 \\ \hline
        900 & 0.8099 & 0.7764 & 0.8015 & 0.8221 & 0.8245 & 0.8104 & 0.8362 & 0.8294 & 0.8585 & 0.9802 \\ \hline
        1000 & 0.8176 & 0.7845 & 0.8099 & 0.8322 & 0.8302 & 0.8197 & 0.8399 & 0.8362 & 0.8617 & 0.9822 \\ \hline
    \end{tabular}
    \label{tab:recall_pygmy_monoT5}
    \end{sideways}
\end{table}

\begin{table}[!ht]
     \caption{Recall Values for the Hardest set Curves of Figure \ref{fig:lesser_monoT5} evaluated against MonoT5 Qrels. No entities is the original BM25 run generated from the non-expanded dataset. Hashed designates the hashed-entity run, RRF1 is the RRF (No entities/ Hashed entities), RRF2 is the	RRF (Hashed entities/ Entities), RRF3 is the RRF (No entities/ Entities), RRF4 is the RRF (No Entities/ Hashed entities/ Entities).} \vspace{0.5cm}
% \hspace*{-2.4cm}
    \centering
    \begin{sideways}
    \scriptsize
    \begin{tabular}{|c|c|c|c|c|c|c|c|c|c|c|}
    \hline
    
    \textbf{Recall} & \textbf{No entities} & \textbf{Hashed} & \textbf{Entities} & \textbf{Classifier} & \textbf{RRF1} & \textbf{RRF2} & \textbf{RRF3} &\textbf{RRF4} & \textbf{Oracle} & \textbf{Dense} \\ [0.5ex]
    \hline
        0 & 0 & 0 & ~ & 0 & 0 & 0 & 0 & 0 & 0 & 0 \\ \hline
        50 & 0.4838 & 0.4188 & 0.4903 & 0.4554 & 0.4908 & 0.4589 & 0.4932 & 0.4891 & 0.5623 & 0.811 \\ \hline
        55 & 0.5009 & 0.4288 & 0.5038 & 0.4666 & 0.5068 & 0.4708 & 0.5115 & 0.5009 & 0.5765 & 0.8187 \\ \hline
        60 & 0.5127 & 0.4454 & 0.5168 & 0.4761 & 0.521 & 0.4838 & 0.521 & 0.5151 & 0.5913 & 0.8246 \\ \hline
        75 & 0.5405 & 0.4761 & 0.5481 & 0.5032 & 0.554 & 0.5204 & 0.5552 & 0.5493 & 0.6232 & 0.8382 \\ \hline
        100 & 0.5812 & 0.5139 & 0.5865 & 0.5428 & 0.5983 & 0.5552 & 0.5966 & 0.5936 & 0.6586 & 0.8571 \\ \hline
        200 & 0.6604 & 0.5978 & 0.6716 & 0.6361 & 0.6722 & 0.6456 & 0.6781 & 0.6781 & 0.7336 & 0.9037 \\ \hline
        300 & 0.7035 & 0.6379 & 0.7094 & 0.6769 & 0.7177 & 0.6899 & 0.7194 & 0.7177 & 0.7667 & 0.9297 \\ \hline
        400 & 0.7389 & 0.6751 & 0.7460 & 0.7112 & 0.7507 & 0.7135 & 0.7566 & 0.749 & 0.7974 & 0.9415 \\ \hline
        500 & 0.762 & 0.7023 & 0.7714 & 0.7377 & 0.7708 & 0.7437 & 0.7797 & 0.7744 & 0.8187 & 0.9545 \\ \hline
        600 & 0.775 & 0.7283 & 0.7856 & 0.7578 & 0.7891 & 0.7608 & 0.7915 & 0.7933 & 0.8317 & 0.9604 \\ \hline
        700 & 0.7891 & 0.7431 & 0.8004 & 0.7708 & 0.8021 & 0.7803 & 0.8092 & 0.8086 & 0.8417 & 0.9669 \\ \hline
        800 & 0.8004 & 0.7578 & 0.8116 & 0.7832 & 0.811 & 0.795 & 0.8198 & 0.8193 & 0.8488 & 0.9734 \\ \hline
        900 & 0.8092 & 0.7726 & 0.8204 & 0.7962 & 0.8187 & 0.8074 & 0.8328 & 0.8275 & 0.8576 & 0.9793 \\ \hline
        1000 & 0.8181 & 0.782 & 0.8323 & 0.8051 & 0.8263 & 0.8157 & 0.8364 & 0.8328 & 0.8606 & 0.9823 \\ \hline
    \end{tabular}
    \label{tab:recall_lesser_monoT5}
    \end{sideways}
\end{table}

\begin{table}[!ht]
     \caption{Recall Values for the Dev set Curves of Figure \ref{fig:dev_duoT5} evaluated against DuoT5 Qrels. No entities is the original BM25 run generated from the non-expanded dataset. Hashed designates the hashed-entity run, RRF1 is the RRF (No entities/ Hashed entities), RRF2 is the	RRF (Hashed entities/ Entities), RRF3 is the RRF (No entities/ Entities), RRF4 is the RRF (No Entities/ Hashed entities/ Entities).} \vspace{0.5cm}
% \hspace*{-2.4cm}
    \centering
    \begin{sideways}
    \scriptsize
    \begin{tabular}{|c|c|c|c|c|c|c|c|c|c|c|}
    \hline
    
    \textbf{Recall} & \textbf{No entities} & \textbf{Hashed} & \textbf{Entities} & \textbf{Classifier} & \textbf{RRF1} & \textbf{RRF2} & \textbf{RRF3} &\textbf{RRF4} & \textbf{Oracle} & \textbf{Dense} \\ [0.5ex]
    \hline
        0 & 0 & 0 & 0 & 0 & 0 & 0 & 0 & 0 & 0 & 0 \\ \hline
        50 & 0.6948 & 0.6383 & 0.6729 & 0.7105 & 0.7006 & 0.6762 & 0.7102 & 0.705 & 0.766 & 0.9236 \\ \hline
        55 & 0.7043 & 0.6506 & 0.683 & 0.7193 & 0.7135 & 0.688 & 0.7205 & 0.7152 & 0.7749 & 0.9277 \\ \hline
        60 & 0.7123 & 0.66 & 0.6915 & 0.7271 & 0.7229 & 0.6983 & 0.7304 & 0.7244 & 0.783 & 0.9308 \\ \hline
        75 & 0.7357 & 0.6837 & 0.7165 & 0.7503 & 0.7484 & 0.7248 & 0.7542 & 0.7517 & 0.8056 & 0.9403 \\ \hline
        100 & 0.7638 & 0.7126 & 0.7431 & 0.7777 & 0.7768 & 0.752 & 0.7824 & 0.7825 & 0.8268 & 0.9519 \\ \hline
        200 & 0.8169 & 0.7765 & 0.8072 & 0.8305 & 0.8312 & 0.8168 & 0.8372 & 0.8395 & 0.8711 & 0.9683 \\ \hline
        300 & 0.8423 & 0.8103 & 0.8338 & 0.8549 & 0.8585 & 0.8461 & 0.8638 & 0.8632 & 0.8917 & 0.9766 \\ \hline
        400 & 0.858 & 0.8309 & 0.8554 & 0.8728 & 0.8749 & 0.864 & 0.8807 & 0.8811 & 0.9062 & 0.9811 \\ \hline
        500 & 0.8698 & 0.8493 & 0.8679 & 0.8854 & 0.8865 & 0.8766 & 0.8915 & 0.8943 & 0.9162 & 0.9837 \\ \hline
        600 & 0.8787 & 0.8616 & 0.8788 & 0.8937 & 0.8968 & 0.8885 & 0.9009 & 0.9024 & 0.9222 & 0.9868 \\ \hline
        700 & 0.8852 & 0.8706 & 0.8857 & 0.8996 & 0.9027 & 0.8961 & 0.908 & 0.9087 & 0.9262 & 0.9877 \\ \hline
        800 & 0.8907 & 0.8779 & 0.8931 & 0.9044 & 0.909 & 0.9036 & 0.9142 & 0.9143 & 0.9295 & 0.9894 \\ \hline
        900 & 0.895 & 0.8848 & 0.8983 & 0.9080 & 0.9139 & 0.9087 & 0.9183 & 0.9186 & 0.9327 & 0.9918 \\ \hline
        1000 & 0.8997 & 0.8904 & 0.9036 & 0.9125 & 0.9175 & 0.9132 & 0.9215 & 0.9219 & 0.9361 & 0.9931 \\ \hline
    \end{tabular}
    \label{tab:recall_dev_duoT5}
    \end{sideways}
\end{table}

\begin{table}[!ht]
    \caption{Recall Values for the Hard set Curves of Figure \ref{fig:veiled_duoT5} evaluated against DuoT5 Qrels. No entities is the original BM25 run generated from the non-expanded dataset. Hashed designates the hashed-entity run, RRF1 is the RRF (No entities/ Hashed entities), RRF2 is the	RRF (Hashed entities/ Entities), RRF3 is the RRF (No entities/ Entities), RRF4 is the RRF (No Entities/ Hashed entities/ Entities).} \vspace{0.5cm}
% \hspace*{-2.4cm}
    \centering
    \begin{sideways}
    \scriptsize
    \begin{tabular}{|c|c|c|c|c|c|c|c|c|c|c|}
    \hline
    
    \textbf{Recall} & \textbf{No entities} & \textbf{Hashed} & \textbf{Entities} & \textbf{Classifier} & \textbf{RRF1} & \textbf{RRF2} & \textbf{RRF3} &\textbf{RRF4} & \textbf{Oracle} & \textbf{Dense} \\ [0.5ex]
    \hline
        0 & 0 & 0 & 0 & 0 & 0 & 0 & 0 & 0 & 0 & 0 \\ \hline
        50 & 0.546 & 0.4861 & 0.5284 & 0.5608 & 0.5527 & 0.5281 & 0.563 & 0.5569 & 0.6374 & 0.8753 \\ \hline
        55 & 0.5592 & 0.4998 & 0.5409 & 0.5723 & 0.5726 & 0.5406 & 0.5765 & 0.5726 & 0.6489 & 0.8811 \\ \hline
        60 & 0.571 & 0.5117 & 0.5515 & 0.5835 & 0.5851 & 0.5531 & 0.5896 & 0.5838 & 0.6624 & 0.8862 \\ \hline
        75 & 0.6034 & 0.5393 & 0.5829 & 0.6172 & 0.6175 & 0.5909 & 0.6255 & 0.622 & 0.6935 & 0.9009 \\ \hline
        100 & 0.6486 & 0.5784 & 0.6181 & 0.6624 & 0.656 & 0.6252 & 0.6653 & 0.6646 & 0.7272 & 0.9205 \\ \hline
        200 & 0.7252 & 0.6678 & 0.706 & 0.7400 & 0.7381 & 0.7159 & 0.7441 & 0.7461 & 0.7987 & 0.9481 \\ \hline
        300 & 0.7615 & 0.7169 & 0.7432 & 0.7769 & 0.782 & 0.7627 & 0.7878 & 0.7845 & 0.8298 & 0.9625 \\ \hline
        400 & 0.7881 & 0.7461 & 0.7785 & 0.8044 & 0.8083 & 0.7887 & 0.8121 & 0.8137 & 0.8541 & 0.9686 \\ \hline
        500 & 0.8051 & 0.7733 & 0.7987 & 0.8262 & 0.8246 & 0.8092 & 0.8304 & 0.8342 & 0.8711 & 0.9727 \\ \hline
        600 & 0.8189 & 0.7929 & 0.8163 & 0.8397 & 0.8413 & 0.8269 & 0.8461 & 0.8471 & 0.8814 & 0.9785 \\ \hline
        700 & 0.8301 & 0.806 & 0.8259 & 0.8506 & 0.8496 & 0.8407 & 0.857 & 0.857 & 0.8884 & 0.9795 \\ \hline
        800 & 0.8378 & 0.8153 & 0.8391 & 0.8592 & 0.8612 & 0.8532 & 0.8676 & 0.8679 & 0.8929 & 0.9824 \\ \hline
        900 & 0.8455 & 0.8262 & 0.8474 & 0.8650 & 0.8685 & 0.8596 & 0.8753 & 0.8743 & 0.8984 & 0.9862 \\ \hline
        1000 & 0.8519 & 0.8352 & 0.8557 & 0.8718 & 0.8743 & 0.8676 & 0.8804 & 0.8804 & 0.9038 & 0.9878 \\ \hline
    \end{tabular}
    \label{tab:recall_veiled_duoT5}
    \end{sideways}
\end{table}

\begin{table}[!ht]
    \caption{Recall Values for the Harder set Curves of Figure \ref{fig:pygmy_duoT5} evaluated against DuoT5 Qrels. No entities is the original BM25 run generated from the non-expanded dataset. Hashed designates the hashed-entity run, RRF1 is the RRF (No entities/ Hashed entities), RRF2 is the	RRF (Hashed entities/ Entities), RRF3 is the RRF (No entities/ Entities), RRF4 is the RRF (No Entities/ Hashed entities/ Entities).} \vspace{0.5cm}
% \hspace*{-2.4cm}
    \centering
    \begin{sideways}
    \scriptsize
    \begin{tabular}{|c|c|c|c|c|c|c|c|c|c|c|}
    \hline
    
    \textbf{Recall} & \textbf{No entities} & \textbf{Hashed} & \textbf{Entities} & \textbf{Classifier} & \textbf{RRF1} & \textbf{RRF2} & \textbf{RRF3} &\textbf{RRF4} & \textbf{Oracle} & \textbf{Dense} \\ [0.5ex]
    \hline
        0 & 0 & 0 & 0 & 0 & 0 & 0 & 0 & 0 & 0 & 0 \\ \hline
        50 & 0.5301 & 0.4695 & 0.5091 & 0.5447 & 0.5354 & 0.5091 & 0.5451 & 0.5374 & 0.6207 & 0.867 \\ \hline
        55 & 0.5447 & 0.4832 & 0.5208 & 0.5572 & 0.5532 & 0.52 & 0.5584 & 0.5532 & 0.6328 & 0.8734 \\ \hline
        60 & 0.5552 & 0.4945 & 0.5321 & 0.5669 & 0.5645 & 0.5326 & 0.5726 & 0.5645 & 0.6458 & 0.8795 \\ \hline
        75 & 0.5871 & 0.5216 & 0.5677 & 0.6017 & 0.5985 & 0.5722 & 0.6082 & 0.6021 & 0.6797 & 0.8953 \\ \hline
        100 & 0.6336 & 0.5592 & 0.6029 & 0.6490 & 0.6389 & 0.607 & 0.6498 & 0.6478 & 0.7169 & 0.9175 \\ \hline
        200 & 0.7093 & 0.6518 & 0.6971 & 0.7258 & 0.725 & 0.7032 & 0.7327 & 0.7351 & 0.7861 & 0.9466 \\ \hline
        300 & 0.7481 & 0.7012 & 0.7335 & 0.7638 & 0.7691 & 0.7513 & 0.7764 & 0.7736 & 0.8192 & 0.9616 \\ \hline
        400 & 0.7756 & 0.7335 & 0.7683 & 0.7922 & 0.797 & 0.7788 & 0.8015 & 0.8043 & 0.8447 & 0.9681 \\ \hline
        500 & 0.795 & 0.7606 & 0.7893 & 0.8172 & 0.814 & 0.8011 & 0.8197 & 0.8229 & 0.8637 & 0.9725 \\ \hline
        600 & 0.8099 & 0.7829 & 0.8083 & 0.8318 & 0.8314 & 0.8172 & 0.8358 & 0.8362 & 0.8751 & 0.9782 \\ \hline
        700 & 0.8229 & 0.797 & 0.8176 & 0.8443 & 0.8415 & 0.8318 & 0.8484 & 0.8484 & 0.8839 & 0.9794 \\ \hline
        800 & 0.8318 & 0.8071 & 0.8326 & 0.8540 & 0.854 & 0.8451 & 0.8605 & 0.8609 & 0.8892 & 0.9818 \\ \hline
        900 & 0.8399 & 0.8197 & 0.8419 & 0.8609 & 0.8633 & 0.8528 & 0.8698 & 0.8678 & 0.8949 & 0.9867 \\ \hline
        1000 & 0.8467 & 0.8281 & 0.8524 & 0.8690 & 0.8702 & 0.8613 & 0.8751 & 0.8751 & 0.9001 & 0.9883 \\ \hline
    \end{tabular}
    \label{tab:recall_pygmy_duoT5}
    
    \end{sideways}
\end{table}

\begin{table}[!ht]
   \caption{Recall Values for the Hardest set Curves of Figure \ref{fig:lesser_duoT5} evaluated against DuoT5 Qrels. No entities is the original BM25 run generated from the non-expanded dataset. Hashed designates the hashed-entity run, RRF1 is the RRF (No entities/ Hashed entities), RRF2 is the	RRF (Hashed entities/ Entities), RRF3 is the RRF (No entities/ Entities), RRF4 is the RRF (No Entities/ Hashed entities/ Entities).} \vspace{0.5cm}
% \hspace*{-2.4cm}
    \centering
    \begin{sideways}
    \scriptsize
    \begin{tabular}{|c|c|c|c|c|c|c|c|c|c|c|}
    \hline
    
    \textbf{Recall} & \textbf{No entities} & \textbf{Hashed} & \textbf{Entities} & \textbf{Classifier} & \textbf{RRF1} & \textbf{RRF2} & \textbf{RRF3} &\textbf{RRF4} & \textbf{Oracle} & \textbf{Dense} \\ [0.5ex]
    \hline
        0 & 0 & 0 & 0 & 0 & 0 & 0 & 0 & 0 & 0 & 0 \\ \hline
        50 & 0.5115 & 0.4536 & 0.4855 & 0.5227 & 0.5162 & 0.4914 & 0.5239 & 0.5127 & 0.6031 & 0.8683 \\ \hline
        55 & 0.5275 & 0.466 & 0.4968 & 0.5369 & 0.5322 & 0.5021 & 0.5399 & 0.5316 & 0.6161 & 0.8742 \\ \hline
        60 & 0.5399 & 0.4773 & 0.5062 & 0.5481 & 0.5452 & 0.5145 & 0.5505 & 0.544 & 0.632 & 0.8789 \\ \hline
        75 & 0.5682 & 0.5056 & 0.544 & 0.5818 & 0.58 & 0.5552 & 0.5865 & 0.5871 & 0.668 & 0.8949 \\ \hline
        100 & 0.6143 & 0.5422 & 0.583 & 0.6291 & 0.6249 & 0.5918 & 0.6314 & 0.6338 & 0.7017 & 0.9155 \\ \hline
        200 & 0.6958 & 0.6385 & 0.684 & 0.7118 & 0.7088 & 0.6905 & 0.7212 & 0.7212 & 0.7744 & 0.9451 \\ \hline
        300 & 0.7383 & 0.6869 & 0.7242 & 0.7519 & 0.7566 & 0.7348 & 0.7631 & 0.7608 & 0.8092 & 0.9592 \\ \hline
        400 & 0.769 & 0.7177 & 0.7566 & 0.7838 & 0.788 & 0.7637 & 0.7897 & 0.7939 & 0.8358 & 0.9646 \\ \hline
        500 & 0.7862 & 0.7484 & 0.7779 & 0.8074 & 0.8069 & 0.7891 & 0.8104 & 0.8128 & 0.8541 & 0.9711 \\ \hline
        600 & 0.8009 & 0.7708 & 0.7992 & 0.8234 & 0.8246 & 0.8057 & 0.8269 & 0.8269 & 0.8653 & 0.9764 \\ \hline
        700 & 0.8128 & 0.7874 & 0.808 & 0.8352 & 0.834 & 0.8216 & 0.8393 & 0.8393 & 0.876 & 0.9776 \\ \hline
        800 & 0.821 & 0.7992 & 0.8246 & 0.8452 & 0.8452 & 0.8376 & 0.8512 & 0.8523 & 0.8819 & 0.9799 \\ \hline
        900 & 0.8275 & 0.8104 & 0.8352 & 0.8529 & 0.8529 & 0.8441 & 0.8612 & 0.8606 & 0.8884 & 0.9846 \\ \hline
        1000 & 0.8358 & 0.8204 & 0.8464 & 0.8618 & 0.8612 & 0.8517 & 0.8677 & 0.8653 & 0.8943 & 0.9864 \\ \hline
    \end{tabular}
    \label{tab:recall_lesser_duoT5}
    \end{sideways}
\end{table}

% GLOSSARIES (Lists of definitions, abbreviations, symbols, etc. provided by the glossaries-extra package)
% -----------------------------
\printglossaries
\cleardoublepage
\phantomsection		% allows hyperref to link to the correct page

%----------------------------------------------------------------------
\end{document}